\documentclass[%
reprint,
superscriptaddress,
 amsmath,amssymb,
]{revtex4-2}

\usepackage[normalem]{ulem}
\usepackage{graphicx}
\usepackage{dcolumn}
\usepackage{bm}
\usepackage{hyperref}
\usepackage{float}
\usepackage[all]{hypcap} 
\usepackage{xcolor}
\newcommand{\beginsupplement}{%
        \setcounter{table}{0}
        \renewcommand{\thetable}{S\arabic{table}}%
        \setcounter{figure}{0}
        \renewcommand{\thefigure}{S\arabic{figure}}%
        \setcounter{equation}{0}
        \renewcommand{\theequation}{S\arabic{equation}}
}
\usepackage[mathlines]{lineno}
\usepackage{braket}
\usepackage{tabularx}
\usepackage{siunitx}
\DeclareSIUnit\angstrom{\text{Å}}
\usepackage{upgreek}
\usepackage{xfrac}

\usepackage{relsize}
\usepackage{bigints}
\usepackage{esvect}
\newcommand{\Nabla}{\raise.15ex\hbox{$\boldsymbol{\nabla}$}}
\newcommand{\ols}[1]{\mskip.5\thinmuskip\overline{\mskip-.5\thinmuskip {#1} \mskip-.5\thinmuskip}\mskip.5\thinmuskip} 

\newcommand{\Beplus}{\ensuremath{{^9}{\rm Be}^{+} \,}}

\renewcommand{\pi}{\ensuremath{\uppi}}
\newcommand{\tx}[1]{\ensuremath{\mathrm{#1}}} 
\newcommand{\mbf}[1]{\ensuremath{\mathbf{#1}}} 
\renewcommand{\i}{\ensuremath{\tx{i}}}
\newcommand{\e}{\ensuremath{\tx{e}}}
\renewcommand{\d}{\ensuremath{\tx{d}}}

\renewcommand{\H}{\ensuremath{\hat{\mathcal{H}}}}

\newcommand{\ms}{\ensuremath{m_{s}}}

\newcommand{\rvecop}{\ensuremath{\hat{\mbf{r}}}}

\begin{document}

\preprint{APS/123-QED}

\title{A 3-dimensional scanning trapped-ion probe}

\author{Tobias Sägesser}
\email{tobiass@phys.ethz.ch}
\thanks{These authors contributed equally.}
\affiliation{%
 Department of Physics, ETH Z{\"u}rich, Z{\"u}rich, Switzerland
}%
\affiliation{%
 Quantum Center, ETH Z{\"u}rich, Z{\"u}rich, Switzerland
}%

\author{Shreyans Jain}
\thanks{These authors contributed equally.}
\affiliation{%
 Department of Physics, ETH Z{\"u}rich, Z{\"u}rich, Switzerland
}%
\affiliation{%
 Quantum Center, ETH Z{\"u}rich, Z{\"u}rich, Switzerland
}%

\author{Pavel Hrmo}%
\affiliation{%
 Department of Physics, ETH Z{\"u}rich, Z{\"u}rich, Switzerland
}%
\affiliation{%
 Quantum Center, ETH Z{\"u}rich, Z{\"u}rich, Switzerland
}%

\author{Alexander Ferk}%
\affiliation{%
 Department of Physics, ETH Z{\"u}rich, Z{\"u}rich, Switzerland
}%
\affiliation{%
 Quantum Center, ETH Z{\"u}rich, Z{\"u}rich, Switzerland
}%

\author{Matteo Simoni}%
\affiliation{%
 Department of Physics, ETH Z{\"u}rich, Z{\"u}rich, Switzerland
}%
\affiliation{%
 Quantum Center, ETH Z{\"u}rich, Z{\"u}rich, Switzerland
}%

\author{Yingying Cui}%
\affiliation{%
 Department of Physics, ETH Z{\"u}rich, Z{\"u}rich, Switzerland
}%
\affiliation{%
 Quantum Center, ETH Z{\"u}rich, Z{\"u}rich, Switzerland
}%

\author{Carmelo Mordini}%
\affiliation{%
 Department of Physics, ETH Z{\"u}rich, Z{\"u}rich, Switzerland
}%
\affiliation{%
 Quantum Center, ETH Z{\"u}rich, Z{\"u}rich, Switzerland
}%
\altaffiliation[Present affiliation:]{%
 Dipartimento di Fisica e Astronomia, Universit{\`a} degli Studi di Padova, Padova, Italy
}%

\author{Daniel Kienzler}%
\affiliation{%
 Department of Physics, ETH Z{\"u}rich, Z{\"u}rich, Switzerland
}%
\affiliation{%
 Quantum Center, ETH Z{\"u}rich, Z{\"u}rich, Switzerland
}%

\author{Jonathan Home}%
\affiliation{%
 Department of Physics, ETH Z{\"u}rich, Z{\"u}rich, Switzerland
}%
\affiliation{%
 Quantum Center, ETH Z{\"u}rich, Z{\"u}rich, Switzerland
}%

\date{\today}

\begin{abstract}
Single-atom quantum sensors offer high spatial resolution and high sensitivity to electric and magnetic fields. Among them, trapped ions offer exceptional performance in sensing electric fields, which has been used in particular to probe these in the proximity of metallic surfaces. However, the flexibility of previous work was limited by the use of radio-frequency trapping fields, which has restricted spatial scanning to linear translations, and calls into question whether observed phenomena are connected to the presence of the radio-frequency fields. Here, using a Penning trap instead, we demonstrate a single ion probe which offers three-dimensional position scanning at distances between \SI{50}{\um} and \SI{450}{\um} from a metallic surface and above a $200\times200~ \si{\micro\meter\squared}$ area, allowing us to reconstruct static and time-varying electric as well as magnetic fields. We use this to map charge distributions on the metallic surface and noise stemming from it. The methods demonstrated here allow similar probing to be carried out on samples with a variety of materials, surface constitutions and geometries, providing a new tool for surface science.
\end{abstract}

\maketitle

The use of quantum systems for sensing provides access to high levels of sensitivity and allows further enhancement through the use of entanglement \cite{Degen_2017}. Magnetic and electric fields are among the environmental signals for which quantum sensors are particularly applicable, providing tools for varied topics such as material science \cite{Aigner_2008, brownnuttIontrapMeasurementsElectricfield2015, Wagner_2015}, biological processes in live samples \cite{Jensen_2016, Webb_2021, Aslam_2023} and in fundamental science \cite{Smiciklas_2011, Fu_2014, Ye_2024}. A broad array of systems has been applied to numerous use cases. While various solid state systems provide high sensitivity \cite{Simmonds_1979, Taylor_2008, Toida_2023, Barry_NV_review_2020, Zhang_2021} as well as the ability to be operated in close proximity to samples or be employed as scanning probes \cite{Balasubramanian_2008, Wagner_2015, Wyss_2022}, atomic systems are interesting for their low background levels of noise, high signal-to-noise ratios and the ability to probe with a high spatial resolution. Ensembles of atoms are among the most sensitive magnetometers \cite{Budker_vapour_review_2007, Wildermuth_2005}, while the highest sensitivities to electric fields over a broad range of frequencies have been achieved with ensembles of Rydberg atoms \cite{Simons_2021, Barredo_2020} and trapped ions \cite{Gilmore_2021, Wu_2024}. Furthermore, arrays of individually controlled quantum systems can leverage parallel operations to directly sense spatial distributions of fields by sampling many locations simultaneously \cite{Schaffner_2024}.

Trapped ions, coupled to external electrical potentials through their net charge and to magnetic fields through the magnetic dipole moment, offer excellent prospects for precision sensing \cite{brownnuttIontrapMeasurementsElectricfield2015, Gilmore_2021, Wunderlich_2016, Wu_2024}. The long confinement times allow for long term averaging, while the use of fluorescence detection of electronic states results in excellent signal-to-noise ratios, which can be extended to motional displacements of the center of mass through laser-ion coupling \cite{winelandExperimentalIssuesCoherent1998}. The high degree of control achieved over these systems allows sensitivities beyond classical limits by harnessing entanglement, engineered quantum states, and optimal control. Examples include dynamical decoupling approaches analogous to lock-in amplifiers \cite{Kotler_2011}, entanglement-enhanced measurement \cite{Leibfried_2004, Ruster_2017, Gilmore_2021} as well as the use of quantum-engineered states such as squeezed \cite{Burd_2019} and Schrödinger-cat states \cite{Hempel2013, Milne_2021}. Sensitivities as low as \SI{19}{\nano\volt\per\meter\per\sqrt\hertz} for electric fields \cite{King_2021} and \SI{4.6}{\pico\tesla\per\sqrt\hertz} for magnetic fields \cite{Wunderlich_2016} have been reported. 

Such results raise the prospect of using single ions as scanning probes of electric and magnetic fields, making use of the ability to accurately place them with nanometer-scale resolution \cite{Ruster_2017, Ricci_2023} as well as to translate them across macroscopic distances \cite{Akhtar2023}. In particular, this capability has been desired as a means to probe electric-field noise stemming from metallic (electrode) surfaces and contaminants, which hinders the ability to operate small ion traps for quantum computing and simulation \cite{Brown_2021, deLeon_2021}. However, previous experiments using ions at distances of a few tens of microns from surfaces used radio-frequency confining fields, which have restricted translation of the ions to one-dimensional displacements \cite{Kaufmann_2018}. The use of a radio-frequency drive also raised the question whether observed behaviors are intrinsic to the materials or result from the drive \cite{Brown_2021}.

In this article, we demonstrate sensing using a single ion translated in 3-d, above a $\SI{200}{\micro\meter}\times\SI{200}{\micro\meter}$ area of the trap and at distances between \SI{50}{\um} and \SI{450}{\um} from the surface. The key to 3-d positioning control is the use of a Penning trap, for which the trap center can be displaced arbitrarily \cite{Jain2020}. We showcase our 3-d trapped-ion scanning probe by measuring static and fluctuating electric and magnetic fields utilizing both classical techniques as well as quantum sensing protocols. The resulting data provides tests of models of electric field noise above metallic surfaces as well as insights into mechanisms of surface charging.

The experimental apparatus consists of a micro-fabricated surface-electrode trap within a cryogenic vacuum apparatus, described by \citeauthor{jain_penning_2024}. The trap chip consists of 25 gold electrodes deposited on a sapphire substrate, placed within a $B=\SI{3}{\tesla}$ static magnetic field oriented along $z$. Suitable voltages applied to all electrodes approximate an electrical potential $\phi(\mbf{r}) = m\omega_z^2\left(2(z-z_0)^2 - (x-x_0)^2- (y-y_0)^2\right)/(4e)$ with a field null at a position $\mbf{r_0}=(x_0, y_0, z_0)$ at which we trap a single \Beplus ion. The confining curvature in the $z$ direction results in an axial motional frequency of $\omega_z$, while the ion is confined in the radial $x$-$y$-plane by the magnetic field. The combined electric and magnetic fields give rise to the magnetron and modified cyclotron modes of motion at angular frequencies $\omega_\pm = \omega_c/2 \pm \sqrt{(\omega_c^2-2\omega_z^2)}/2$, where $\omega_c=eB/m$ is the bare cyclotron frequency.

\begin{figure}[ht]
\resizebox{246pt}{!}{\includegraphics{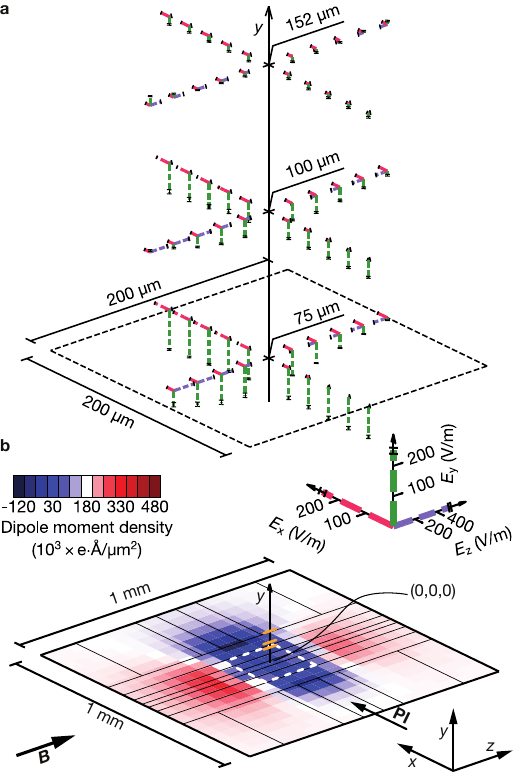}}
\caption{\label{fig:StrayFields}\textbf{Sensing of static electric fields. }\textbf{a.} Stray electric fields measured in 3-d above a 
$\SI{200}{\um}\times\SI{200}{\um}$ area. The three spatial components of the fields are drawn as shown in the magnified scale indication, with one dash corresponding to \SI{100}{\volt\per\meter} (\SI{200}{\volt\per\meter}) in the $x$ and $y$ ($z$) directions. Error bars are as detailed in the SI. At a given ion--electrode distance (not to scale), measurement locations are spaced by \SI{20}{\micro\meter}. \textbf{b.} Inferred distribution of the dipole-moment density on a $\SI{1}{\mm}\times\SI{1}{\mm}$ surface area, overlaid over the trap electrode layout. Note that the white color corresponds to \SI{180e3}{\elementarycharge\angstrom\per\um\squared}. The white dashed rectangle indicates the measurement region, while the yellow dashes show the three ion heights to scale. The coordinate system, its origin, the magnetic field and the path of the PI laser are indicated.}
\end{figure}

We first characterize the static electric fields present in our trap across a 3-d grid of positions. The method involves transporting the ion to a desired location and subsequently performing fluorescence detection using a camera, recording the resulting position on a camera and the shape of the point spread function. If electric fields beyond those included in our trap model are present, varying the curvature of the applied potential $\phi$ results in a shift of the observed ion position and a distortion of its shape. To calibrate these additional fields, we iteratively apply correction fields until both the shift and distortion are minimized (more details are included in the SI). Fig.~\ref{fig:StrayFields}a shows the stray electric fields measured above the chip surface. We achieve accuracies in the range of \SIrange{1}{20}{\volt\per\meter}, with the measurement being most precise for fields along $x$ while out-of-plane fields are determined least accurately. This magnitude is comparable to the fields of tens of elementary charges at a distance of \SI{100}{\micro\meter}. The sensitivity of this method improves with low ion mass, high imaging magnification, small camera pixels and weak electrical confinement.

Assuming that the observed fields are generated by charges on the trap surface, we use a model consisting of electric dipoles, which could be imagined to arise from photo-induced charges interacting with the surface \cite{Wang_2011, Harlander_2010}. Contact potentials between the gold surface and adsorbed contaminants also result in dipolar fields \cite{Leung_2003, Tauschinsky_2010}. The electron distribution at the surface is distorted by the adsorbates, in particular if the involved species differ in electronegativity \cite{McGuirk_2004}. We infer the dipole distribution which yields the best fit to the measured fields and present the result in fig.~\ref{fig:StrayFields}b. In the region where the \SI{235}{\nano\meter} photo-ionization (PI) laser passes over the trap along $x$, a strip of negative dipoles is observed, which is in line with previously observed laser-induced charging \cite{Harlander_2010}. The data is best explained when allowing a uniform background dipole moment density $D$ of \SI{180e3}{\elementarycharge\angstrom\per\um\squared}. We speculate that surface dipoles due to deposited beryllium atoms or other contaminants may be the cause. Converting $D$ to a difference in work function $\Delta\phi = -eD/\epsilon_0$ yields $\Delta\phi\approx\SI{-0.33}{\eV}$, compatible with the difference between beryllium and gold, but also with sub-monolayer hydrocarbon contamination \cite{Hite_2021}.

Similar visualization of electrical potentials is enabled at the atomic scale by scanning-probe sensors using NV centers \cite{Bian_2021} or quantum dots \cite{Wagner_2019}, but is restricted to working distances on the order of nanometers and measurement precision on the order of \SI{1e6}{\V\per\meter} \cite{Zhang2023}. Trapped ions in Penning micro-traps as presented here enable nanometer-scale spatial resolution in 3-d, precision of a few \si{\V\per\meter} as well as macroscopic ranges, enabling the study of charge distributions of 2-d or 3-d structures with micrometer-scale extents \cite{Qiu_2022, Ong_2020}.

\begin{figure*}[ht]
\centering
\resizebox{510pt}{!}{\includegraphics{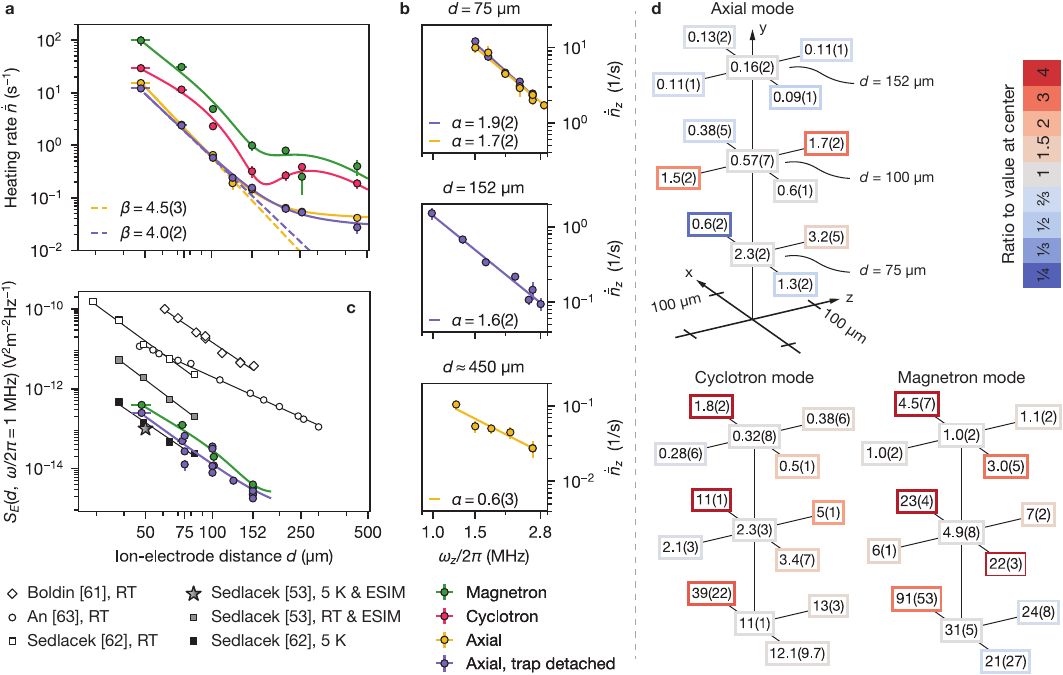}}
\caption{\label{fig:hr}\textbf{Sensing of electric-field noise. }\textbf{a. }Heating rates of all three modes depending on the ion--electrode distance $d$, measured above the trap center. Solid lines are fits to the data as explained in the main text, with dashed lines showing the surface-noise component. Mode frequencies are kept at $\omega_z = 2\pi\times\SI{2.6}{\mega\hertz}$, $\omega_- = 2\pi\times\SI{0.845}{\mega\hertz}$ and $\omega_+ = 2\pi\times\SI{4.32}{\mega\hertz}$, with the radial modes chosen to avoid noise at discrete frequencies (see SI). Axial data is taken with and without the trap electrodes detached. Below \SI{450}{\um}, $d$ was verified independently, resulting in the shown error bars (see SI). \textbf{b. }Frequency scaling of the axial heating rates measured at $d=\SI{75}{\micro\meter}$, \SI{152}{\micro\meter} and $\approx\SI{450}{\micro\meter}$. Solid lines are power law fits, yielding the exponent $\alpha$. \textbf{c. }Spectral noise densities inferred from the heating rates (see SI) and rescaled to $\omega=2\pi\times\SI{1}{\MHz}$ using $\alpha=1.7$. Data where surface noise likely dominates is shown, including from \textbf{d}. Previous investigations with variable distance $d$ are added for comparison, rescaled using the reported exponents $\alpha$. Open symbols indicate data taken with traps at room temperature (RT), while closed symbols correspond to cryogenic traps. Grey symbols show data taken after \textit{ex-situ} ion-milling. \textbf{d. }Heating rates measured on a 3-d grid of positions above a $\SI{200}{\um}\times\SI{200}{\um}$ area at heights $d=\SI{152}{\micro\meter}$, \SI{100}{\micro\meter} and \SI{75}{\micro\meter}, stated in phonons per second. Border colors indicate the ratio to the heating rate measured at the central position at the same distance $d$. The mode frequencies are the same as in \textbf{a}. Error bars correspond to the standard error in all panels.}
\end{figure*}

Besides being affected by static fields, trapped ions also exhibit enhanced sensitivity to fluctuating electric fields at frequencies resonant with the ion motion. Such electric fields excite the oscillatory motion, which can be measured by coupling it to the electronic states using suitable laser or microwave fields, before detecting the latter using state-dependent fluorescence \cite{winelandExperimentalIssuesCoherent1998}. A long-standing challenge for surface science is that measured noise spectral densities above metallic electrodes are generally orders of magnitude higher than expected from Johnson noise \cite{Monroe1995}, even after surface cleaning \cite{Sedlacek_2018_Evidence}. There has been much effort to reveal the source of such noise \cite{brownnuttIontrapMeasurementsElectricfield2015, Brown_2021}, with candidate models including fluctuating patch potentials \cite{Low_2011}, dielectric layers \cite{Kumph2016,Martinetz_2022}, surface contaminants and adatoms \cite{Safavi-Naini_2011, Kim_2017, Foulon_2022} as well as two-level fluctuators with unknown physical realization \cite{Noel_2019}. However, no single model fits the vast body of data, which is complicated by a large number of combinations of surface materials, temperatures, fabrication methods, geometries and ion species. The electric-field noise spectral density $S_\mathrm{E}$ is typically assumed to show power-law scaling with the angular frequency $\omega$, the distance to the surface $d$ and the surface temperature $T$: $S_\mathrm{E} \propto \omega^{-\alpha} d^{-\beta} T^{\gamma}$. Models predict specific values for some or all of the scaling exponents $\alpha$, $\beta$ and $\gamma$. Aside from the ability to vary multiple of these parameters as well as the location in the plane above the surface, the Penning trap measurements reported here are notable for the absence of high-voltage radio-frequency trapping fields and for the ability to detach all trap electrodes from the voltage sources during measurements, removing 
external noise present in the out-of-vacuum wiring \cite{jain_penning_2024}.

To measure electric-field noise, an ion is first prepared in the ground state of the motional mode of interest $\lambda$ and then transported to a target location for a variable wait time. A sideband probe pulse and subsequent fluorescence detection determine the average phonon number $\bar{n}_\lambda$ in the mode \cite{Monroe1995}. A linear fit to the phonon number as a function of the wait time yields the heating rate $\dot{\bar{n}}_\lambda$, which can be converted to the power spectral density of the electric field noise through $S_{\mathrm{E},z}(\omega_z)=4\hbar m \omega_z\Dot{\Bar{n}}_z/e^2$ for the axial mode and through $S_{\mathrm{E},r}(\omega_\pm)=4\hbar m (\omega_+ - \omega_-)\Dot{\Bar{n}}_\pm/e^2$ for the radial modes.

We investigate the scaling of the noise with the ion--electrode distance $d$ and make a set of measurements at 8 different heights above the center of the trap. As expected from work in radio-frequency traps, the data presented in fig \ref{fig:hr}a shows increased noise affecting all modes as the ion is moved closer to the surface. At ion--electrode distances below \SI{152}{\um}, this increase broadly takes a power-law form while the dependence on the height is less strong at larger separations. We explain these observations by assuming that the noise stems from several sources which may differ in their position dependence. One component is assumed to arise from surface effects with power-law scaling $d^{-\beta}$, identified with noise from microscopic processes on the trap surface. In addition we add sources due to thermal noise in resistive elements of the trap circuitry as well as noise from technical equipment external to the apparatus. For the axial mode, little difference is found in the heating rates when additionally detaching the electrodes from the voltage sources. This indicates that external noise affecting the axial mode may be picked up by elements between the trap and the switches or be caused by direct electromagnetic interference. Finding the strengths of the various contributions which explain the data best leads to the fits shown in fig.~\ref{fig:hr}a (see SI). For the axial mode, surface noise is the dominant source below heights of \SI{152}{\um}, with the extracted scaling being $\beta=4.0(2)$. This scaling exponent is consistent with $\beta=4$, as would be expected for noise originating from microscopic fluctuators on the surface \cite{brownnuttIontrapMeasurementsElectricfield2015}. We further determine the frequency-scaling exponent $\alpha$ at three different heights by measuring the axial heating rate $\Dot{\Bar{n}}_z$ at a range of different frequencies $\omega_z$, for which the results are shown in fig.~\ref{fig:hr}b. At $d=\SI{75}{\um}$ and $\SI{152}{\um}$, the scaling exponent is consistent with $\alpha \approx 1.7$ while $\alpha=0.6(3)$ is measured at $d\approx\SI{450}{\um}$, again indicating the dominance of different noise sources depending on location. While many models predict $1/f$-scaling which corresponds to $\alpha \approx1$, the bulk of experimental evidence \cite{brownnuttIontrapMeasurementsElectricfield2015} finds $\alpha$ between 0 and 2. 

A similar treatment is carried out for the radial modes. For both of these the heating is well above that of the axial mode. We find that this is likely due to voltage fluctuations of technical origin related to different filtering between the electrodes used for axialization (which are placed in the radial direction, but extend along the axis of the trap) and the other axial electrodes close to the ion - this produces negligible heating at a symmetric point $\SI{152}{\um}$ above the surface, and increases away from this point. This technical noise is comparable in strength to the surface noise component, for which we obtain scaling exponents $\beta=4.1(6)$ for the magnetron data and $\beta=3.5(6)$ for the cyclotron data. 

Converting the measured heating rates to noise spectral densities and using the determined value of $\alpha$ to rescale to a common frequency $\omega = 2\pi\times\SI{1}{\MHz}$ allows a comparison of our results to prior work in radio-frequency traps. Fig.~\ref{fig:hr}c shows the portion of our data which is dominated by surface noise, as well as results obtained in previous experimental efforts in rf traps with adjustable ion--electrode distance. The noise levels reported here are comparable to the lowest values found in other experiments and the distance-scaling exponent is well in line with previous experimental investigations in rf traps \cite{Boldin_2018, Sedlacek_2018_Distance, Sedlacek_2018_Evidence, An_2019, McKay_2021}. It is apparent from this comparison that the absence of a high-voltage rf drive in our apparatus does not necessarily yield a reduction in noise beyond what is observed in rf traps.

We further utilize 3-d transport to sample the heating rates of all motional modes on the same grid of positions used for the stray-field measurements, with the results shown in fig.~\ref{fig:hr}d. Besides the increase in heating rates when approaching the surface, some structure is also visible within the planes parallel to the electrodes. As similarly concluded using cantilever probes sampling at distances around \SI{100}{\nano\meter} \cite{Heritier_2021}, surface noise can have significant spatial variation, which may further depend on the frequency and polarization of interest. Focusing on the axial mode, the heating rates at $d=\SI{75}{\um}$ increase by half when moving \SI{100}{\um} along $z$, while reducing to one-fourth when displaced by the same amount along $x$. Using the cyclotron and magnetron modes, increased noise is observed when displaced perpendicular to the axial direction, however the majority of this can be accounted for by thermal and technical noise (see SI). This 3-d spatial resolution paves the way for investigations into surface science where electric-field noise is probed depending on all three relevant variables $\omega$, $d$ and $T$ and involving multiple materials and surface constitutions.

\begin{figure}[ht]
\resizebox{246pt}{!}{\includegraphics{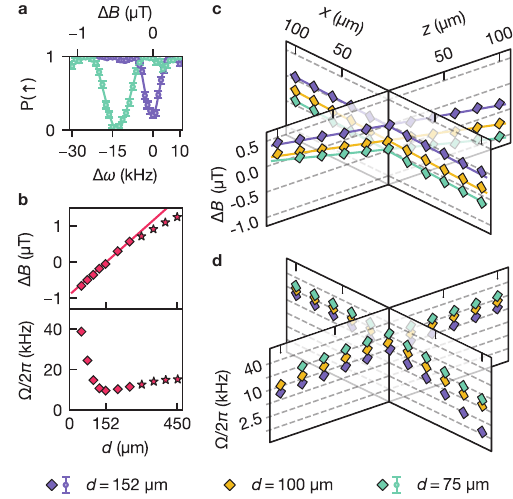}}
\caption{\label{fig:mmw}\textbf{Sensing of magnetic fields. }\textbf{a.} Population $\mathrm{P}(\uparrow )$ after applying a microwave pulse to an ion initially in $\ket{\uparrow}$. Data in purple (turquoise) correspond to the ion being \SI{152}{\micro\meter} (\SI{75}{\micro\meter}) from the surface. Frequency shifts are measured relative to the resonance frequency $\omega_0$ at position $(0, \SI{152}{\micro\meter}, 0)$. \textbf{b. } Change in the magnetic field and Rabi rate of the microwave drive measured for several ion--electrode distances $d$ between \SI{50}{\um} and \SI{450}{\um}. Star symbols indicate that $d$ was not verified and no stray-field correction was applied. The solid line is a linear fit to the data with verified distance. \textbf{c. }Magnetic field shifts within planes parallel to the trap surface at three heights. Solid lines are linear fits, revealing field gradients between \SI{-3.59(0.06)}{\nano\tesla\per\micro\meter} and \SI{-5.26(0.08)}{\nano\tesla\per\micro\meter} along $z$ and less than \SI{0.74(0.04)}{\nano\tesla\per\micro\meter} along $x$. \textbf{d. }Spatial dependence of the Rabi rate of the microwave drive. Error bars are smaller than the markers for panels \textbf{b}-\textbf{d}.}
\end{figure}

In addition to electric field sensing, we use the sensitivity of the electron spin of the ion to probe magnetic fields. This is performed using Rabi spectroscopy with a microwave field close to 83.2 GHz. Simultaneous with probing of the frequency of the spin-flip resonance, we also measure the Rabi frequency of the transition, probing in addition the microwave field strength perpendicular to the quantization axis. Example data for positions (0, \SI{152}{\um}, 0) and (0, \SI{75}{\um}, 0) are shown in fig.~\ref{fig:mmw}a, indicating a frequency shift as well as broadening due to a variation in the microwave field strength. Fig.~\ref{fig:mmw}b shows the variation of both the static and microwave field strengths at (x, z) = (0, 0) as a function of the height of the ion above the surface. The magnetic field shows a gradient of \SI{5.87(0.16)}{\nano\tesla\per\micro\meter}, while the microwave field shows a rapid increase in strength towards the surface. In fig.~\ref{fig:mmw}c and d we show variations in the three planes at different heights above the surface. Here we see that the static field shows a gradient mainly along $z$, while the microwave field strength is greatly reduced at position (\SI{-100}{\um}, \SI{152}{\um}, 0). The measured static field gradients are about two orders of magnitude larger than the empty-bore magnet inhomogeneity specified by the manufacturer, with materials placed in the bore a likely cause of the field distortions. While the microwave radiation is expected to show standing-wave behavior along its propagation direction ($z$) due to a Faraday cage surrounding the trap, the rapid variation of the microwave field strength along the $x$ and $y$ directions are unexpected and likely caused by interaction of the microwave radiation with the metal surface. Such measurement capability may aid the design and characterization of near-field microwave devices \cite{Warring_2013, Fan_14, Weber_2024}.

The 3-d positioning available in Penning micro-traps extends trapped-ion sensing into a versatile tool with a variety of applications where spatially resolved signals are of interest. Several routes of development can be pursued. Combining such a trapped-ion sensor with controlled deposition of contaminants as well as \textit{in-situ} surface cleaning would result in a general-purpose apparatus for surface science investigations. Efforts at realizing quantum computing, both using trapped ions and other platforms, are stifled by issues of material selection \cite{deLeon_2021}. Probing surfaces and validating manufacturing processes, as well as characterizing complex structures such as ion traps with integrated photonics and microwave sources stands to be an important technological step toward scaled quantum computers. More immediately, the interaction of laser light with a trap surface can be studied in a targeted way. Correlated sensing of noise using multiple ions can be realized \cite{Galve_2017} as well as further investigations into surface noise, such as the effect on motional coherence \cite{Talukdar_2016}.

\section*{Acknowledgements}
This project received funding from the European Research Council (ERC) under the European Union’s Horizon 2020 research and innovation program grant agreement No 818195, the Swiss National Science Foundation (SNF) under Grant No. 200020\_207334 and from the Swiss State Secretariat for Education, Research and Innovation (SERI), under contract UeM029-6.1.
T. S. and S. J. thank Michael Marti for designing the trap-voltage amplifier board, Robin Oswald for designing the ex-cryo filterboard, Martin Stadler for assistance with the control system and Andreas Stuker for manufacturing of mechanical components. The authors thank Amado Bautista-Salvador and Christian Ospelkaus for fabricating the trap chip. The authors express their gratitude towards Jeremy Flannery for careful assessment of the manuscript, as well as to Alexander Eichler and Antonius Armanious for helpful discussions.

\section*{Author contributions}
Data taking was performed by T. S., S. J. and Y. C. Data analysis was performed by T. S., S. J. and M. S. The apparatus was primarily built by S. J. and T. S. with contributions from A. F., C. M. and M. S. The manuscript was written by T. S., S. J. and J. H. with input from all authors. The work was supervised by P. H., D. K. and J. H.

\section{Competing interests}
T. S., S. J., P. H., D. K. and J. H. are associated with ZuriQ AG, a commercially oriented quantum computing company. The other authors declare no competing interests.

\bibliography{refs.bib}

\begin{thebibliography}{84}%
\makeatletter
\providecommand \@ifxundefined [1]{%
 \@ifx{#1\undefined}
}%
\providecommand \@ifnum [1]{%
 \ifnum #1\expandafter \@firstoftwo
 \else \expandafter \@secondoftwo
 \fi
}%
\providecommand \@ifx [1]{%
 \ifx #1\expandafter \@firstoftwo
 \else \expandafter \@secondoftwo
 \fi
}%
\providecommand \natexlab [1]{#1}%
\providecommand \enquote  [1]{``#1''}%
\providecommand \bibnamefont  [1]{#1}%
\providecommand \bibfnamefont [1]{#1}%
\providecommand \citenamefont [1]{#1}%
\providecommand \href@noop [0]{\@secondoftwo}%
\providecommand \href [0]{\begingroup \@sanitize@url \@href}%
\providecommand \@href[1]{\@@startlink{#1}\@@href}%
\providecommand \@@href[1]{\endgroup#1\@@endlink}%
\providecommand \@sanitize@url [0]{\catcode `\\12\catcode `\$12\catcode
  `\&12\catcode `\#12\catcode `\^12\catcode `\_12\catcode `\%12\relax}%
\providecommand \@@startlink[1]{}%
\providecommand \@@endlink[0]{}%
\providecommand \url  [0]{\begingroup\@sanitize@url \@url }%
\providecommand \@url [1]{\endgroup\@href {#1}{\urlprefix }}%
\providecommand \urlprefix  [0]{URL }%
\providecommand \Eprint [0]{\href }%
\providecommand \doibase [0]{https://doi.org/}%
\providecommand \selectlanguage [0]{\@gobble}%
\providecommand \bibinfo  [0]{\@secondoftwo}%
\providecommand \bibfield  [0]{\@secondoftwo}%
\providecommand \translation [1]{[#1]}%
\providecommand \BibitemOpen [0]{}%
\providecommand \bibitemStop [0]{}%
\providecommand \bibitemNoStop [0]{.\EOS\space}%
\providecommand \EOS [0]{\spacefactor3000\relax}%
\providecommand \BibitemShut  [1]{\csname bibitem#1\endcsname}%
\let\auto@bib@innerbib\@empty
\bibitem [{\citenamefont {Degen}\ \emph {et~al.}(2017)\citenamefont {Degen},
  \citenamefont {Reinhard},\ and\ \citenamefont {Cappellaro}}]{Degen_2017}%
  \BibitemOpen
  \bibfield  {author} {\bibinfo {author} {\bibfnamefont {C.~L.}\ \bibnamefont
  {Degen}}, \bibinfo {author} {\bibfnamefont {F.}~\bibnamefont {Reinhard}},\
  and\ \bibinfo {author} {\bibfnamefont {P.}~\bibnamefont {Cappellaro}},\
  }\bibfield  {title} {\bibinfo {title} {Quantum sensing},\ }\href
  {https://doi.org/10.1103/RevModPhys.89.035002} {\bibfield  {journal}
  {\bibinfo  {journal} {Rev. Mod. Phys.}\ }\textbf {\bibinfo {volume} {89}},\
  \bibinfo {pages} {035002} (\bibinfo {year} {2017})}\BibitemShut {NoStop}%
\bibitem [{\citenamefont {Aigner}\ \emph {et~al.}(2008)\citenamefont {Aigner},
  \citenamefont {Pietra}, \citenamefont {Japha}, \citenamefont {Entin-Wohlman},
  \citenamefont {David}, \citenamefont {Salem}, \citenamefont {Folman},\ and\
  \citenamefont {Schmiedmayer}}]{Aigner_2008}%
  \BibitemOpen
  \bibfield  {author} {\bibinfo {author} {\bibfnamefont {S.}~\bibnamefont
  {Aigner}}, \bibinfo {author} {\bibfnamefont {L.~D.}\ \bibnamefont {Pietra}},
  \bibinfo {author} {\bibfnamefont {Y.}~\bibnamefont {Japha}}, \bibinfo
  {author} {\bibfnamefont {O.}~\bibnamefont {Entin-Wohlman}}, \bibinfo {author}
  {\bibfnamefont {T.}~\bibnamefont {David}}, \bibinfo {author} {\bibfnamefont
  {R.}~\bibnamefont {Salem}}, \bibinfo {author} {\bibfnamefont
  {R.}~\bibnamefont {Folman}},\ and\ \bibinfo {author} {\bibfnamefont
  {J.}~\bibnamefont {Schmiedmayer}},\ }\bibfield  {title} {\bibinfo {title}
  {Long-range order in electronic transport through disordered metal films},\
  }\href {https://doi.org/10.1126/science.1152458} {\bibfield  {journal}
  {\bibinfo  {journal} {Science}\ }\textbf {\bibinfo {volume} {319}},\ \bibinfo
  {pages} {1226} (\bibinfo {year} {2008})}\BibitemShut {NoStop}%
\bibitem [{\citenamefont {Brownnutt}\ \emph {et~al.}(2015)\citenamefont
  {Brownnutt}, \citenamefont {Kumph}, \citenamefont {Rabl},\ and\ \citenamefont
  {Blatt}}]{brownnuttIontrapMeasurementsElectricfield2015}%
  \BibitemOpen
  \bibfield  {author} {\bibinfo {author} {\bibfnamefont {M.}~\bibnamefont
  {Brownnutt}}, \bibinfo {author} {\bibfnamefont {M.}~\bibnamefont {Kumph}},
  \bibinfo {author} {\bibfnamefont {P.}~\bibnamefont {Rabl}},\ and\ \bibinfo
  {author} {\bibfnamefont {R.}~\bibnamefont {Blatt}},\ }\bibfield  {title}
  {\bibinfo {title} {Ion-trap measurements of electric-field noise near
  surfaces},\ }\href {https://doi.org/10.1103/RevModPhys.87.1419} {\bibfield
  {journal} {\bibinfo  {journal} {Reviews of Modern Physics}\ }\textbf
  {\bibinfo {volume} {87}},\ \bibinfo {pages} {1419} (\bibinfo {year}
  {2015})}\BibitemShut {NoStop}%
\bibitem [{\citenamefont {Wagner}\ \emph {et~al.}(2015)\citenamefont {Wagner},
  \citenamefont {Green}, \citenamefont {Leinen}, \citenamefont {Deilmann},
  \citenamefont {Kr\"uger}, \citenamefont {Rohlfing}, \citenamefont {Temirov},\
  and\ \citenamefont {Tautz}}]{Wagner_2015}%
  \BibitemOpen
  \bibfield  {author} {\bibinfo {author} {\bibfnamefont {C.}~\bibnamefont
  {Wagner}}, \bibinfo {author} {\bibfnamefont {M.~F.~B.}\ \bibnamefont
  {Green}}, \bibinfo {author} {\bibfnamefont {P.}~\bibnamefont {Leinen}},
  \bibinfo {author} {\bibfnamefont {T.}~\bibnamefont {Deilmann}}, \bibinfo
  {author} {\bibfnamefont {P.}~\bibnamefont {Kr\"uger}}, \bibinfo {author}
  {\bibfnamefont {M.}~\bibnamefont {Rohlfing}}, \bibinfo {author}
  {\bibfnamefont {R.}~\bibnamefont {Temirov}},\ and\ \bibinfo {author}
  {\bibfnamefont {F.~S.}\ \bibnamefont {Tautz}},\ }\bibfield  {title} {\bibinfo
  {title} {Scanning quantum dot microscopy},\ }\href
  {https://doi.org/10.1103/PhysRevLett.115.026101} {\bibfield  {journal}
  {\bibinfo  {journal} {Phys. Rev. Lett.}\ }\textbf {\bibinfo {volume} {115}},\
  \bibinfo {pages} {026101} (\bibinfo {year} {2015})}\BibitemShut {NoStop}%
\bibitem [{\citenamefont {Jensen}\ \emph {et~al.}(2016)\citenamefont {Jensen},
  \citenamefont {Budvytyte}, \citenamefont {Thomas}, \citenamefont {Wang},
  \citenamefont {Fuchs}, \citenamefont {Balabas}, \citenamefont {Vasilakis},
  \citenamefont {Mosgaard}, \citenamefont {St{\ae}rkind}, \citenamefont
  {M{\"u}ller}, \citenamefont {Heimburg}, \citenamefont {Olesen},\ and\
  \citenamefont {Polzik}}]{Jensen_2016}%
  \BibitemOpen
  \bibfield  {author} {\bibinfo {author} {\bibfnamefont {K.}~\bibnamefont
  {Jensen}}, \bibinfo {author} {\bibfnamefont {R.}~\bibnamefont {Budvytyte}},
  \bibinfo {author} {\bibfnamefont {R.~A.}\ \bibnamefont {Thomas}}, \bibinfo
  {author} {\bibfnamefont {T.}~\bibnamefont {Wang}}, \bibinfo {author}
  {\bibfnamefont {A.~M.}\ \bibnamefont {Fuchs}}, \bibinfo {author}
  {\bibfnamefont {M.~V.}\ \bibnamefont {Balabas}}, \bibinfo {author}
  {\bibfnamefont {G.}~\bibnamefont {Vasilakis}}, \bibinfo {author}
  {\bibfnamefont {L.~D.}\ \bibnamefont {Mosgaard}}, \bibinfo {author}
  {\bibfnamefont {H.~C.}\ \bibnamefont {St{\ae}rkind}}, \bibinfo {author}
  {\bibfnamefont {J.~H.}\ \bibnamefont {M{\"u}ller}}, \bibinfo {author}
  {\bibfnamefont {T.}~\bibnamefont {Heimburg}}, \bibinfo {author}
  {\bibfnamefont {S.-P.}\ \bibnamefont {Olesen}},\ and\ \bibinfo {author}
  {\bibfnamefont {E.~S.}\ \bibnamefont {Polzik}},\ }\bibfield  {title}
  {\bibinfo {title} {Non-invasive detection of animal nerve impulses with an
  atomic magnetometer operating near quantum limited sensitivity},\ }\href
  {https://doi.org/10.1038/srep29638} {\bibfield  {journal} {\bibinfo
  {journal} {Scientific Reports}\ }\textbf {\bibinfo {volume} {6}},\ \bibinfo
  {pages} {29638} (\bibinfo {year} {2016})}\BibitemShut {NoStop}%
\bibitem [{\citenamefont {Webb}\ \emph {et~al.}(2021)\citenamefont {Webb},
  \citenamefont {Troise}, \citenamefont {Hansen}, \citenamefont {Olsson},
  \citenamefont {Wojciechowski}, \citenamefont {Achard}, \citenamefont
  {Brinza}, \citenamefont {Staacke}, \citenamefont {Kieschnick}, \citenamefont
  {Meijer}, \citenamefont {Thielscher}, \citenamefont {Perrier}, \citenamefont
  {Berg-S{\o}rensen}, \citenamefont {Huck},\ and\ \citenamefont
  {Andersen}}]{Webb_2021}%
  \BibitemOpen
  \bibfield  {author} {\bibinfo {author} {\bibfnamefont {J.~L.}\ \bibnamefont
  {Webb}}, \bibinfo {author} {\bibfnamefont {L.}~\bibnamefont {Troise}},
  \bibinfo {author} {\bibfnamefont {N.~W.}\ \bibnamefont {Hansen}}, \bibinfo
  {author} {\bibfnamefont {C.}~\bibnamefont {Olsson}}, \bibinfo {author}
  {\bibfnamefont {A.~M.}\ \bibnamefont {Wojciechowski}}, \bibinfo {author}
  {\bibfnamefont {J.}~\bibnamefont {Achard}}, \bibinfo {author} {\bibfnamefont
  {O.}~\bibnamefont {Brinza}}, \bibinfo {author} {\bibfnamefont
  {R.}~\bibnamefont {Staacke}}, \bibinfo {author} {\bibfnamefont
  {M.}~\bibnamefont {Kieschnick}}, \bibinfo {author} {\bibfnamefont
  {J.}~\bibnamefont {Meijer}}, \bibinfo {author} {\bibfnamefont
  {A.}~\bibnamefont {Thielscher}}, \bibinfo {author} {\bibfnamefont {J.-F.}\
  \bibnamefont {Perrier}}, \bibinfo {author} {\bibfnamefont {K.}~\bibnamefont
  {Berg-S{\o}rensen}}, \bibinfo {author} {\bibfnamefont {A.}~\bibnamefont
  {Huck}},\ and\ \bibinfo {author} {\bibfnamefont {U.~L.}\ \bibnamefont
  {Andersen}},\ }\bibfield  {title} {\bibinfo {title} {Detection of biological
  signals from a live mammalian muscle using an early stage diamond quantum
  sensor},\ }\href {https://doi.org/10.1038/s41598-021-81828-x} {\bibfield
  {journal} {\bibinfo  {journal} {Scientific Reports}\ }\textbf {\bibinfo
  {volume} {11}},\ \bibinfo {pages} {2412} (\bibinfo {year}
  {2021})}\BibitemShut {NoStop}%
\bibitem [{\citenamefont {Aslam}\ \emph {et~al.}(2023)\citenamefont {Aslam},
  \citenamefont {Zhou}, \citenamefont {Urbach}, \citenamefont {Turner},
  \citenamefont {Walsworth}, \citenamefont {Lukin},\ and\ \citenamefont
  {Park}}]{Aslam_2023}%
  \BibitemOpen
  \bibfield  {author} {\bibinfo {author} {\bibfnamefont {N.}~\bibnamefont
  {Aslam}}, \bibinfo {author} {\bibfnamefont {H.}~\bibnamefont {Zhou}},
  \bibinfo {author} {\bibfnamefont {E.~K.}\ \bibnamefont {Urbach}}, \bibinfo
  {author} {\bibfnamefont {M.~J.}\ \bibnamefont {Turner}}, \bibinfo {author}
  {\bibfnamefont {R.~L.}\ \bibnamefont {Walsworth}}, \bibinfo {author}
  {\bibfnamefont {M.~D.}\ \bibnamefont {Lukin}},\ and\ \bibinfo {author}
  {\bibfnamefont {H.}~\bibnamefont {Park}},\ }\bibfield  {title} {\bibinfo
  {title} {Quantum sensors for biomedical applications},\ }\href
  {https://doi.org/10.1038/s42254-023-00558-3} {\bibfield  {journal} {\bibinfo
  {journal} {Nature Reviews Physics}\ }\textbf {\bibinfo {volume} {5}},\
  \bibinfo {pages} {157} (\bibinfo {year} {2023})}\BibitemShut {NoStop}%
\bibitem [{\citenamefont {Smiciklas}\ \emph {et~al.}(2011)\citenamefont
  {Smiciklas}, \citenamefont {Brown}, \citenamefont {Cheuk}, \citenamefont
  {Smullin},\ and\ \citenamefont {Romalis}}]{Smiciklas_2011}%
  \BibitemOpen
  \bibfield  {author} {\bibinfo {author} {\bibfnamefont {M.}~\bibnamefont
  {Smiciklas}}, \bibinfo {author} {\bibfnamefont {J.~M.}\ \bibnamefont
  {Brown}}, \bibinfo {author} {\bibfnamefont {L.~W.}\ \bibnamefont {Cheuk}},
  \bibinfo {author} {\bibfnamefont {S.~J.}\ \bibnamefont {Smullin}},\ and\
  \bibinfo {author} {\bibfnamefont {M.~V.}\ \bibnamefont {Romalis}},\
  }\bibfield  {title} {\bibinfo {title} {New test of local lorentz invariance
  using a
  $^{21}\mathrm{Ne}\mathrm{\text{\ensuremath{-}}}\mathrm{Rb}\mathrm{\text{\ensuremath{-}}}\mathbf{K}$
  comagnetometer},\ }\href {https://doi.org/10.1103/PhysRevLett.107.171604}
  {\bibfield  {journal} {\bibinfo  {journal} {Phys. Rev. Lett.}\ }\textbf
  {\bibinfo {volume} {107}},\ \bibinfo {pages} {171604} (\bibinfo {year}
  {2011})}\BibitemShut {NoStop}%
\bibitem [{\citenamefont {Fu}\ \emph {et~al.}(2014)\citenamefont {Fu},
  \citenamefont {Weiss}, \citenamefont {Lima}, \citenamefont {Harrison},
  \citenamefont {Bai}, \citenamefont {Desch}, \citenamefont {Ebel},
  \citenamefont {Suavet}, \citenamefont {Wang}, \citenamefont {Glenn},
  \citenamefont {Sage}, \citenamefont {Kasama}, \citenamefont {Walsworth},\
  and\ \citenamefont {Kuan}}]{Fu_2014}%
  \BibitemOpen
  \bibfield  {author} {\bibinfo {author} {\bibfnamefont {R.~R.}\ \bibnamefont
  {Fu}}, \bibinfo {author} {\bibfnamefont {B.~P.}\ \bibnamefont {Weiss}},
  \bibinfo {author} {\bibfnamefont {E.~A.}\ \bibnamefont {Lima}}, \bibinfo
  {author} {\bibfnamefont {R.~J.}\ \bibnamefont {Harrison}}, \bibinfo {author}
  {\bibfnamefont {X.-N.}\ \bibnamefont {Bai}}, \bibinfo {author} {\bibfnamefont
  {S.~J.}\ \bibnamefont {Desch}}, \bibinfo {author} {\bibfnamefont {D.~S.}\
  \bibnamefont {Ebel}}, \bibinfo {author} {\bibfnamefont {C.}~\bibnamefont
  {Suavet}}, \bibinfo {author} {\bibfnamefont {H.}~\bibnamefont {Wang}},
  \bibinfo {author} {\bibfnamefont {D.}~\bibnamefont {Glenn}}, \bibinfo
  {author} {\bibfnamefont {D.~L.}\ \bibnamefont {Sage}}, \bibinfo {author}
  {\bibfnamefont {T.}~\bibnamefont {Kasama}}, \bibinfo {author} {\bibfnamefont
  {R.~L.}\ \bibnamefont {Walsworth}},\ and\ \bibinfo {author} {\bibfnamefont
  {A.~T.}\ \bibnamefont {Kuan}},\ }\bibfield  {title} {\bibinfo {title} {Solar
  nebula magnetic fields recorded in the semarkona meteorite},\ }\href
  {https://doi.org/10.1126/science.1258022} {\bibfield  {journal} {\bibinfo
  {journal} {Science}\ }\textbf {\bibinfo {volume} {346}},\ \bibinfo {pages}
  {1089} (\bibinfo {year} {2014})}\BibitemShut {NoStop}%
\bibitem [{\citenamefont {Ye}\ and\ \citenamefont {Zoller}(2024)}]{Ye_2024}%
  \BibitemOpen
  \bibfield  {author} {\bibinfo {author} {\bibfnamefont {J.}~\bibnamefont
  {Ye}}\ and\ \bibinfo {author} {\bibfnamefont {P.}~\bibnamefont {Zoller}},\
  }\bibfield  {title} {\bibinfo {title} {Essay: Quantum sensing with atomic,
  molecular, and optical platforms for fundamental physics},\ }\href
  {https://doi.org/10.1103/PhysRevLett.132.190001} {\bibfield  {journal}
  {\bibinfo  {journal} {Phys. Rev. Lett.}\ }\textbf {\bibinfo {volume} {132}},\
  \bibinfo {pages} {190001} (\bibinfo {year} {2024})}\BibitemShut {NoStop}%
\bibitem [{\citenamefont {Simmonds}\ \emph {et~al.}(1979)\citenamefont
  {Simmonds}, \citenamefont {Fertig},\ and\ \citenamefont
  {Giffard}}]{Simmonds_1979}%
  \BibitemOpen
  \bibfield  {author} {\bibinfo {author} {\bibfnamefont {M.}~\bibnamefont
  {Simmonds}}, \bibinfo {author} {\bibfnamefont {W.}~\bibnamefont {Fertig}},\
  and\ \bibinfo {author} {\bibfnamefont {R.}~\bibnamefont {Giffard}},\
  }\bibfield  {title} {\bibinfo {title} {Performance of a resonant input squid
  amplifier system},\ }\href {https://doi.org/10.1109/TMAG.1979.1060155}
  {\bibfield  {journal} {\bibinfo  {journal} {IEEE Transactions on Magnetics}\
  }\textbf {\bibinfo {volume} {15}},\ \bibinfo {pages} {478} (\bibinfo {year}
  {1979})}\BibitemShut {NoStop}%
\bibitem [{\citenamefont {Taylor}\ \emph {et~al.}(2008)\citenamefont {Taylor},
  \citenamefont {Cappellaro}, \citenamefont {Childress}, \citenamefont {Jiang},
  \citenamefont {Budker}, \citenamefont {Hemmer}, \citenamefont {Yacoby},
  \citenamefont {Walsworth},\ and\ \citenamefont {Lukin}}]{Taylor_2008}%
  \BibitemOpen
  \bibfield  {author} {\bibinfo {author} {\bibfnamefont {J.~M.}\ \bibnamefont
  {Taylor}}, \bibinfo {author} {\bibfnamefont {P.}~\bibnamefont {Cappellaro}},
  \bibinfo {author} {\bibfnamefont {L.}~\bibnamefont {Childress}}, \bibinfo
  {author} {\bibfnamefont {L.}~\bibnamefont {Jiang}}, \bibinfo {author}
  {\bibfnamefont {D.}~\bibnamefont {Budker}}, \bibinfo {author} {\bibfnamefont
  {P.~R.}\ \bibnamefont {Hemmer}}, \bibinfo {author} {\bibfnamefont
  {A.}~\bibnamefont {Yacoby}}, \bibinfo {author} {\bibfnamefont
  {R.}~\bibnamefont {Walsworth}},\ and\ \bibinfo {author} {\bibfnamefont
  {M.~D.}\ \bibnamefont {Lukin}},\ }\bibfield  {title} {\bibinfo {title}
  {High-sensitivity diamond magnetometer with nanoscale resolution},\ }\href
  {High-sensitivity diamond magnetometer with nanoscale resolution} {\bibfield
  {journal} {\bibinfo  {journal} {Nature Physics}\ }\textbf {\bibinfo {volume}
  {4}},\ \bibinfo {pages} {810} (\bibinfo {year} {2008})}\BibitemShut {NoStop}%
\bibitem [{\citenamefont {Toida}\ \emph {et~al.}(2023)\citenamefont {Toida},
  \citenamefont {Sakai}, \citenamefont {Teshima}, \citenamefont {Hori},
  \citenamefont {Kakuyanagi}, \citenamefont {Mahboob}, \citenamefont {Ono},\
  and\ \citenamefont {Saito}}]{Toida_2023}%
  \BibitemOpen
  \bibfield  {author} {\bibinfo {author} {\bibfnamefont {H.}~\bibnamefont
  {Toida}}, \bibinfo {author} {\bibfnamefont {K.}~\bibnamefont {Sakai}},
  \bibinfo {author} {\bibfnamefont {T.~F.}\ \bibnamefont {Teshima}}, \bibinfo
  {author} {\bibfnamefont {M.}~\bibnamefont {Hori}}, \bibinfo {author}
  {\bibfnamefont {K.}~\bibnamefont {Kakuyanagi}}, \bibinfo {author}
  {\bibfnamefont {I.}~\bibnamefont {Mahboob}}, \bibinfo {author} {\bibfnamefont
  {Y.}~\bibnamefont {Ono}},\ and\ \bibinfo {author} {\bibfnamefont
  {S.}~\bibnamefont {Saito}},\ }\bibfield  {title} {\bibinfo {title}
  {Magnetometry of neurons using a superconducting qubit},\ }\href
  {https://doi.org/10.1038/s42005-023-01133-z} {\bibfield  {journal} {\bibinfo
  {journal} {Communications Physics}\ }\textbf {\bibinfo {volume} {6}},\
  \bibinfo {pages} {19} (\bibinfo {year} {2023})}\BibitemShut {NoStop}%
\bibitem [{\citenamefont {Barry}\ \emph {et~al.}(2020)\citenamefont {Barry},
  \citenamefont {Schloss}, \citenamefont {Bauch}, \citenamefont {Turner},
  \citenamefont {Hart}, \citenamefont {Pham},\ and\ \citenamefont
  {Walsworth}}]{Barry_NV_review_2020}%
  \BibitemOpen
  \bibfield  {author} {\bibinfo {author} {\bibfnamefont {J.~F.}\ \bibnamefont
  {Barry}}, \bibinfo {author} {\bibfnamefont {J.~M.}\ \bibnamefont {Schloss}},
  \bibinfo {author} {\bibfnamefont {E.}~\bibnamefont {Bauch}}, \bibinfo
  {author} {\bibfnamefont {M.~J.}\ \bibnamefont {Turner}}, \bibinfo {author}
  {\bibfnamefont {C.~A.}\ \bibnamefont {Hart}}, \bibinfo {author}
  {\bibfnamefont {L.~M.}\ \bibnamefont {Pham}},\ and\ \bibinfo {author}
  {\bibfnamefont {R.~L.}\ \bibnamefont {Walsworth}},\ }\bibfield  {title}
  {\bibinfo {title} {Sensitivity optimization for nv-diamond magnetometry},\
  }\href {https://doi.org/10.1103/RevModPhys.92.015004} {\bibfield  {journal}
  {\bibinfo  {journal} {Rev. Mod. Phys.}\ }\textbf {\bibinfo {volume} {92}},\
  \bibinfo {pages} {015004} (\bibinfo {year} {2020})}\BibitemShut {NoStop}%
\bibitem [{\citenamefont {Zhang}\ \emph {et~al.}(2021)\citenamefont {Zhang},
  \citenamefont {Shagieva}, \citenamefont {Widmann}, \citenamefont {K\"ubler},
  \citenamefont {Vorobyov}, \citenamefont {Kapitanova}, \citenamefont
  {Nenasheva}, \citenamefont {Corkill}, \citenamefont {Rhrle}, \citenamefont
  {Nakamura}, \citenamefont {Sumiya}, \citenamefont {Onoda}, \citenamefont
  {Isoya},\ and\ \citenamefont {Wrachtrup}}]{Zhang_2021}%
  \BibitemOpen
  \bibfield  {author} {\bibinfo {author} {\bibfnamefont {C.}~\bibnamefont
  {Zhang}}, \bibinfo {author} {\bibfnamefont {F.}~\bibnamefont {Shagieva}},
  \bibinfo {author} {\bibfnamefont {M.}~\bibnamefont {Widmann}}, \bibinfo
  {author} {\bibfnamefont {M.}~\bibnamefont {K\"ubler}}, \bibinfo {author}
  {\bibfnamefont {V.}~\bibnamefont {Vorobyov}}, \bibinfo {author}
  {\bibfnamefont {P.}~\bibnamefont {Kapitanova}}, \bibinfo {author}
  {\bibfnamefont {E.}~\bibnamefont {Nenasheva}}, \bibinfo {author}
  {\bibfnamefont {R.}~\bibnamefont {Corkill}}, \bibinfo {author} {\bibfnamefont
  {O.}~\bibnamefont {Rhrle}}, \bibinfo {author} {\bibfnamefont
  {K.}~\bibnamefont {Nakamura}}, \bibinfo {author} {\bibfnamefont
  {H.}~\bibnamefont {Sumiya}}, \bibinfo {author} {\bibfnamefont
  {S.}~\bibnamefont {Onoda}}, \bibinfo {author} {\bibfnamefont
  {J.}~\bibnamefont {Isoya}},\ and\ \bibinfo {author} {\bibfnamefont
  {J.}~\bibnamefont {Wrachtrup}},\ }\bibfield  {title} {\bibinfo {title}
  {Diamond magnetometry and gradiometry towards subpicotesla dc field
  measurement},\ }\href {https://doi.org/10.1103/PhysRevApplied.15.064075}
  {\bibfield  {journal} {\bibinfo  {journal} {Phys. Rev. Appl.}\ }\textbf
  {\bibinfo {volume} {15}},\ \bibinfo {pages} {064075} (\bibinfo {year}
  {2021})}\BibitemShut {NoStop}%
\bibitem [{\citenamefont {Balasubramanian}\ \emph {et~al.}(2008)\citenamefont
  {Balasubramanian}, \citenamefont {Chan}, \citenamefont {Kolesov},
  \citenamefont {Al-Hmoud}, \citenamefont {Tisler}, \citenamefont {Shin},
  \citenamefont {Kim}, \citenamefont {Wojcik}, \citenamefont {Hemmer},
  \citenamefont {Krueger}, \citenamefont {Hanke}, \citenamefont
  {Leitenstorfer}, \citenamefont {Bratschitsch}, \citenamefont {Jelezko},\ and\
  \citenamefont {Wrachtrup}}]{Balasubramanian_2008}%
  \BibitemOpen
  \bibfield  {author} {\bibinfo {author} {\bibfnamefont {G.}~\bibnamefont
  {Balasubramanian}}, \bibinfo {author} {\bibfnamefont {I.~Y.}\ \bibnamefont
  {Chan}}, \bibinfo {author} {\bibfnamefont {R.}~\bibnamefont {Kolesov}},
  \bibinfo {author} {\bibfnamefont {M.}~\bibnamefont {Al-Hmoud}}, \bibinfo
  {author} {\bibfnamefont {J.}~\bibnamefont {Tisler}}, \bibinfo {author}
  {\bibfnamefont {C.}~\bibnamefont {Shin}}, \bibinfo {author} {\bibfnamefont
  {C.}~\bibnamefont {Kim}}, \bibinfo {author} {\bibfnamefont {A.}~\bibnamefont
  {Wojcik}}, \bibinfo {author} {\bibfnamefont {P.~R.}\ \bibnamefont {Hemmer}},
  \bibinfo {author} {\bibfnamefont {A.}~\bibnamefont {Krueger}}, \bibinfo
  {author} {\bibfnamefont {T.}~\bibnamefont {Hanke}}, \bibinfo {author}
  {\bibfnamefont {A.}~\bibnamefont {Leitenstorfer}}, \bibinfo {author}
  {\bibfnamefont {R.}~\bibnamefont {Bratschitsch}}, \bibinfo {author}
  {\bibfnamefont {F.}~\bibnamefont {Jelezko}},\ and\ \bibinfo {author}
  {\bibfnamefont {J.}~\bibnamefont {Wrachtrup}},\ }\bibfield  {title} {\bibinfo
  {title} {Nanoscale imaging magnetometry with diamond spins under ambient
  conditions},\ }\href {https://doi.org/10.1038/nature07278} {\bibfield
  {journal} {\bibinfo  {journal} {Nature}\ }\textbf {\bibinfo {volume} {455}},\
  \bibinfo {pages} {648} (\bibinfo {year} {2008})}\BibitemShut {NoStop}%
\bibitem [{\citenamefont {Wyss}\ \emph {et~al.}(2022)\citenamefont {Wyss},
  \citenamefont {Bagani}, \citenamefont {Jetter}, \citenamefont {Marchiori},
  \citenamefont {Vervelaki}, \citenamefont {Gross}, \citenamefont {Ridderbos},
  \citenamefont {Gliga}, \citenamefont {Sch\"onenberger},\ and\ \citenamefont
  {Poggio}}]{Wyss_2022}%
  \BibitemOpen
  \bibfield  {author} {\bibinfo {author} {\bibfnamefont {M.}~\bibnamefont
  {Wyss}}, \bibinfo {author} {\bibfnamefont {K.}~\bibnamefont {Bagani}},
  \bibinfo {author} {\bibfnamefont {D.}~\bibnamefont {Jetter}}, \bibinfo
  {author} {\bibfnamefont {E.}~\bibnamefont {Marchiori}}, \bibinfo {author}
  {\bibfnamefont {A.}~\bibnamefont {Vervelaki}}, \bibinfo {author}
  {\bibfnamefont {B.}~\bibnamefont {Gross}}, \bibinfo {author} {\bibfnamefont
  {J.}~\bibnamefont {Ridderbos}}, \bibinfo {author} {\bibfnamefont
  {S.}~\bibnamefont {Gliga}}, \bibinfo {author} {\bibfnamefont
  {C.}~\bibnamefont {Sch\"onenberger}},\ and\ \bibinfo {author} {\bibfnamefont
  {M.}~\bibnamefont {Poggio}},\ }\bibfield  {title} {\bibinfo {title}
  {Magnetic, thermal, and topographic imaging with a nanometer-scale
  squid-on-lever scanning probe},\ }\href
  {https://doi.org/10.1103/PhysRevApplied.17.034002} {\bibfield  {journal}
  {\bibinfo  {journal} {Phys. Rev. Appl.}\ }\textbf {\bibinfo {volume} {17}},\
  \bibinfo {pages} {034002} (\bibinfo {year} {2022})}\BibitemShut {NoStop}%
\bibitem [{\citenamefont {Budker}\ and\ \citenamefont
  {Romalis}(2007)}]{Budker_vapour_review_2007}%
  \BibitemOpen
  \bibfield  {author} {\bibinfo {author} {\bibfnamefont {D.}~\bibnamefont
  {Budker}}\ and\ \bibinfo {author} {\bibfnamefont {M.}~\bibnamefont
  {Romalis}},\ }\bibfield  {title} {\bibinfo {title} {Optical magnetometry},\
  }\href {https://doi.org/10.1038/nphys566} {\bibfield  {journal} {\bibinfo
  {journal} {Nature Physics}\ }\textbf {\bibinfo {volume} {3}},\ \bibinfo
  {pages} {227} (\bibinfo {year} {2007})}\BibitemShut {NoStop}%
\bibitem [{\citenamefont {Wildermuth}\ \emph {et~al.}(2005)\citenamefont
  {Wildermuth}, \citenamefont {Hofferberth}, \citenamefont {Lesanovsky},
  \citenamefont {Haller}, \citenamefont {Andersson}, \citenamefont {Groth},
  \citenamefont {Bar-Joseph}, \citenamefont {Kr{\"u}ger},\ and\ \citenamefont
  {Schmiedmayer}}]{Wildermuth_2005}%
  \BibitemOpen
  \bibfield  {author} {\bibinfo {author} {\bibfnamefont {S.}~\bibnamefont
  {Wildermuth}}, \bibinfo {author} {\bibfnamefont {S.}~\bibnamefont
  {Hofferberth}}, \bibinfo {author} {\bibfnamefont {I.}~\bibnamefont
  {Lesanovsky}}, \bibinfo {author} {\bibfnamefont {E.}~\bibnamefont {Haller}},
  \bibinfo {author} {\bibfnamefont {L.~M.}\ \bibnamefont {Andersson}}, \bibinfo
  {author} {\bibfnamefont {S.}~\bibnamefont {Groth}}, \bibinfo {author}
  {\bibfnamefont {I.}~\bibnamefont {Bar-Joseph}}, \bibinfo {author}
  {\bibfnamefont {P.}~\bibnamefont {Kr{\"u}ger}},\ and\ \bibinfo {author}
  {\bibfnamefont {J.}~\bibnamefont {Schmiedmayer}},\ }\bibfield  {title}
  {\bibinfo {title} {Microscopic magnetic-field imaging},\ }\href@noop {}
  {\bibfield  {journal} {\bibinfo  {journal} {Nature}\ }\textbf {\bibinfo
  {volume} {435}},\ \bibinfo {pages} {440} (\bibinfo {year}
  {2005})}\BibitemShut {NoStop}%
\bibitem [{\citenamefont {Simons}\ \emph {et~al.}(2021)\citenamefont {Simons},
  \citenamefont {Artusio-Glimpse}, \citenamefont {Robinson}, \citenamefont
  {Prajapati},\ and\ \citenamefont {Holloway}}]{Simons_2021}%
  \BibitemOpen
  \bibfield  {author} {\bibinfo {author} {\bibfnamefont {M.~T.}\ \bibnamefont
  {Simons}}, \bibinfo {author} {\bibfnamefont {A.~B.}\ \bibnamefont
  {Artusio-Glimpse}}, \bibinfo {author} {\bibfnamefont {A.~K.}\ \bibnamefont
  {Robinson}}, \bibinfo {author} {\bibfnamefont {N.}~\bibnamefont
  {Prajapati}},\ and\ \bibinfo {author} {\bibfnamefont {C.~L.}\ \bibnamefont
  {Holloway}},\ }\bibfield  {title} {\bibinfo {title} {Rydberg atom-based
  sensors for radio-frequency electric field metrology, sensing, and
  communications},\ }\href
  {https://doi.org/https://doi.org/10.1016/j.measen.2021.100273} {\bibfield
  {journal} {\bibinfo  {journal} {Measurement: Sensors}\ }\textbf {\bibinfo
  {volume} {18}},\ \bibinfo {pages} {100273} (\bibinfo {year}
  {2021})}\BibitemShut {NoStop}%
\bibitem [{\citenamefont {Barredo}\ \emph {et~al.}(2020)\citenamefont
  {Barredo}, \citenamefont {Lienhard}, \citenamefont {Scholl}, \citenamefont
  {de~L\'es\'eleuc}, \citenamefont {Boulier}, \citenamefont {Browaeys},\ and\
  \citenamefont {Lahaye}}]{Barredo_2020}%
  \BibitemOpen
  \bibfield  {author} {\bibinfo {author} {\bibfnamefont {D.}~\bibnamefont
  {Barredo}}, \bibinfo {author} {\bibfnamefont {V.}~\bibnamefont {Lienhard}},
  \bibinfo {author} {\bibfnamefont {P.}~\bibnamefont {Scholl}}, \bibinfo
  {author} {\bibfnamefont {S.}~\bibnamefont {de~L\'es\'eleuc}}, \bibinfo
  {author} {\bibfnamefont {T.}~\bibnamefont {Boulier}}, \bibinfo {author}
  {\bibfnamefont {A.}~\bibnamefont {Browaeys}},\ and\ \bibinfo {author}
  {\bibfnamefont {T.}~\bibnamefont {Lahaye}},\ }\bibfield  {title} {\bibinfo
  {title} {Three-dimensional trapping of individual rydberg atoms in
  ponderomotive bottle beam traps},\ }\href
  {https://doi.org/10.1103/PhysRevLett.124.023201} {\bibfield  {journal}
  {\bibinfo  {journal} {Phys. Rev. Lett.}\ }\textbf {\bibinfo {volume} {124}},\
  \bibinfo {pages} {023201} (\bibinfo {year} {2020})}\BibitemShut {NoStop}%
\bibitem [{\citenamefont {Gilmore}\ \emph {et~al.}(2021)\citenamefont
  {Gilmore}, \citenamefont {Affolter}, \citenamefont {Lewis-Swan},
  \citenamefont {Barberena}, \citenamefont {Jordan}, \citenamefont {Rey},\ and\
  \citenamefont {Bollinger}}]{Gilmore_2021}%
  \BibitemOpen
  \bibfield  {author} {\bibinfo {author} {\bibfnamefont {K.~A.}\ \bibnamefont
  {Gilmore}}, \bibinfo {author} {\bibfnamefont {M.}~\bibnamefont {Affolter}},
  \bibinfo {author} {\bibfnamefont {R.~J.}\ \bibnamefont {Lewis-Swan}},
  \bibinfo {author} {\bibfnamefont {D.}~\bibnamefont {Barberena}}, \bibinfo
  {author} {\bibfnamefont {E.}~\bibnamefont {Jordan}}, \bibinfo {author}
  {\bibfnamefont {A.~M.}\ \bibnamefont {Rey}},\ and\ \bibinfo {author}
  {\bibfnamefont {J.~J.}\ \bibnamefont {Bollinger}},\ }\bibfield  {title}
  {\bibinfo {title} {Quantum-enhanced sensing of displacements and electric
  fields with two-dimensional trapped-ion crystals},\ }\href
  {https://doi.org/10.1126/science.abi5226} {\bibfield  {journal} {\bibinfo
  {journal} {Science}\ }\textbf {\bibinfo {volume} {373}},\ \bibinfo {pages}
  {673} (\bibinfo {year} {2021})}\BibitemShut {NoStop}%
\bibitem [{\citenamefont {Wu}\ \emph {et~al.}(2024)\citenamefont {Wu},
  \citenamefont {Mitts}, \citenamefont {Ho}, \citenamefont {Rabinowitz},\ and\
  \citenamefont {Hudson}}]{Wu_2024}%
  \BibitemOpen
  \bibfield  {author} {\bibinfo {author} {\bibfnamefont {H.}~\bibnamefont
  {Wu}}, \bibinfo {author} {\bibfnamefont {G.}~\bibnamefont {Mitts}}, \bibinfo
  {author} {\bibfnamefont {C.}~\bibnamefont {Ho}}, \bibinfo {author}
  {\bibfnamefont {J.}~\bibnamefont {Rabinowitz}},\ and\ \bibinfo {author}
  {\bibfnamefont {E.~R.}\ \bibnamefont {Hudson}},\ }\href
  {https://arxiv.org/abs/2311.12263} {\bibinfo {title} {Quantum vector signal
  analyzer: Wideband electric field sensing via motional raman transitions}}
  (\bibinfo {year} {2024}),\ \Eprint {https://arxiv.org/abs/2311.12263}
  {arXiv:2311.12263 [physics.atom-ph]} \BibitemShut {NoStop}%
\bibitem [{\citenamefont {Sch\"affner}\ \emph {et~al.}(2024)\citenamefont
  {Sch\"affner}, \citenamefont {Schreiber}, \citenamefont {Lenz}, \citenamefont
  {Schlosser},\ and\ \citenamefont {Birkl}}]{Schaffner_2024}%
  \BibitemOpen
  \bibfield  {author} {\bibinfo {author} {\bibfnamefont {D.}~\bibnamefont
  {Sch\"affner}}, \bibinfo {author} {\bibfnamefont {T.}~\bibnamefont
  {Schreiber}}, \bibinfo {author} {\bibfnamefont {F.}~\bibnamefont {Lenz}},
  \bibinfo {author} {\bibfnamefont {M.}~\bibnamefont {Schlosser}},\ and\
  \bibinfo {author} {\bibfnamefont {G.}~\bibnamefont {Birkl}},\ }\bibfield
  {title} {\bibinfo {title} {Quantum sensing in tweezer arrays: Optical
  magnetometry on an individual-atom sensor grid},\ }\href
  {https://doi.org/10.1103/PRXQuantum.5.010311} {\bibfield  {journal} {\bibinfo
   {journal} {PRX Quantum}\ }\textbf {\bibinfo {volume} {5}},\ \bibinfo {pages}
  {010311} (\bibinfo {year} {2024})}\BibitemShut {NoStop}%
\bibitem [{\citenamefont {Baumgart}\ \emph {et~al.}(2016)\citenamefont
  {Baumgart}, \citenamefont {Cai}, \citenamefont {Retzker}, \citenamefont
  {Plenio},\ and\ \citenamefont {Wunderlich}}]{Wunderlich_2016}%
  \BibitemOpen
  \bibfield  {author} {\bibinfo {author} {\bibfnamefont {I.}~\bibnamefont
  {Baumgart}}, \bibinfo {author} {\bibfnamefont {J.-M.}\ \bibnamefont {Cai}},
  \bibinfo {author} {\bibfnamefont {A.}~\bibnamefont {Retzker}}, \bibinfo
  {author} {\bibfnamefont {M.~B.}\ \bibnamefont {Plenio}},\ and\ \bibinfo
  {author} {\bibfnamefont {C.}~\bibnamefont {Wunderlich}},\ }\bibfield  {title}
  {\bibinfo {title} {Ultrasensitive magnetometer using a single atom},\ }\href
  {https://doi.org/10.1103/PhysRevLett.116.240801} {\bibfield  {journal}
  {\bibinfo  {journal} {Phys. Rev. Lett.}\ }\textbf {\bibinfo {volume} {116}},\
  \bibinfo {pages} {240801} (\bibinfo {year} {2016})}\BibitemShut {NoStop}%
\bibitem [{\citenamefont {Wineland}\ \emph {et~al.}(1998)\citenamefont
  {Wineland}, \citenamefont {Monroe}, \citenamefont {Itano}, \citenamefont
  {Leibfried}, \citenamefont {King},\ and\ \citenamefont
  {Meekhof}}]{winelandExperimentalIssuesCoherent1998}%
  \BibitemOpen
  \bibfield  {author} {\bibinfo {author} {\bibfnamefont {D.}~\bibnamefont
  {Wineland}}, \bibinfo {author} {\bibfnamefont {C.}~\bibnamefont {Monroe}},
  \bibinfo {author} {\bibfnamefont {W.}~\bibnamefont {Itano}}, \bibinfo
  {author} {\bibfnamefont {D.}~\bibnamefont {Leibfried}}, \bibinfo {author}
  {\bibfnamefont {B.}~\bibnamefont {King}},\ and\ \bibinfo {author}
  {\bibfnamefont {D.}~\bibnamefont {Meekhof}},\ }\bibfield  {title} {\bibinfo
  {title} {Experimental issues in coherent quantum-state manipulation of
  trapped atomic ions},\ }\href {https://doi.org/10.6028/jres.103.019}
  {\bibfield  {journal} {\bibinfo  {journal} {Journal of Research of the
  National Institute of Standards and Technology}\ }\textbf {\bibinfo {volume}
  {103}},\ \bibinfo {pages} {259} (\bibinfo {year} {1998})}\BibitemShut
  {NoStop}%
\bibitem [{\citenamefont {Kotler}\ \emph {et~al.}(2011)\citenamefont {Kotler},
  \citenamefont {Akerman}, \citenamefont {Glickman}, \citenamefont {Keselman},\
  and\ \citenamefont {Ozeri}}]{Kotler_2011}%
  \BibitemOpen
  \bibfield  {author} {\bibinfo {author} {\bibfnamefont {S.}~\bibnamefont
  {Kotler}}, \bibinfo {author} {\bibfnamefont {N.}~\bibnamefont {Akerman}},
  \bibinfo {author} {\bibfnamefont {Y.}~\bibnamefont {Glickman}}, \bibinfo
  {author} {\bibfnamefont {A.}~\bibnamefont {Keselman}},\ and\ \bibinfo
  {author} {\bibfnamefont {R.}~\bibnamefont {Ozeri}},\ }\bibfield  {title}
  {\bibinfo {title} {Single-ion quantum lock-in amplifier},\ }\href
  {https://doi.org/10.1038/nature10010} {\bibfield  {journal} {\bibinfo
  {journal} {Nature}\ }\textbf {\bibinfo {volume} {473}},\ \bibinfo {pages}
  {61} (\bibinfo {year} {2011})}\BibitemShut {NoStop}%
\bibitem [{\citenamefont {Leibfried}\ \emph {et~al.}(2004)\citenamefont
  {Leibfried}, \citenamefont {Barrett}, \citenamefont {Schaetz}, \citenamefont
  {Britton}, \citenamefont {Chiaverini}, \citenamefont {Itano}, \citenamefont
  {Jost}, \citenamefont {Langer},\ and\ \citenamefont
  {Wineland}}]{Leibfried_2004}%
  \BibitemOpen
  \bibfield  {author} {\bibinfo {author} {\bibfnamefont {D.}~\bibnamefont
  {Leibfried}}, \bibinfo {author} {\bibfnamefont {M.~D.}\ \bibnamefont
  {Barrett}}, \bibinfo {author} {\bibfnamefont {T.}~\bibnamefont {Schaetz}},
  \bibinfo {author} {\bibfnamefont {J.}~\bibnamefont {Britton}}, \bibinfo
  {author} {\bibfnamefont {J.}~\bibnamefont {Chiaverini}}, \bibinfo {author}
  {\bibfnamefont {W.~M.}\ \bibnamefont {Itano}}, \bibinfo {author}
  {\bibfnamefont {J.~D.}\ \bibnamefont {Jost}}, \bibinfo {author}
  {\bibfnamefont {C.}~\bibnamefont {Langer}},\ and\ \bibinfo {author}
  {\bibfnamefont {D.~J.}\ \bibnamefont {Wineland}},\ }\bibfield  {title}
  {\bibinfo {title} {Toward heisenberg-limited spectroscopy with multiparticle
  entangled states},\ }\href {https://doi.org/10.1126/science.1097576}
  {\bibfield  {journal} {\bibinfo  {journal} {Science}\ }\textbf {\bibinfo
  {volume} {304}},\ \bibinfo {pages} {1476} (\bibinfo {year}
  {2004})}\BibitemShut {NoStop}%
\bibitem [{\citenamefont {Ruster}\ \emph {et~al.}(2017)\citenamefont {Ruster},
  \citenamefont {Kaufmann}, \citenamefont {Luda}, \citenamefont {Kaushal},
  \citenamefont {Schmiegelow}, \citenamefont {Schmidt-Kaler},\ and\
  \citenamefont {Poschinger}}]{Ruster_2017}%
  \BibitemOpen
  \bibfield  {author} {\bibinfo {author} {\bibfnamefont {T.}~\bibnamefont
  {Ruster}}, \bibinfo {author} {\bibfnamefont {H.}~\bibnamefont {Kaufmann}},
  \bibinfo {author} {\bibfnamefont {M.~A.}\ \bibnamefont {Luda}}, \bibinfo
  {author} {\bibfnamefont {V.}~\bibnamefont {Kaushal}}, \bibinfo {author}
  {\bibfnamefont {C.~T.}\ \bibnamefont {Schmiegelow}}, \bibinfo {author}
  {\bibfnamefont {F.}~\bibnamefont {Schmidt-Kaler}},\ and\ \bibinfo {author}
  {\bibfnamefont {U.~G.}\ \bibnamefont {Poschinger}},\ }\bibfield  {title}
  {\bibinfo {title} {Entanglement-based dc magnetometry with separated ions},\
  }\href {https://doi.org/10.1103/PhysRevX.7.031050} {\bibfield  {journal}
  {\bibinfo  {journal} {Phys. Rev. X}\ }\textbf {\bibinfo {volume} {7}},\
  \bibinfo {pages} {031050} (\bibinfo {year} {2017})}\BibitemShut {NoStop}%
\bibitem [{\citenamefont {Burd}\ \emph {et~al.}(2019)\citenamefont {Burd},
  \citenamefont {Srinivas}, \citenamefont {Bollinger}, \citenamefont {Wilson},
  \citenamefont {Wineland}, \citenamefont {Leibfried}, \citenamefont
  {Slichter},\ and\ \citenamefont {Allcock}}]{Burd_2019}%
  \BibitemOpen
  \bibfield  {author} {\bibinfo {author} {\bibfnamefont {S.~C.}\ \bibnamefont
  {Burd}}, \bibinfo {author} {\bibfnamefont {R.}~\bibnamefont {Srinivas}},
  \bibinfo {author} {\bibfnamefont {J.~J.}\ \bibnamefont {Bollinger}}, \bibinfo
  {author} {\bibfnamefont {A.~C.}\ \bibnamefont {Wilson}}, \bibinfo {author}
  {\bibfnamefont {D.~J.}\ \bibnamefont {Wineland}}, \bibinfo {author}
  {\bibfnamefont {D.}~\bibnamefont {Leibfried}}, \bibinfo {author}
  {\bibfnamefont {D.~H.}\ \bibnamefont {Slichter}},\ and\ \bibinfo {author}
  {\bibfnamefont {D.~T.~C.}\ \bibnamefont {Allcock}},\ }\bibfield  {title}
  {\bibinfo {title} {Quantum amplification of mechanical oscillator motion},\
  }\href {https://doi.org/10.1126/science.aaw2884} {\bibfield  {journal}
  {\bibinfo  {journal} {Science}\ }\textbf {\bibinfo {volume} {364}},\ \bibinfo
  {pages} {1163} (\bibinfo {year} {2019})}\BibitemShut {NoStop}%
\bibitem [{\citenamefont {Hempel}\ \emph {et~al.}(2013)\citenamefont {Hempel},
  \citenamefont {Lanyon}, \citenamefont {Jurcevic}, \citenamefont {Gerritsma},
  \citenamefont {Blatt},\ and\ \citenamefont {Roos}}]{Hempel2013}%
  \BibitemOpen
  \bibfield  {author} {\bibinfo {author} {\bibfnamefont {C.}~\bibnamefont
  {Hempel}}, \bibinfo {author} {\bibfnamefont {B.~P.}\ \bibnamefont {Lanyon}},
  \bibinfo {author} {\bibfnamefont {P.}~\bibnamefont {Jurcevic}}, \bibinfo
  {author} {\bibfnamefont {R.}~\bibnamefont {Gerritsma}}, \bibinfo {author}
  {\bibfnamefont {R.}~\bibnamefont {Blatt}},\ and\ \bibinfo {author}
  {\bibfnamefont {C.~F.}\ \bibnamefont {Roos}},\ }\bibfield  {title} {\bibinfo
  {title} {Entanglement-enhanced detection of single-photon scattering
  events},\ }\href {https://doi.org/10.1038/nphoton.2013.172} {\bibfield
  {journal} {\bibinfo  {journal} {Nature Photonics}\ }\textbf {\bibinfo
  {volume} {7}},\ \bibinfo {pages} {630} (\bibinfo {year} {2013})}\BibitemShut
  {NoStop}%
\bibitem [{\citenamefont {Milne}\ \emph {et~al.}(2021)\citenamefont {Milne},
  \citenamefont {Hempel}, \citenamefont {Li}, \citenamefont {Edmunds},
  \citenamefont {Slatyer}, \citenamefont {Ball}, \citenamefont {Hush},\ and\
  \citenamefont {Biercuk}}]{Milne_2021}%
  \BibitemOpen
  \bibfield  {author} {\bibinfo {author} {\bibfnamefont {A.~R.}\ \bibnamefont
  {Milne}}, \bibinfo {author} {\bibfnamefont {C.}~\bibnamefont {Hempel}},
  \bibinfo {author} {\bibfnamefont {L.}~\bibnamefont {Li}}, \bibinfo {author}
  {\bibfnamefont {C.~L.}\ \bibnamefont {Edmunds}}, \bibinfo {author}
  {\bibfnamefont {H.~J.}\ \bibnamefont {Slatyer}}, \bibinfo {author}
  {\bibfnamefont {H.}~\bibnamefont {Ball}}, \bibinfo {author} {\bibfnamefont
  {M.~R.}\ \bibnamefont {Hush}},\ and\ \bibinfo {author} {\bibfnamefont
  {M.~J.}\ \bibnamefont {Biercuk}},\ }\bibfield  {title} {\bibinfo {title}
  {Quantum oscillator noise spectroscopy via displaced cat states},\ }\href
  {https://doi.org/10.1103/PhysRevLett.126.250506} {\bibfield  {journal}
  {\bibinfo  {journal} {Phys. Rev. Lett.}\ }\textbf {\bibinfo {volume} {126}},\
  \bibinfo {pages} {250506} (\bibinfo {year} {2021})}\BibitemShut {NoStop}%
\bibitem [{\citenamefont {King}\ \emph {et~al.}(2021)\citenamefont {King},
  \citenamefont {Spie\ss{}}, \citenamefont {Micke}, \citenamefont {Wilzewski},
  \citenamefont {Leopold}, \citenamefont {Crespo L\'opez-Urrutia},\ and\
  \citenamefont {Schmidt}}]{King_2021}%
  \BibitemOpen
  \bibfield  {author} {\bibinfo {author} {\bibfnamefont {S.~A.}\ \bibnamefont
  {King}}, \bibinfo {author} {\bibfnamefont {L.~J.}\ \bibnamefont {Spie\ss{}}},
  \bibinfo {author} {\bibfnamefont {P.}~\bibnamefont {Micke}}, \bibinfo
  {author} {\bibfnamefont {A.}~\bibnamefont {Wilzewski}}, \bibinfo {author}
  {\bibfnamefont {T.}~\bibnamefont {Leopold}}, \bibinfo {author} {\bibfnamefont
  {J.~R.}\ \bibnamefont {Crespo L\'opez-Urrutia}},\ and\ \bibinfo {author}
  {\bibfnamefont {P.~O.}\ \bibnamefont {Schmidt}},\ }\bibfield  {title}
  {\bibinfo {title} {Algorithmic ground-state cooling of weakly coupled
  oscillators using quantum logic},\ }\href
  {https://doi.org/10.1103/PhysRevX.11.041049} {\bibfield  {journal} {\bibinfo
  {journal} {Phys. Rev. X}\ }\textbf {\bibinfo {volume} {11}},\ \bibinfo
  {pages} {041049} (\bibinfo {year} {2021})}\BibitemShut {NoStop}%
\bibitem [{\citenamefont {Vasquez}\ \emph {et~al.}(2023)\citenamefont
  {Vasquez}, \citenamefont {Mordini}, \citenamefont {Verni\`ere}, \citenamefont
  {Stadler}, \citenamefont {Malinowski}, \citenamefont {Zhang}, \citenamefont
  {Kienzler}, \citenamefont {Mehta},\ and\ \citenamefont {Home}}]{Ricci_2023}%
  \BibitemOpen
  \bibfield  {author} {\bibinfo {author} {\bibfnamefont {A.~R.}\ \bibnamefont
  {Vasquez}}, \bibinfo {author} {\bibfnamefont {C.}~\bibnamefont {Mordini}},
  \bibinfo {author} {\bibfnamefont {C.}~\bibnamefont {Verni\`ere}}, \bibinfo
  {author} {\bibfnamefont {M.}~\bibnamefont {Stadler}}, \bibinfo {author}
  {\bibfnamefont {M.}~\bibnamefont {Malinowski}}, \bibinfo {author}
  {\bibfnamefont {C.}~\bibnamefont {Zhang}}, \bibinfo {author} {\bibfnamefont
  {D.}~\bibnamefont {Kienzler}}, \bibinfo {author} {\bibfnamefont {K.~K.}\
  \bibnamefont {Mehta}},\ and\ \bibinfo {author} {\bibfnamefont {J.~P.}\
  \bibnamefont {Home}},\ }\bibfield  {title} {\bibinfo {title} {Control of an
  atomic quadrupole transition in a phase-stable standing wave},\ }\href
  {https://doi.org/10.1103/PhysRevLett.130.133201} {\bibfield  {journal}
  {\bibinfo  {journal} {Phys. Rev. Lett.}\ }\textbf {\bibinfo {volume} {130}},\
  \bibinfo {pages} {133201} (\bibinfo {year} {2023})}\BibitemShut {NoStop}%
\bibitem [{\citenamefont {Akhtar}\ \emph {et~al.}(2023)\citenamefont {Akhtar},
  \citenamefont {Bonus}, \citenamefont {Lebrun-Gallagher}, \citenamefont
  {Johnson}, \citenamefont {Siegele-Brown}, \citenamefont {Hong}, \citenamefont
  {Hile}, \citenamefont {Kulmiya}, \citenamefont {Weidt},\ and\ \citenamefont
  {Hensinger}}]{Akhtar2023}%
  \BibitemOpen
  \bibfield  {author} {\bibinfo {author} {\bibfnamefont {M.}~\bibnamefont
  {Akhtar}}, \bibinfo {author} {\bibfnamefont {F.}~\bibnamefont {Bonus}},
  \bibinfo {author} {\bibfnamefont {F.~R.}\ \bibnamefont {Lebrun-Gallagher}},
  \bibinfo {author} {\bibfnamefont {N.~I.}\ \bibnamefont {Johnson}}, \bibinfo
  {author} {\bibfnamefont {M.}~\bibnamefont {Siegele-Brown}}, \bibinfo {author}
  {\bibfnamefont {S.}~\bibnamefont {Hong}}, \bibinfo {author} {\bibfnamefont
  {S.~J.}\ \bibnamefont {Hile}}, \bibinfo {author} {\bibfnamefont {S.~A.}\
  \bibnamefont {Kulmiya}}, \bibinfo {author} {\bibfnamefont {S.}~\bibnamefont
  {Weidt}},\ and\ \bibinfo {author} {\bibfnamefont {W.~K.}\ \bibnamefont
  {Hensinger}},\ }\bibfield  {title} {\bibinfo {title} {A high-fidelity quantum
  matter-link between ion-trap microchip modules},\ }\href
  {https://doi.org/10.1038/s41467-022-35285-3} {\bibfield  {journal} {\bibinfo
  {journal} {Nature Communications}\ }\textbf {\bibinfo {volume} {14}},\
  \bibinfo {pages} {531} (\bibinfo {year} {2023})}\BibitemShut {NoStop}%
\bibitem [{\citenamefont {Brown}\ \emph {et~al.}(2021)\citenamefont {Brown},
  \citenamefont {Chiaverini}, \citenamefont {Sage},\ and\ \citenamefont
  {Häffner}}]{Brown_2021}%
  \BibitemOpen
  \bibfield  {author} {\bibinfo {author} {\bibfnamefont {K.~R.}\ \bibnamefont
  {Brown}}, \bibinfo {author} {\bibfnamefont {J.}~\bibnamefont {Chiaverini}},
  \bibinfo {author} {\bibfnamefont {J.~M.}\ \bibnamefont {Sage}},\ and\
  \bibinfo {author} {\bibfnamefont {H.}~\bibnamefont {Häffner}},\ }\bibfield
  {title} {\bibinfo {title} {Materials challenges for trapped-ion quantum
  computers},\ }\href {https://doi.org/10.1038/s41578-021-00292-1} {\bibfield
  {journal} {\bibinfo  {journal} {Nature Reviews Materials}\ }\textbf {\bibinfo
  {volume} {6}},\ \bibinfo {pages} {892} (\bibinfo {year} {2021})}\BibitemShut
  {NoStop}%
\bibitem [{\citenamefont {de~Leon}\ \emph {et~al.}(2021)\citenamefont
  {de~Leon}, \citenamefont {Itoh}, \citenamefont {Kim}, \citenamefont {Mehta},
  \citenamefont {Northup}, \citenamefont {Paik}, \citenamefont {Palmer},
  \citenamefont {Samarth}, \citenamefont {Sangtawesin},\ and\ \citenamefont
  {Steuerman}}]{deLeon_2021}%
  \BibitemOpen
  \bibfield  {author} {\bibinfo {author} {\bibfnamefont {N.~P.}\ \bibnamefont
  {de~Leon}}, \bibinfo {author} {\bibfnamefont {K.~M.}\ \bibnamefont {Itoh}},
  \bibinfo {author} {\bibfnamefont {D.}~\bibnamefont {Kim}}, \bibinfo {author}
  {\bibfnamefont {K.~K.}\ \bibnamefont {Mehta}}, \bibinfo {author}
  {\bibfnamefont {T.~E.}\ \bibnamefont {Northup}}, \bibinfo {author}
  {\bibfnamefont {H.}~\bibnamefont {Paik}}, \bibinfo {author} {\bibfnamefont
  {B.~S.}\ \bibnamefont {Palmer}}, \bibinfo {author} {\bibfnamefont
  {N.}~\bibnamefont {Samarth}}, \bibinfo {author} {\bibfnamefont
  {S.}~\bibnamefont {Sangtawesin}},\ and\ \bibinfo {author} {\bibfnamefont
  {D.~W.}\ \bibnamefont {Steuerman}},\ }\bibfield  {title} {\bibinfo {title}
  {Materials challenges and opportunities for quantum computing hardware},\
  }\href {https://doi.org/10.1126/science.abb2823} {\bibfield  {journal}
  {\bibinfo  {journal} {Science}\ }\textbf {\bibinfo {volume} {372}},\ \bibinfo
  {pages} {eabb2823} (\bibinfo {year} {2021})}\BibitemShut {NoStop}%
\bibitem [{\citenamefont {Kaufmann}\ \emph {et~al.}(2018)\citenamefont
  {Kaufmann}, \citenamefont {Gloger}, \citenamefont {Kaufmann}, \citenamefont
  {Johanning},\ and\ \citenamefont {Wunderlich}}]{Kaufmann_2018}%
  \BibitemOpen
  \bibfield  {author} {\bibinfo {author} {\bibfnamefont {P.}~\bibnamefont
  {Kaufmann}}, \bibinfo {author} {\bibfnamefont {T.~F.}\ \bibnamefont
  {Gloger}}, \bibinfo {author} {\bibfnamefont {D.}~\bibnamefont {Kaufmann}},
  \bibinfo {author} {\bibfnamefont {M.}~\bibnamefont {Johanning}},\ and\
  \bibinfo {author} {\bibfnamefont {C.}~\bibnamefont {Wunderlich}},\ }\bibfield
   {title} {\bibinfo {title} {High-fidelity preservation of quantum information
  during trapped-ion transport},\ }\href
  {https://doi.org/10.1103/PhysRevLett.120.010501} {\bibfield  {journal}
  {\bibinfo  {journal} {Phys. Rev. Lett.}\ }\textbf {\bibinfo {volume} {120}},\
  \bibinfo {pages} {010501} (\bibinfo {year} {2018})}\BibitemShut {NoStop}%
\bibitem [{\citenamefont {Jain}\ \emph {et~al.}(2020)\citenamefont {Jain},
  \citenamefont {Alonso}, \citenamefont {Grau},\ and\ \citenamefont
  {Home}}]{Jain2020}%
  \BibitemOpen
  \bibfield  {author} {\bibinfo {author} {\bibfnamefont {S.}~\bibnamefont
  {Jain}}, \bibinfo {author} {\bibfnamefont {J.}~\bibnamefont {Alonso}},
  \bibinfo {author} {\bibfnamefont {M.}~\bibnamefont {Grau}},\ and\ \bibinfo
  {author} {\bibfnamefont {J.~P.}\ \bibnamefont {Home}},\ }\bibfield  {title}
  {\bibinfo {title} {Scalable arrays of micro-penning traps for quantum
  computing and simulation},\ }\href
  {https://doi.org/10.1103/PhysRevX.10.031027} {\bibfield  {journal} {\bibinfo
  {journal} {Physical Review X}\ }\textbf {\bibinfo {volume} {10}},\ \bibinfo
  {pages} {031027} (\bibinfo {year} {2020})}\BibitemShut {NoStop}%
\bibitem [{\citenamefont {Jain}\ \emph {et~al.}(2024)\citenamefont {Jain},
  \citenamefont {Sägesser}, \citenamefont {Hrmo}, \citenamefont {Torkzaban},
  \citenamefont {Stadler}, \citenamefont {Oswald}, \citenamefont {Axline},
  \citenamefont {Bautista-Salvador}, \citenamefont {Ospelkaus}, \citenamefont
  {Kienzler},\ and\ \citenamefont {Home}}]{jain_penning_2024}%
  \BibitemOpen
  \bibfield  {author} {\bibinfo {author} {\bibfnamefont {S.}~\bibnamefont
  {Jain}}, \bibinfo {author} {\bibfnamefont {T.}~\bibnamefont {Sägesser}},
  \bibinfo {author} {\bibfnamefont {P.}~\bibnamefont {Hrmo}}, \bibinfo {author}
  {\bibfnamefont {C.}~\bibnamefont {Torkzaban}}, \bibinfo {author}
  {\bibfnamefont {M.}~\bibnamefont {Stadler}}, \bibinfo {author} {\bibfnamefont
  {R.}~\bibnamefont {Oswald}}, \bibinfo {author} {\bibfnamefont
  {C.}~\bibnamefont {Axline}}, \bibinfo {author} {\bibfnamefont
  {A.}~\bibnamefont {Bautista-Salvador}}, \bibinfo {author} {\bibfnamefont
  {C.}~\bibnamefont {Ospelkaus}}, \bibinfo {author} {\bibfnamefont
  {D.}~\bibnamefont {Kienzler}},\ and\ \bibinfo {author} {\bibfnamefont
  {J.}~\bibnamefont {Home}},\ }\bibfield  {title} {\bibinfo {title} {Penning
  micro-trap for quantum computing},\ }\href
  {https://doi.org/10.1038/s41586-024-07111-x} {\bibfield  {journal} {\bibinfo
  {journal} {Nature}\ }\textbf {\bibinfo {volume} {627}},\ \bibinfo {pages}
  {510} (\bibinfo {year} {2024})}\BibitemShut {NoStop}%
\bibitem [{\citenamefont {Wang}\ \emph {et~al.}(2011)\citenamefont {Wang},
  \citenamefont {Hao~Low}, \citenamefont {Lachenmyer}, \citenamefont {Ge},
  \citenamefont {Herskind},\ and\ \citenamefont {Chuang}}]{Wang_2011}%
  \BibitemOpen
  \bibfield  {author} {\bibinfo {author} {\bibfnamefont {S.~X.}\ \bibnamefont
  {Wang}}, \bibinfo {author} {\bibfnamefont {G.}~\bibnamefont {Hao~Low}},
  \bibinfo {author} {\bibfnamefont {N.~S.}\ \bibnamefont {Lachenmyer}},
  \bibinfo {author} {\bibfnamefont {Y.}~\bibnamefont {Ge}}, \bibinfo {author}
  {\bibfnamefont {P.~F.}\ \bibnamefont {Herskind}},\ and\ \bibinfo {author}
  {\bibfnamefont {I.~L.}\ \bibnamefont {Chuang}},\ }\bibfield  {title}
  {\bibinfo {title} {{Laser-induced charging of microfabricated ion traps}},\
  }\href {https://doi.org/10.1063/1.3662118} {\bibfield  {journal} {\bibinfo
  {journal} {Journal of Applied Physics}\ }\textbf {\bibinfo {volume} {110}},\
  \bibinfo {pages} {104901} (\bibinfo {year} {2011})}\BibitemShut {NoStop}%
\bibitem [{\citenamefont {Harlander}\ \emph {et~al.}(2010)\citenamefont
  {Harlander}, \citenamefont {Brownnutt}, \citenamefont {Hänsel},\ and\
  \citenamefont {Blatt}}]{Harlander_2010}%
  \BibitemOpen
  \bibfield  {author} {\bibinfo {author} {\bibfnamefont {M.}~\bibnamefont
  {Harlander}}, \bibinfo {author} {\bibfnamefont {M.}~\bibnamefont
  {Brownnutt}}, \bibinfo {author} {\bibfnamefont {W.}~\bibnamefont {Hänsel}},\
  and\ \bibinfo {author} {\bibfnamefont {R.}~\bibnamefont {Blatt}},\ }\bibfield
   {title} {\bibinfo {title} {Trapped-ion probing of light-induced charging
  effects on dielectrics},\ }\href
  {https://doi.org/10.1088/1367-2630/12/9/093035} {\bibfield  {journal}
  {\bibinfo  {journal} {New Journal of Physics}\ }\textbf {\bibinfo {volume}
  {12}},\ \bibinfo {pages} {093035} (\bibinfo {year} {2010})}\BibitemShut
  {NoStop}%
\bibitem [{\citenamefont {Leung}\ \emph {et~al.}(2003)\citenamefont {Leung},
  \citenamefont {Kao}, \citenamefont {Su}, \citenamefont {Feng},\ and\
  \citenamefont {Chan}}]{Leung_2003}%
  \BibitemOpen
  \bibfield  {author} {\bibinfo {author} {\bibfnamefont {T.~C.}\ \bibnamefont
  {Leung}}, \bibinfo {author} {\bibfnamefont {C.~L.}\ \bibnamefont {Kao}},
  \bibinfo {author} {\bibfnamefont {W.~S.}\ \bibnamefont {Su}}, \bibinfo
  {author} {\bibfnamefont {Y.~J.}\ \bibnamefont {Feng}},\ and\ \bibinfo
  {author} {\bibfnamefont {C.~T.}\ \bibnamefont {Chan}},\ }\bibfield  {title}
  {\bibinfo {title} {Relationship between surface dipole, work function and
  charge transfer: Some exceptions to an established rule},\ }\href
  {https://doi.org/10.1103/PhysRevB.68.195408} {\bibfield  {journal} {\bibinfo
  {journal} {Phys. Rev. B}\ }\textbf {\bibinfo {volume} {68}},\ \bibinfo
  {pages} {195408} (\bibinfo {year} {2003})}\BibitemShut {NoStop}%
\bibitem [{\citenamefont {Tauschinsky}\ \emph {et~al.}(2010)\citenamefont
  {Tauschinsky}, \citenamefont {Thijssen}, \citenamefont {Whitlock},
  \citenamefont {van Linden van~den Heuvell},\ and\ \citenamefont
  {Spreeuw}}]{Tauschinsky_2010}%
  \BibitemOpen
  \bibfield  {author} {\bibinfo {author} {\bibfnamefont {A.}~\bibnamefont
  {Tauschinsky}}, \bibinfo {author} {\bibfnamefont {R.~M.~T.}\ \bibnamefont
  {Thijssen}}, \bibinfo {author} {\bibfnamefont {S.}~\bibnamefont {Whitlock}},
  \bibinfo {author} {\bibfnamefont {H.~B.}\ \bibnamefont {van Linden van~den
  Heuvell}},\ and\ \bibinfo {author} {\bibfnamefont {R.~J.~C.}\ \bibnamefont
  {Spreeuw}},\ }\bibfield  {title} {\bibinfo {title} {Spatially resolved
  excitation of rydberg atoms and surface effects on an atom chip},\ }\href
  {https://doi.org/10.1103/PhysRevA.81.063411} {\bibfield  {journal} {\bibinfo
  {journal} {Phys. Rev. A}\ }\textbf {\bibinfo {volume} {81}},\ \bibinfo
  {pages} {063411} (\bibinfo {year} {2010})}\BibitemShut {NoStop}%
\bibitem [{\citenamefont {McGuirk}\ \emph {et~al.}(2004)\citenamefont
  {McGuirk}, \citenamefont {Harber}, \citenamefont {Obrecht},\ and\
  \citenamefont {Cornell}}]{McGuirk_2004}%
  \BibitemOpen
  \bibfield  {author} {\bibinfo {author} {\bibfnamefont {J.~M.}\ \bibnamefont
  {McGuirk}}, \bibinfo {author} {\bibfnamefont {D.~M.}\ \bibnamefont {Harber}},
  \bibinfo {author} {\bibfnamefont {J.~M.}\ \bibnamefont {Obrecht}},\ and\
  \bibinfo {author} {\bibfnamefont {E.~A.}\ \bibnamefont {Cornell}},\
  }\bibfield  {title} {\bibinfo {title} {Alkali-metal adsorbate polarization on
  conducting and insulating surfaces probed with bose-einstein condensates},\
  }\href {https://doi.org/10.1103/PhysRevA.69.062905} {\bibfield  {journal}
  {\bibinfo  {journal} {Phys. Rev. A}\ }\textbf {\bibinfo {volume} {69}},\
  \bibinfo {pages} {062905} (\bibinfo {year} {2004})}\BibitemShut {NoStop}%
\bibitem [{\citenamefont {Hite}\ \emph {et~al.}(2021)\citenamefont {Hite},
  \citenamefont {McKay},\ and\ \citenamefont {Pappas}}]{Hite_2021}%
  \BibitemOpen
  \bibfield  {author} {\bibinfo {author} {\bibfnamefont {D.~A.}\ \bibnamefont
  {Hite}}, \bibinfo {author} {\bibfnamefont {K.~S.}\ \bibnamefont {McKay}},\
  and\ \bibinfo {author} {\bibfnamefont {D.~P.}\ \bibnamefont {Pappas}},\
  }\bibfield  {title} {\bibinfo {title} {Surface science motivated by heating
  of trapped ions from the quantum ground state},\ }\href
  {https://doi.org/10.1088/1367-2630/ac2c2c} {\bibfield  {journal} {\bibinfo
  {journal} {New Journal of Physics}\ }\textbf {\bibinfo {volume} {23}},\
  \bibinfo {pages} {103028} (\bibinfo {year} {2021})}\BibitemShut {NoStop}%
\bibitem [{\citenamefont {Bian}\ \emph {et~al.}(2021)\citenamefont {Bian},
  \citenamefont {Zheng}, \citenamefont {Zeng}, \citenamefont {Chen},
  \citenamefont {St{\"o}hr}, \citenamefont {Denisenko}, \citenamefont {Yang},
  \citenamefont {Wrachtrup},\ and\ \citenamefont {Jiang}}]{Bian_2021}%
  \BibitemOpen
  \bibfield  {author} {\bibinfo {author} {\bibfnamefont {K.}~\bibnamefont
  {Bian}}, \bibinfo {author} {\bibfnamefont {W.}~\bibnamefont {Zheng}},
  \bibinfo {author} {\bibfnamefont {X.}~\bibnamefont {Zeng}}, \bibinfo {author}
  {\bibfnamefont {X.}~\bibnamefont {Chen}}, \bibinfo {author} {\bibfnamefont
  {R.}~\bibnamefont {St{\"o}hr}}, \bibinfo {author} {\bibfnamefont
  {A.}~\bibnamefont {Denisenko}}, \bibinfo {author} {\bibfnamefont
  {S.}~\bibnamefont {Yang}}, \bibinfo {author} {\bibfnamefont {J.}~\bibnamefont
  {Wrachtrup}},\ and\ \bibinfo {author} {\bibfnamefont {Y.}~\bibnamefont
  {Jiang}},\ }\bibfield  {title} {\bibinfo {title} {Nanoscale electric-field
  imaging based on a quantum sensor and its charge-state control under ambient
  condition},\ }\href {https://doi.org/10.1038/s41467-021-22709-9} {\bibfield
  {journal} {\bibinfo  {journal} {Nature Communications}\ }\textbf {\bibinfo
  {volume} {12}},\ \bibinfo {pages} {2457} (\bibinfo {year}
  {2021})}\BibitemShut {NoStop}%
\bibitem [{\citenamefont {Wagner}\ \emph {et~al.}(2019)\citenamefont {Wagner},
  \citenamefont {Green}, \citenamefont {Maiworm}, \citenamefont {Leinen},
  \citenamefont {Esat}, \citenamefont {Ferri}, \citenamefont {Friedrich},
  \citenamefont {Findeisen}, \citenamefont {Tkatchenko}, \citenamefont
  {Temirov},\ and\ \citenamefont {Tautz}}]{Wagner_2019}%
  \BibitemOpen
  \bibfield  {author} {\bibinfo {author} {\bibfnamefont {C.}~\bibnamefont
  {Wagner}}, \bibinfo {author} {\bibfnamefont {M.~F.~B.}\ \bibnamefont
  {Green}}, \bibinfo {author} {\bibfnamefont {M.}~\bibnamefont {Maiworm}},
  \bibinfo {author} {\bibfnamefont {P.}~\bibnamefont {Leinen}}, \bibinfo
  {author} {\bibfnamefont {T.}~\bibnamefont {Esat}}, \bibinfo {author}
  {\bibfnamefont {N.}~\bibnamefont {Ferri}}, \bibinfo {author} {\bibfnamefont
  {N.}~\bibnamefont {Friedrich}}, \bibinfo {author} {\bibfnamefont
  {R.}~\bibnamefont {Findeisen}}, \bibinfo {author} {\bibfnamefont
  {A.}~\bibnamefont {Tkatchenko}}, \bibinfo {author} {\bibfnamefont
  {R.}~\bibnamefont {Temirov}},\ and\ \bibinfo {author} {\bibfnamefont {F.~S.}\
  \bibnamefont {Tautz}},\ }\bibfield  {title} {\bibinfo {title} {Quantitative
  imaging of electric surface potentials with single-atom sensitivity},\ }\href
  {https://doi.org/10.1038/s41563-019-0382-8} {\bibfield  {journal} {\bibinfo
  {journal} {Nature Materials}\ }\textbf {\bibinfo {volume} {18}},\ \bibinfo
  {pages} {853} (\bibinfo {year} {2019})}\BibitemShut {NoStop}%
\bibitem [{\citenamefont {Zhang}\ \emph {et~al.}(2023)\citenamefont {Zhang},
  \citenamefont {Bian},\ and\ \citenamefont {Jiang}}]{Zhang2023}%
  \BibitemOpen
  \bibfield  {author} {\bibinfo {author} {\bibfnamefont {S.}~\bibnamefont
  {Zhang}}, \bibinfo {author} {\bibfnamefont {K.}~\bibnamefont {Bian}},\ and\
  \bibinfo {author} {\bibfnamefont {Y.}~\bibnamefont {Jiang}},\ }\bibfield
  {title} {\bibinfo {title} {Perspective: nanoscale electric sensing and
  imaging based on quantum sensors},\ }\href@noop {} {\bibfield  {journal}
  {\bibinfo  {journal} {Quantum Frontiers}\ }\textbf {\bibinfo {volume} {2}},\
  \bibinfo {pages} {19} (\bibinfo {year} {2023})}\BibitemShut {NoStop}%
\bibitem [{\citenamefont {Qiu}\ \emph {et~al.}(2022)\citenamefont {Qiu},
  \citenamefont {Hamo}, \citenamefont {Vool}, \citenamefont {Zhou},\ and\
  \citenamefont {Yacoby}}]{Qiu_2022}%
  \BibitemOpen
  \bibfield  {author} {\bibinfo {author} {\bibfnamefont {Z.}~\bibnamefont
  {Qiu}}, \bibinfo {author} {\bibfnamefont {A.}~\bibnamefont {Hamo}}, \bibinfo
  {author} {\bibfnamefont {U.}~\bibnamefont {Vool}}, \bibinfo {author}
  {\bibfnamefont {T.~X.}\ \bibnamefont {Zhou}},\ and\ \bibinfo {author}
  {\bibfnamefont {A.}~\bibnamefont {Yacoby}},\ }\bibfield  {title} {\bibinfo
  {title} {Nanoscale electric field imaging with an ambient scanning quantum
  sensor microscope},\ }\href {https://doi.org/10.1038/s41534-022-00622-3}
  {\bibfield  {journal} {\bibinfo  {journal} {npj Quantum Information}\
  }\textbf {\bibinfo {volume} {8}},\ \bibinfo {pages} {107} (\bibinfo {year}
  {2022})}\BibitemShut {NoStop}%
\bibitem [{\citenamefont {Ong}\ \emph {et~al.}(2020)\citenamefont {Ong},
  \citenamefont {Schüppert}, \citenamefont {Jobez}, \citenamefont {Teller},
  \citenamefont {Ames}, \citenamefont {Fioretto}, \citenamefont {Friebe},
  \citenamefont {Lee}, \citenamefont {Colombe}, \citenamefont {Blatt},\ and\
  \citenamefont {Northup}}]{Ong_2020}%
  \BibitemOpen
  \bibfield  {author} {\bibinfo {author} {\bibfnamefont {F.~R.}\ \bibnamefont
  {Ong}}, \bibinfo {author} {\bibfnamefont {K.}~\bibnamefont {Schüppert}},
  \bibinfo {author} {\bibfnamefont {P.}~\bibnamefont {Jobez}}, \bibinfo
  {author} {\bibfnamefont {M.}~\bibnamefont {Teller}}, \bibinfo {author}
  {\bibfnamefont {B.}~\bibnamefont {Ames}}, \bibinfo {author} {\bibfnamefont
  {D.~A.}\ \bibnamefont {Fioretto}}, \bibinfo {author} {\bibfnamefont
  {K.}~\bibnamefont {Friebe}}, \bibinfo {author} {\bibfnamefont
  {M.}~\bibnamefont {Lee}}, \bibinfo {author} {\bibfnamefont {Y.}~\bibnamefont
  {Colombe}}, \bibinfo {author} {\bibfnamefont {R.}~\bibnamefont {Blatt}},\
  and\ \bibinfo {author} {\bibfnamefont {T.~E.}\ \bibnamefont {Northup}},\
  }\bibfield  {title} {\bibinfo {title} {Probing surface charge densities on
  optical fibers with a trapped ion},\ }\href
  {https://doi.org/10.1088/1367-2630/ab8af9} {\bibfield  {journal} {\bibinfo
  {journal} {New Journal of Physics}\ }\textbf {\bibinfo {volume} {22}},\
  \bibinfo {pages} {063018} (\bibinfo {year} {2020})}\BibitemShut {NoStop}%
\bibitem [{\citenamefont {Monroe}\ \emph {et~al.}(1995)\citenamefont {Monroe},
  \citenamefont {Meekhof}, \citenamefont {King}, \citenamefont {Jefferts},
  \citenamefont {Itano}, \citenamefont {Wineland},\ and\ \citenamefont
  {Gould}}]{Monroe1995}%
  \BibitemOpen
  \bibfield  {author} {\bibinfo {author} {\bibfnamefont {C.}~\bibnamefont
  {Monroe}}, \bibinfo {author} {\bibfnamefont {D.~M.}\ \bibnamefont {Meekhof}},
  \bibinfo {author} {\bibfnamefont {B.~E.}\ \bibnamefont {King}}, \bibinfo
  {author} {\bibfnamefont {S.~R.}\ \bibnamefont {Jefferts}}, \bibinfo {author}
  {\bibfnamefont {W.~M.}\ \bibnamefont {Itano}}, \bibinfo {author}
  {\bibfnamefont {D.~J.}\ \bibnamefont {Wineland}},\ and\ \bibinfo {author}
  {\bibfnamefont {P.}~\bibnamefont {Gould}},\ }\bibfield  {title} {\bibinfo
  {title} {{Resolved-Sideband Raman Cooling of a Bound Atom to the 3D
  Zero-Point Energy}},\ }\href {https://doi.org/10.1103/PhysRevLett.75.4011}
  {\bibfield  {journal} {\bibinfo  {journal} {Physical Review Letters}\
  }\textbf {\bibinfo {volume} {75}},\ \bibinfo {pages} {4011} (\bibinfo {year}
  {1995})}\BibitemShut {NoStop}%
\bibitem [{\citenamefont {Sedlacek}\ \emph
  {et~al.}(2018{\natexlab{a}})\citenamefont {Sedlacek}, \citenamefont {Stuart},
  \citenamefont {Slichter}, \citenamefont {Bruzewicz}, \citenamefont
  {McConnell}, \citenamefont {Sage},\ and\ \citenamefont
  {Chiaverini}}]{Sedlacek_2018_Evidence}%
  \BibitemOpen
  \bibfield  {author} {\bibinfo {author} {\bibfnamefont {J.~A.}\ \bibnamefont
  {Sedlacek}}, \bibinfo {author} {\bibfnamefont {J.}~\bibnamefont {Stuart}},
  \bibinfo {author} {\bibfnamefont {D.~H.}\ \bibnamefont {Slichter}}, \bibinfo
  {author} {\bibfnamefont {C.~D.}\ \bibnamefont {Bruzewicz}}, \bibinfo {author}
  {\bibfnamefont {R.}~\bibnamefont {McConnell}}, \bibinfo {author}
  {\bibfnamefont {J.~M.}\ \bibnamefont {Sage}},\ and\ \bibinfo {author}
  {\bibfnamefont {J.}~\bibnamefont {Chiaverini}},\ }\bibfield  {title}
  {\bibinfo {title} {Evidence for multiple mechanisms underlying surface
  electric-field noise in ion traps},\ }\href
  {https://doi.org/10.1103/PhysRevA.98.063430} {\bibfield  {journal} {\bibinfo
  {journal} {Phys. Rev. A}\ }\textbf {\bibinfo {volume} {98}},\ \bibinfo
  {pages} {063430} (\bibinfo {year} {2018}{\natexlab{a}})}\BibitemShut
  {NoStop}%
\bibitem [{\citenamefont {Low}\ \emph {et~al.}(2011)\citenamefont {Low},
  \citenamefont {Herskind},\ and\ \citenamefont {Chuang}}]{Low_2011}%
  \BibitemOpen
  \bibfield  {author} {\bibinfo {author} {\bibfnamefont {G.~H.}\ \bibnamefont
  {Low}}, \bibinfo {author} {\bibfnamefont {P.~F.}\ \bibnamefont {Herskind}},\
  and\ \bibinfo {author} {\bibfnamefont {I.~L.}\ \bibnamefont {Chuang}},\
  }\bibfield  {title} {\bibinfo {title} {Finite-geometry models of electric
  field noise from patch potentials in ion traps},\ }\href
  {https://doi.org/10.1103/PhysRevA.84.053425} {\bibfield  {journal} {\bibinfo
  {journal} {Phys. Rev. A}\ }\textbf {\bibinfo {volume} {84}},\ \bibinfo
  {pages} {053425} (\bibinfo {year} {2011})}\BibitemShut {NoStop}%
\bibitem [{\citenamefont {Kumph}\ \emph {et~al.}(2016)\citenamefont {Kumph},
  \citenamefont {Holz}, \citenamefont {Langer}, \citenamefont {Meraner},
  \citenamefont {Niedermayr}, \citenamefont {Brownnutt},\ and\ \citenamefont
  {Blatt}}]{Kumph2016}%
  \BibitemOpen
  \bibfield  {author} {\bibinfo {author} {\bibfnamefont {M.}~\bibnamefont
  {Kumph}}, \bibinfo {author} {\bibfnamefont {P.}~\bibnamefont {Holz}},
  \bibinfo {author} {\bibfnamefont {K.}~\bibnamefont {Langer}}, \bibinfo
  {author} {\bibfnamefont {M.}~\bibnamefont {Meraner}}, \bibinfo {author}
  {\bibfnamefont {M.}~\bibnamefont {Niedermayr}}, \bibinfo {author}
  {\bibfnamefont {M.}~\bibnamefont {Brownnutt}},\ and\ \bibinfo {author}
  {\bibfnamefont {R.}~\bibnamefont {Blatt}},\ }\bibfield  {title} {\bibinfo
  {title} {Operation of a planar-electrode ion-trap array with adjustable rf
  electrodes},\ }\href {https://doi.org/10.1088/1367-2630/18/2/023047}
  {\bibfield  {journal} {\bibinfo  {journal} {New Journal of Physics}\ }\textbf
  {\bibinfo {volume} {18}},\ \bibinfo {pages} {023047} (\bibinfo {year}
  {2016})}\BibitemShut {NoStop}%
\bibitem [{\citenamefont {Martinetz}\ \emph {et~al.}(2022)\citenamefont
  {Martinetz}, \citenamefont {Hornberger},\ and\ \citenamefont
  {Stickler}}]{Martinetz_2022}%
  \BibitemOpen
  \bibfield  {author} {\bibinfo {author} {\bibfnamefont {L.}~\bibnamefont
  {Martinetz}}, \bibinfo {author} {\bibfnamefont {K.}~\bibnamefont
  {Hornberger}},\ and\ \bibinfo {author} {\bibfnamefont {B.~A.}\ \bibnamefont
  {Stickler}},\ }\bibfield  {title} {\bibinfo {title} {Surface-induced
  decoherence and heating of charged particles},\ }\href
  {https://doi.org/10.1103/PRXQuantum.3.030327} {\bibfield  {journal} {\bibinfo
   {journal} {PRX Quantum}\ }\textbf {\bibinfo {volume} {3}},\ \bibinfo {pages}
  {030327} (\bibinfo {year} {2022})}\BibitemShut {NoStop}%
\bibitem [{\citenamefont {Safavi-Naini}\ \emph {et~al.}(2011)\citenamefont
  {Safavi-Naini}, \citenamefont {Rabl}, \citenamefont {Weck},\ and\
  \citenamefont {Sadeghpour}}]{Safavi-Naini_2011}%
  \BibitemOpen
  \bibfield  {author} {\bibinfo {author} {\bibfnamefont {A.}~\bibnamefont
  {Safavi-Naini}}, \bibinfo {author} {\bibfnamefont {P.}~\bibnamefont {Rabl}},
  \bibinfo {author} {\bibfnamefont {P.~F.}\ \bibnamefont {Weck}},\ and\
  \bibinfo {author} {\bibfnamefont {H.~R.}\ \bibnamefont {Sadeghpour}},\
  }\bibfield  {title} {\bibinfo {title} {Microscopic model of
  electric-field-noise heating in ion traps},\ }\href
  {https://doi.org/10.1103/PhysRevA.84.023412} {\bibfield  {journal} {\bibinfo
  {journal} {Phys. Rev. A}\ }\textbf {\bibinfo {volume} {84}},\ \bibinfo
  {pages} {023412} (\bibinfo {year} {2011})}\BibitemShut {NoStop}%
\bibitem [{\citenamefont {Kim}\ \emph {et~al.}(2017)\citenamefont {Kim},
  \citenamefont {Safavi-Naini}, \citenamefont {Hite}, \citenamefont {McKay},
  \citenamefont {Pappas}, \citenamefont {Weck},\ and\ \citenamefont
  {Sadeghpour}}]{Kim_2017}%
  \BibitemOpen
  \bibfield  {author} {\bibinfo {author} {\bibfnamefont {E.}~\bibnamefont
  {Kim}}, \bibinfo {author} {\bibfnamefont {A.}~\bibnamefont {Safavi-Naini}},
  \bibinfo {author} {\bibfnamefont {D.~A.}\ \bibnamefont {Hite}}, \bibinfo
  {author} {\bibfnamefont {K.~S.}\ \bibnamefont {McKay}}, \bibinfo {author}
  {\bibfnamefont {D.~P.}\ \bibnamefont {Pappas}}, \bibinfo {author}
  {\bibfnamefont {P.~F.}\ \bibnamefont {Weck}},\ and\ \bibinfo {author}
  {\bibfnamefont {H.~R.}\ \bibnamefont {Sadeghpour}},\ }\bibfield  {title}
  {\bibinfo {title} {Electric-field noise from carbon-adatom diffusion on a
  au(110) surface: First-principles calculations and experiments},\ }\href
  {https://doi.org/10.1103/PhysRevA.95.033407} {\bibfield  {journal} {\bibinfo
  {journal} {Phys. Rev. A}\ }\textbf {\bibinfo {volume} {95}},\ \bibinfo
  {pages} {033407} (\bibinfo {year} {2017})}\BibitemShut {NoStop}%
\bibitem [{\citenamefont {Foulon}\ \emph {et~al.}(2022)\citenamefont {Foulon},
  \citenamefont {Ray}, \citenamefont {Kim}, \citenamefont {Liu}, \citenamefont
  {Rubenstein},\ and\ \citenamefont {Lordi}}]{Foulon_2022}%
  \BibitemOpen
  \bibfield  {author} {\bibinfo {author} {\bibfnamefont {B.~L.}\ \bibnamefont
  {Foulon}}, \bibinfo {author} {\bibfnamefont {K.~G.}\ \bibnamefont {Ray}},
  \bibinfo {author} {\bibfnamefont {C.-E.}\ \bibnamefont {Kim}}, \bibinfo
  {author} {\bibfnamefont {Y.}~\bibnamefont {Liu}}, \bibinfo {author}
  {\bibfnamefont {B.~M.}\ \bibnamefont {Rubenstein}},\ and\ \bibinfo {author}
  {\bibfnamefont {V.}~\bibnamefont {Lordi}},\ }\bibfield  {title} {\bibinfo
  {title} {$1/\ensuremath{\omega}$ electric-field noise in surface ion traps
  from correlated adsorbate dynamics},\ }\href
  {https://doi.org/10.1103/PhysRevA.105.013107} {\bibfield  {journal} {\bibinfo
   {journal} {Phys. Rev. A}\ }\textbf {\bibinfo {volume} {105}},\ \bibinfo
  {pages} {013107} (\bibinfo {year} {2022})}\BibitemShut {NoStop}%
\bibitem [{\citenamefont {Noel}\ \emph {et~al.}(2019)\citenamefont {Noel},
  \citenamefont {Berlin-Udi}, \citenamefont {Matthiesen}, \citenamefont {Yu},
  \citenamefont {Zhou}, \citenamefont {Lordi},\ and\ \citenamefont
  {H\"affner}}]{Noel_2019}%
  \BibitemOpen
  \bibfield  {author} {\bibinfo {author} {\bibfnamefont {C.}~\bibnamefont
  {Noel}}, \bibinfo {author} {\bibfnamefont {M.}~\bibnamefont {Berlin-Udi}},
  \bibinfo {author} {\bibfnamefont {C.}~\bibnamefont {Matthiesen}}, \bibinfo
  {author} {\bibfnamefont {J.}~\bibnamefont {Yu}}, \bibinfo {author}
  {\bibfnamefont {Y.}~\bibnamefont {Zhou}}, \bibinfo {author} {\bibfnamefont
  {V.}~\bibnamefont {Lordi}},\ and\ \bibinfo {author} {\bibfnamefont
  {H.}~\bibnamefont {H\"affner}},\ }\bibfield  {title} {\bibinfo {title}
  {Electric-field noise from thermally activated fluctuators in a surface ion
  trap},\ }\href {https://doi.org/10.1103/PhysRevA.99.063427} {\bibfield
  {journal} {\bibinfo  {journal} {Phys. Rev. A}\ }\textbf {\bibinfo {volume}
  {99}},\ \bibinfo {pages} {063427} (\bibinfo {year} {2019})}\BibitemShut
  {NoStop}%
\bibitem [{\citenamefont {Boldin}\ \emph {et~al.}(2018)\citenamefont {Boldin},
  \citenamefont {Kraft},\ and\ \citenamefont {Wunderlich}}]{Boldin_2018}%
  \BibitemOpen
  \bibfield  {author} {\bibinfo {author} {\bibfnamefont {I.~A.}\ \bibnamefont
  {Boldin}}, \bibinfo {author} {\bibfnamefont {A.}~\bibnamefont {Kraft}},\ and\
  \bibinfo {author} {\bibfnamefont {C.}~\bibnamefont {Wunderlich}},\ }\bibfield
   {title} {\bibinfo {title} {Measuring anomalous heating in a planar ion trap
  with variable ion-surface separation},\ }\href
  {https://doi.org/10.1103/PhysRevLett.120.023201} {\bibfield  {journal}
  {\bibinfo  {journal} {Phys. Rev. Lett.}\ }\textbf {\bibinfo {volume} {120}},\
  \bibinfo {pages} {023201} (\bibinfo {year} {2018})}\BibitemShut {NoStop}%
\bibitem [{\citenamefont {Sedlacek}\ \emph
  {et~al.}(2018{\natexlab{b}})\citenamefont {Sedlacek}, \citenamefont {Greene},
  \citenamefont {Stuart}, \citenamefont {McConnell}, \citenamefont {Bruzewicz},
  \citenamefont {Sage},\ and\ \citenamefont
  {Chiaverini}}]{Sedlacek_2018_Distance}%
  \BibitemOpen
  \bibfield  {author} {\bibinfo {author} {\bibfnamefont {J.~A.}\ \bibnamefont
  {Sedlacek}}, \bibinfo {author} {\bibfnamefont {A.}~\bibnamefont {Greene}},
  \bibinfo {author} {\bibfnamefont {J.}~\bibnamefont {Stuart}}, \bibinfo
  {author} {\bibfnamefont {R.}~\bibnamefont {McConnell}}, \bibinfo {author}
  {\bibfnamefont {C.~D.}\ \bibnamefont {Bruzewicz}}, \bibinfo {author}
  {\bibfnamefont {J.~M.}\ \bibnamefont {Sage}},\ and\ \bibinfo {author}
  {\bibfnamefont {J.}~\bibnamefont {Chiaverini}},\ }\bibfield  {title}
  {\bibinfo {title} {Distance scaling of electric-field noise in a
  surface-electrode ion trap},\ }\href
  {https://doi.org/10.1103/PhysRevA.97.020302} {\bibfield  {journal} {\bibinfo
  {journal} {Phys. Rev. A}\ }\textbf {\bibinfo {volume} {97}},\ \bibinfo
  {pages} {020302} (\bibinfo {year} {2018}{\natexlab{b}})}\BibitemShut
  {NoStop}%
\bibitem [{\citenamefont {An}\ \emph {et~al.}(2019)\citenamefont {An},
  \citenamefont {Matthiesen}, \citenamefont {Urban},\ and\ \citenamefont
  {H\"affner}}]{An_2019}%
  \BibitemOpen
  \bibfield  {author} {\bibinfo {author} {\bibfnamefont {D.}~\bibnamefont
  {An}}, \bibinfo {author} {\bibfnamefont {C.}~\bibnamefont {Matthiesen}},
  \bibinfo {author} {\bibfnamefont {E.}~\bibnamefont {Urban}},\ and\ \bibinfo
  {author} {\bibfnamefont {H.}~\bibnamefont {H\"affner}},\ }\bibfield  {title}
  {\bibinfo {title} {Distance scaling and polarization of electric-field noise
  in a surface ion trap},\ }\href {https://doi.org/10.1103/PhysRevA.100.063405}
  {\bibfield  {journal} {\bibinfo  {journal} {Phys. Rev. A}\ }\textbf {\bibinfo
  {volume} {100}},\ \bibinfo {pages} {063405} (\bibinfo {year}
  {2019})}\BibitemShut {NoStop}%
\bibitem [{\citenamefont {McKay}\ \emph {et~al.}(2021)\citenamefont {McKay},
  \citenamefont {Hite}, \citenamefont {Kent}, \citenamefont {Kotler},
  \citenamefont {Leibfried}, \citenamefont {Slichter}, \citenamefont {Wilson},\
  and\ \citenamefont {Pappas}}]{McKay_2021}%
  \BibitemOpen
  \bibfield  {author} {\bibinfo {author} {\bibfnamefont {K.~S.}\ \bibnamefont
  {McKay}}, \bibinfo {author} {\bibfnamefont {D.~A.}\ \bibnamefont {Hite}},
  \bibinfo {author} {\bibfnamefont {P.~D.}\ \bibnamefont {Kent}}, \bibinfo
  {author} {\bibfnamefont {S.}~\bibnamefont {Kotler}}, \bibinfo {author}
  {\bibfnamefont {D.}~\bibnamefont {Leibfried}}, \bibinfo {author}
  {\bibfnamefont {D.~H.}\ \bibnamefont {Slichter}}, \bibinfo {author}
  {\bibfnamefont {A.~C.}\ \bibnamefont {Wilson}},\ and\ \bibinfo {author}
  {\bibfnamefont {D.~P.}\ \bibnamefont {Pappas}},\ }\bibfield  {title}
  {\bibinfo {title} {Measurement of electric-field noise from interchangeable
  samples with a trapped-ion sensor},\ }\href
  {https://doi.org/10.1103/PhysRevA.104.052610} {\bibfield  {journal} {\bibinfo
   {journal} {Phys. Rev. A}\ }\textbf {\bibinfo {volume} {104}},\ \bibinfo
  {pages} {052610} (\bibinfo {year} {2021})}\BibitemShut {NoStop}%
\bibitem [{\citenamefont {H\'eritier}\ \emph {et~al.}(2021)\citenamefont
  {H\'eritier}, \citenamefont {Pachlatko}, \citenamefont {Tao}, \citenamefont
  {Abendroth}, \citenamefont {Degen},\ and\ \citenamefont
  {Eichler}}]{Heritier_2021}%
  \BibitemOpen
  \bibfield  {author} {\bibinfo {author} {\bibfnamefont {M.}~\bibnamefont
  {H\'eritier}}, \bibinfo {author} {\bibfnamefont {R.}~\bibnamefont
  {Pachlatko}}, \bibinfo {author} {\bibfnamefont {Y.}~\bibnamefont {Tao}},
  \bibinfo {author} {\bibfnamefont {J.~M.}\ \bibnamefont {Abendroth}}, \bibinfo
  {author} {\bibfnamefont {C.~L.}\ \bibnamefont {Degen}},\ and\ \bibinfo
  {author} {\bibfnamefont {A.}~\bibnamefont {Eichler}},\ }\bibfield  {title}
  {\bibinfo {title} {Spatial correlation between fluctuating and static fields
  over metal and dielectric substrates},\ }\href
  {https://doi.org/10.1103/PhysRevLett.127.216101} {\bibfield  {journal}
  {\bibinfo  {journal} {Phys. Rev. Lett.}\ }\textbf {\bibinfo {volume} {127}},\
  \bibinfo {pages} {216101} (\bibinfo {year} {2021})}\BibitemShut {NoStop}%
\bibitem [{\citenamefont {Warring}\ \emph {et~al.}(2013)\citenamefont
  {Warring}, \citenamefont {Ospelkaus}, \citenamefont {Colombe}, \citenamefont
  {Brown}, \citenamefont {Amini}, \citenamefont {Carsjens}, \citenamefont
  {Leibfried},\ and\ \citenamefont {Wineland}}]{Warring_2013}%
  \BibitemOpen
  \bibfield  {author} {\bibinfo {author} {\bibfnamefont {U.}~\bibnamefont
  {Warring}}, \bibinfo {author} {\bibfnamefont {C.}~\bibnamefont {Ospelkaus}},
  \bibinfo {author} {\bibfnamefont {Y.}~\bibnamefont {Colombe}}, \bibinfo
  {author} {\bibfnamefont {K.~R.}\ \bibnamefont {Brown}}, \bibinfo {author}
  {\bibfnamefont {J.~M.}\ \bibnamefont {Amini}}, \bibinfo {author}
  {\bibfnamefont {M.}~\bibnamefont {Carsjens}}, \bibinfo {author}
  {\bibfnamefont {D.}~\bibnamefont {Leibfried}},\ and\ \bibinfo {author}
  {\bibfnamefont {D.~J.}\ \bibnamefont {Wineland}},\ }\bibfield  {title}
  {\bibinfo {title} {Techniques for microwave near-field quantum control of
  trapped ions},\ }\href {https://doi.org/10.1103/PhysRevA.87.013437}
  {\bibfield  {journal} {\bibinfo  {journal} {Phys. Rev. A}\ }\textbf {\bibinfo
  {volume} {87}},\ \bibinfo {pages} {013437} (\bibinfo {year}
  {2013})}\BibitemShut {NoStop}%
\bibitem [{\citenamefont {Fan}\ \emph {et~al.}(2014)\citenamefont {Fan},
  \citenamefont {Kumar}, \citenamefont {Daschner}, \citenamefont {K\"{u}bler},\
  and\ \citenamefont {Shaffer}}]{Fan_14}%
  \BibitemOpen
  \bibfield  {author} {\bibinfo {author} {\bibfnamefont {H.~Q.}\ \bibnamefont
  {Fan}}, \bibinfo {author} {\bibfnamefont {S.}~\bibnamefont {Kumar}}, \bibinfo
  {author} {\bibfnamefont {R.}~\bibnamefont {Daschner}}, \bibinfo {author}
  {\bibfnamefont {H.}~\bibnamefont {K\"{u}bler}},\ and\ \bibinfo {author}
  {\bibfnamefont {J.~P.}\ \bibnamefont {Shaffer}},\ }\bibfield  {title}
  {\bibinfo {title} {Subwavelength microwave electric-field imaging using
  rydberg atoms inside atomic vapor cells},\ }\href
  {https://doi.org/10.1364/OL.39.003030} {\bibfield  {journal} {\bibinfo
  {journal} {Opt. Lett.}\ }\textbf {\bibinfo {volume} {39}},\ \bibinfo {pages}
  {3030} (\bibinfo {year} {2014})}\BibitemShut {NoStop}%
\bibitem [{\citenamefont {Weber}\ \emph {et~al.}(2023)\citenamefont {Weber},
  \citenamefont {Löschnauer}, \citenamefont {Wolf}, \citenamefont {Gely},
  \citenamefont {Hanley}, \citenamefont {Goodwin}, \citenamefont {Ballance},
  \citenamefont {Harty},\ and\ \citenamefont {Lucas}}]{Weber_2024}%
  \BibitemOpen
  \bibfield  {author} {\bibinfo {author} {\bibfnamefont {M.~A.}\ \bibnamefont
  {Weber}}, \bibinfo {author} {\bibfnamefont {C.}~\bibnamefont {Löschnauer}},
  \bibinfo {author} {\bibfnamefont {J.}~\bibnamefont {Wolf}}, \bibinfo {author}
  {\bibfnamefont {M.~F.}\ \bibnamefont {Gely}}, \bibinfo {author}
  {\bibfnamefont {R.~K.}\ \bibnamefont {Hanley}}, \bibinfo {author}
  {\bibfnamefont {J.~F.}\ \bibnamefont {Goodwin}}, \bibinfo {author}
  {\bibfnamefont {C.~J.}\ \bibnamefont {Ballance}}, \bibinfo {author}
  {\bibfnamefont {T.~P.}\ \bibnamefont {Harty}},\ and\ \bibinfo {author}
  {\bibfnamefont {D.~M.}\ \bibnamefont {Lucas}},\ }\bibfield  {title} {\bibinfo
  {title} {Cryogenic ion trap system for high-fidelity near-field
  microwave-driven quantum logic},\ }\href
  {https://doi.org/10.1088/2058-9565/acfba8} {\bibfield  {journal} {\bibinfo
  {journal} {Quantum Science and Technology}\ }\textbf {\bibinfo {volume}
  {9}},\ \bibinfo {pages} {015007} (\bibinfo {year} {2023})}\BibitemShut
  {NoStop}%
\bibitem [{\citenamefont {Galve}\ \emph {et~al.}(2017)\citenamefont {Galve},
  \citenamefont {Alonso},\ and\ \citenamefont {Zambrini}}]{Galve_2017}%
  \BibitemOpen
  \bibfield  {author} {\bibinfo {author} {\bibfnamefont {F.}~\bibnamefont
  {Galve}}, \bibinfo {author} {\bibfnamefont {J.}~\bibnamefont {Alonso}},\ and\
  \bibinfo {author} {\bibfnamefont {R.}~\bibnamefont {Zambrini}},\ }\bibfield
  {title} {\bibinfo {title} {Multi-ion sensing of dipolar noise sources in ion
  traps},\ }\href {https://doi.org/10.1103/PhysRevA.96.033409} {\bibfield
  {journal} {\bibinfo  {journal} {Phys. Rev. A}\ }\textbf {\bibinfo {volume}
  {96}},\ \bibinfo {pages} {033409} (\bibinfo {year} {2017})}\BibitemShut
  {NoStop}%
\bibitem [{\citenamefont {Talukdar}\ \emph {et~al.}(2016)\citenamefont
  {Talukdar}, \citenamefont {Gorman}, \citenamefont {Daniilidis}, \citenamefont
  {Schindler}, \citenamefont {Ebadi}, \citenamefont {Kaufmann}, \citenamefont
  {Zhang},\ and\ \citenamefont {H\"affner}}]{Talukdar_2016}%
  \BibitemOpen
  \bibfield  {author} {\bibinfo {author} {\bibfnamefont {I.}~\bibnamefont
  {Talukdar}}, \bibinfo {author} {\bibfnamefont {D.~J.}\ \bibnamefont
  {Gorman}}, \bibinfo {author} {\bibfnamefont {N.}~\bibnamefont {Daniilidis}},
  \bibinfo {author} {\bibfnamefont {P.}~\bibnamefont {Schindler}}, \bibinfo
  {author} {\bibfnamefont {S.}~\bibnamefont {Ebadi}}, \bibinfo {author}
  {\bibfnamefont {H.}~\bibnamefont {Kaufmann}}, \bibinfo {author}
  {\bibfnamefont {T.}~\bibnamefont {Zhang}},\ and\ \bibinfo {author}
  {\bibfnamefont {H.}~\bibnamefont {H\"affner}},\ }\bibfield  {title} {\bibinfo
  {title} {Implications of surface noise for the motional coherence of trapped
  ions},\ }\href {https://doi.org/10.1103/PhysRevA.93.043415} {\bibfield
  {journal} {\bibinfo  {journal} {Phys. Rev. A}\ }\textbf {\bibinfo {volume}
  {93}},\ \bibinfo {pages} {043415} (\bibinfo {year} {2016})}\BibitemShut
  {NoStop}%
\bibitem [{\citenamefont {Mordini}\ \emph {et~al.}(2023)\citenamefont
  {Mordini}, \citenamefont {Lancellotti}, \citenamefont {Negnevitsky},
  \citenamefont {Marinelli}, \citenamefont {Oswald},\ and\ \citenamefont
  {Saegesser}}]{Mordini_pytrans}%
  \BibitemOpen
  \bibfield  {author} {\bibinfo {author} {\bibfnamefont {C.}~\bibnamefont
  {Mordini}}, \bibinfo {author} {\bibfnamefont {F.}~\bibnamefont
  {Lancellotti}}, \bibinfo {author} {\bibfnamefont {V.}~\bibnamefont
  {Negnevitsky}}, \bibinfo {author} {\bibfnamefont {M.}~\bibnamefont
  {Marinelli}}, \bibinfo {author} {\bibfnamefont {R.}~\bibnamefont {Oswald}},\
  and\ \bibinfo {author} {\bibfnamefont {T.}~\bibnamefont {Saegesser}},\ }\href
  {https://doi.org/10.5281/zenodo.10204606} {\bibinfo {title} {pytrans}}
  (\bibinfo {year} {2023})\BibitemShut {NoStop}%
\bibitem [{\citenamefont {Bradski}(2000)}]{opencv_library}%
  \BibitemOpen
  \bibfield  {author} {\bibinfo {author} {\bibfnamefont {G.}~\bibnamefont
  {Bradski}},\ }\bibfield  {title} {\bibinfo {title} {{The OpenCV Library}},\
  }\href@noop {} {\bibfield  {journal} {\bibinfo  {journal} {Dr. Dobb's Journal
  of Software Tools}\ } (\bibinfo {year} {2000})}\BibitemShut {NoStop}%
\bibitem [{Fun(2012)}]{Fundamentals_Light_Microscopy}%
  \BibitemOpen
  \bibinfo {title} {Diffraction and spatial resolution},\ in\ \href
  {https://doi.org/https://doi.org/10.1002/9781118382905} {\emph {\bibinfo
  {booktitle} {Fundamentals of Light Microscopy and Electronic Imaging}}}\
  (\bibinfo  {publisher} {John Wiley \& Sons, Ltd},\ \bibinfo {year} {2012})\
  Chap.~\bibinfo {chapter} {6}, p.\ \bibinfo {pages} {109}\BibitemShut
  {NoStop}%
\bibitem [{\citenamefont {Lee}\ \emph {et~al.}(2024)\citenamefont {Lee},
  \citenamefont {Chung}, \citenamefont {Jeon}, \citenamefont {Cho},
  \citenamefont {Choi}, \citenamefont {Yoo}, \citenamefont {Jung},
  \citenamefont {Jeong}, \citenamefont {Kim}, \citenamefont {Cho},\ and\
  \citenamefont {Kim}}]{Lee_2024}%
  \BibitemOpen
  \bibfield  {author} {\bibinfo {author} {\bibfnamefont {W.}~\bibnamefont
  {Lee}}, \bibinfo {author} {\bibfnamefont {D.}~\bibnamefont {Chung}}, \bibinfo
  {author} {\bibfnamefont {H.}~\bibnamefont {Jeon}}, \bibinfo {author}
  {\bibfnamefont {B.}~\bibnamefont {Cho}}, \bibinfo {author} {\bibfnamefont
  {K.}~\bibnamefont {Choi}}, \bibinfo {author} {\bibfnamefont {S.}~\bibnamefont
  {Yoo}}, \bibinfo {author} {\bibfnamefont {C.}~\bibnamefont {Jung}}, \bibinfo
  {author} {\bibfnamefont {J.}~\bibnamefont {Jeong}}, \bibinfo {author}
  {\bibfnamefont {C.}~\bibnamefont {Kim}}, \bibinfo {author} {\bibfnamefont
  {D.-I.~D.}\ \bibnamefont {Cho}},\ and\ \bibinfo {author} {\bibfnamefont
  {T.}~\bibnamefont {Kim}},\ }\bibfield  {title} {\bibinfo {title}
  {Photoinduced charge-carrier dynamics in a semiconductor-based ion trap
  investigated via motion-sensitive qubit transitions},\ }\href
  {https://doi.org/10.1103/PhysRevA.109.043106} {\bibfield  {journal} {\bibinfo
   {journal} {Phys. Rev. A}\ }\textbf {\bibinfo {volume} {109}},\ \bibinfo
  {pages} {043106} (\bibinfo {year} {2024})}\BibitemShut {NoStop}%
\bibitem [{\citenamefont {Auchter}\ \emph {et~al.}(2022)\citenamefont
  {Auchter}, \citenamefont {Axline}, \citenamefont {Decaroli}, \citenamefont
  {Valentini}, \citenamefont {Purwin}, \citenamefont {Oswald}, \citenamefont
  {Matt}, \citenamefont {Aschauer}, \citenamefont {Colombe}, \citenamefont
  {Holz}, \citenamefont {Monz}, \citenamefont {Blatt}, \citenamefont
  {Schindler}, \citenamefont {R{\"o}ssler},\ and\ \citenamefont
  {Home}}]{auchterIndustriallyMicrofabricatedIon2022}%
  \BibitemOpen
  \bibfield  {author} {\bibinfo {author} {\bibfnamefont {S.}~\bibnamefont
  {Auchter}}, \bibinfo {author} {\bibfnamefont {C.}~\bibnamefont {Axline}},
  \bibinfo {author} {\bibfnamefont {C.}~\bibnamefont {Decaroli}}, \bibinfo
  {author} {\bibfnamefont {M.}~\bibnamefont {Valentini}}, \bibinfo {author}
  {\bibfnamefont {L.}~\bibnamefont {Purwin}}, \bibinfo {author} {\bibfnamefont
  {R.}~\bibnamefont {Oswald}}, \bibinfo {author} {\bibfnamefont
  {R.}~\bibnamefont {Matt}}, \bibinfo {author} {\bibfnamefont {E.}~\bibnamefont
  {Aschauer}}, \bibinfo {author} {\bibfnamefont {Y.}~\bibnamefont {Colombe}},
  \bibinfo {author} {\bibfnamefont {P.}~\bibnamefont {Holz}}, \bibinfo {author}
  {\bibfnamefont {T.}~\bibnamefont {Monz}}, \bibinfo {author} {\bibfnamefont
  {R.}~\bibnamefont {Blatt}}, \bibinfo {author} {\bibfnamefont
  {P.}~\bibnamefont {Schindler}}, \bibinfo {author} {\bibfnamefont
  {C.}~\bibnamefont {R{\"o}ssler}},\ and\ \bibinfo {author} {\bibfnamefont
  {J.}~\bibnamefont {Home}},\ }\bibfield  {title} {\bibinfo {title}
  {Industrially microfabricated ion trap with 1 {{eV}} trap depth},\ }\href
  {https://doi.org/10.1088/2058-9565/ac7072} {\bibfield  {journal} {\bibinfo
  {journal} {Quantum Science and Technology}\ }\textbf {\bibinfo {volume}
  {7}},\ \bibinfo {pages} {035015} (\bibinfo {year} {2022})}\BibitemShut
  {NoStop}%
\bibitem [{\citenamefont {H{\"a}rter}\ \emph {et~al.}(2014)\citenamefont
  {H{\"a}rter}, \citenamefont {Kr{\"u}kow}, \citenamefont {Brunner},\ and\
  \citenamefont {Hecker~Denschlag}}]{Harter_2014}%
  \BibitemOpen
  \bibfield  {author} {\bibinfo {author} {\bibfnamefont {A.}~\bibnamefont
  {H{\"a}rter}}, \bibinfo {author} {\bibfnamefont {A.}~\bibnamefont
  {Kr{\"u}kow}}, \bibinfo {author} {\bibfnamefont {A.}~\bibnamefont
  {Brunner}},\ and\ \bibinfo {author} {\bibfnamefont {J.}~\bibnamefont
  {Hecker~Denschlag}},\ }\bibfield  {title} {\bibinfo {title} {Long-term drifts
  of stray electric fields in a paul trap},\ }\href@noop {} {\bibfield
  {journal} {\bibinfo  {journal} {Applied Physics B}\ }\textbf {\bibinfo
  {volume} {114}},\ \bibinfo {pages} {275} (\bibinfo {year}
  {2014})}\BibitemShut {NoStop}%
\bibitem [{\citenamefont {Doret}\ \emph {et~al.}(2012)\citenamefont {Doret},
  \citenamefont {Amini}, \citenamefont {Wright}, \citenamefont {Volin},
  \citenamefont {Killian}, \citenamefont {Ozakin}, \citenamefont {Denison},
  \citenamefont {Hayden}, \citenamefont {Pai}, \citenamefont {Slusher},\ and\
  \citenamefont {Harter}}]{Doret_2012}%
  \BibitemOpen
  \bibfield  {author} {\bibinfo {author} {\bibfnamefont {S.~C.}\ \bibnamefont
  {Doret}}, \bibinfo {author} {\bibfnamefont {J.~M.}\ \bibnamefont {Amini}},
  \bibinfo {author} {\bibfnamefont {K.}~\bibnamefont {Wright}}, \bibinfo
  {author} {\bibfnamefont {C.}~\bibnamefont {Volin}}, \bibinfo {author}
  {\bibfnamefont {T.}~\bibnamefont {Killian}}, \bibinfo {author} {\bibfnamefont
  {A.}~\bibnamefont {Ozakin}}, \bibinfo {author} {\bibfnamefont
  {D.}~\bibnamefont {Denison}}, \bibinfo {author} {\bibfnamefont
  {H.}~\bibnamefont {Hayden}}, \bibinfo {author} {\bibfnamefont {C.-S.}\
  \bibnamefont {Pai}}, \bibinfo {author} {\bibfnamefont {R.~E.}\ \bibnamefont
  {Slusher}},\ and\ \bibinfo {author} {\bibfnamefont {A.~W.}\ \bibnamefont
  {Harter}},\ }\bibfield  {title} {\bibinfo {title} {Controlling trapping
  potentials and stray electric fields in a microfabricated ion trap through
  design and compensation},\ }\href
  {https://doi.org/10.1088/1367-2630/14/7/073012} {\bibfield  {journal}
  {\bibinfo  {journal} {New Journal of Physics}\ }\textbf {\bibinfo {volume}
  {14}},\ \bibinfo {pages} {073012} (\bibinfo {year} {2012})}\BibitemShut
  {NoStop}%
\bibitem [{\citenamefont {Narayanan}\ \emph {et~al.}(2011)\citenamefont
  {Narayanan}, \citenamefont {Daniilidis}, \citenamefont {Möller},
  \citenamefont {Clark}, \citenamefont {Ziesel}, \citenamefont {Singer},
  \citenamefont {Schmidt-Kaler},\ and\ \citenamefont
  {Häffner}}]{Narayanan_2011}%
  \BibitemOpen
  \bibfield  {author} {\bibinfo {author} {\bibfnamefont {S.}~\bibnamefont
  {Narayanan}}, \bibinfo {author} {\bibfnamefont {N.}~\bibnamefont
  {Daniilidis}}, \bibinfo {author} {\bibfnamefont {S.~A.}\ \bibnamefont
  {Möller}}, \bibinfo {author} {\bibfnamefont {R.}~\bibnamefont {Clark}},
  \bibinfo {author} {\bibfnamefont {F.}~\bibnamefont {Ziesel}}, \bibinfo
  {author} {\bibfnamefont {K.}~\bibnamefont {Singer}}, \bibinfo {author}
  {\bibfnamefont {F.}~\bibnamefont {Schmidt-Kaler}},\ and\ \bibinfo {author}
  {\bibfnamefont {H.}~\bibnamefont {Häffner}},\ }\bibfield  {title} {\bibinfo
  {title} {Electric field compensation and sensing with a single ion in a
  planar trap},\ }\href {https://doi.org/10.1063/1.3665647} {\bibfield
  {journal} {\bibinfo  {journal} {Journal of Applied Physics}\ }\textbf
  {\bibinfo {volume} {110}},\ \bibinfo {pages} {114909} (\bibinfo {year}
  {2011})}\BibitemShut {NoStop}%
\bibitem [{\citenamefont {Davtyan}\ \emph {et~al.}(2018)\citenamefont
  {Davtyan}, \citenamefont {Machluf}, \citenamefont {Soudijn}, \citenamefont
  {Naber}, \citenamefont {van Druten}, \citenamefont {van Linden van~den
  Heuvell},\ and\ \citenamefont {Spreeuw}}]{Davtyan_2018}%
  \BibitemOpen
  \bibfield  {author} {\bibinfo {author} {\bibfnamefont {D.}~\bibnamefont
  {Davtyan}}, \bibinfo {author} {\bibfnamefont {S.}~\bibnamefont {Machluf}},
  \bibinfo {author} {\bibfnamefont {M.~L.}\ \bibnamefont {Soudijn}}, \bibinfo
  {author} {\bibfnamefont {J.~B.}\ \bibnamefont {Naber}}, \bibinfo {author}
  {\bibfnamefont {N.~J.}\ \bibnamefont {van Druten}}, \bibinfo {author}
  {\bibfnamefont {H.~B.}\ \bibnamefont {van Linden van~den Heuvell}},\ and\
  \bibinfo {author} {\bibfnamefont {R.~J.~C.}\ \bibnamefont {Spreeuw}},\
  }\bibfield  {title} {\bibinfo {title} {Controlling stray electric fields on
  an atom chip for experiments on rydberg atoms},\ }\href
  {https://doi.org/10.1103/PhysRevA.97.023418} {\bibfield  {journal} {\bibinfo
  {journal} {Phys. Rev. A}\ }\textbf {\bibinfo {volume} {97}},\ \bibinfo
  {pages} {023418} (\bibinfo {year} {2018})}\BibitemShut {NoStop}%
\bibitem [{\citenamefont {Sedlacek}\ \emph {et~al.}(2016)\citenamefont
  {Sedlacek}, \citenamefont {Kim}, \citenamefont {Rittenhouse}, \citenamefont
  {Weck}, \citenamefont {Sadeghpour},\ and\ \citenamefont
  {Shaffer}}]{Sedlacek_2016}%
  \BibitemOpen
  \bibfield  {author} {\bibinfo {author} {\bibfnamefont {J.~A.}\ \bibnamefont
  {Sedlacek}}, \bibinfo {author} {\bibfnamefont {E.}~\bibnamefont {Kim}},
  \bibinfo {author} {\bibfnamefont {S.~T.}\ \bibnamefont {Rittenhouse}},
  \bibinfo {author} {\bibfnamefont {P.~F.}\ \bibnamefont {Weck}}, \bibinfo
  {author} {\bibfnamefont {H.~R.}\ \bibnamefont {Sadeghpour}},\ and\ \bibinfo
  {author} {\bibfnamefont {J.~P.}\ \bibnamefont {Shaffer}},\ }\bibfield
  {title} {\bibinfo {title} {Electric field cancellation on quartz by rb
  adsorbate-induced negative electron affinity},\ }\href
  {https://doi.org/10.1103/PhysRevLett.116.133201} {\bibfield  {journal}
  {\bibinfo  {journal} {Phys. Rev. Lett.}\ }\textbf {\bibinfo {volume} {116}},\
  \bibinfo {pages} {133201} (\bibinfo {year} {2016})}\BibitemShut {NoStop}%
\bibitem [{\citenamefont {Thiele}\ \emph {et~al.}(2014)\citenamefont {Thiele},
  \citenamefont {Filipp}, \citenamefont {Agner}, \citenamefont {Schmutz},
  \citenamefont {Deiglmayr}, \citenamefont {Stammeier}, \citenamefont
  {Allmendinger}, \citenamefont {Merkt},\ and\ \citenamefont
  {Wallraff}}]{Thiele_2014}%
  \BibitemOpen
  \bibfield  {author} {\bibinfo {author} {\bibfnamefont {T.}~\bibnamefont
  {Thiele}}, \bibinfo {author} {\bibfnamefont {S.}~\bibnamefont {Filipp}},
  \bibinfo {author} {\bibfnamefont {J.~A.}\ \bibnamefont {Agner}}, \bibinfo
  {author} {\bibfnamefont {H.}~\bibnamefont {Schmutz}}, \bibinfo {author}
  {\bibfnamefont {J.}~\bibnamefont {Deiglmayr}}, \bibinfo {author}
  {\bibfnamefont {M.}~\bibnamefont {Stammeier}}, \bibinfo {author}
  {\bibfnamefont {P.}~\bibnamefont {Allmendinger}}, \bibinfo {author}
  {\bibfnamefont {F.}~\bibnamefont {Merkt}},\ and\ \bibinfo {author}
  {\bibfnamefont {A.}~\bibnamefont {Wallraff}},\ }\bibfield  {title} {\bibinfo
  {title} {Manipulating rydberg atoms close to surfaces at cryogenic
  temperatures},\ }\href {https://doi.org/10.1103/PhysRevA.90.013414}
  {\bibfield  {journal} {\bibinfo  {journal} {Phys. Rev. A}\ }\textbf {\bibinfo
  {volume} {90}},\ \bibinfo {pages} {013414} (\bibinfo {year}
  {2014})}\BibitemShut {NoStop}%
\bibitem [{\citenamefont {Obrecht}\ \emph {et~al.}(2007)\citenamefont
  {Obrecht}, \citenamefont {Wild},\ and\ \citenamefont
  {Cornell}}]{Obrecht_2007}%
  \BibitemOpen
  \bibfield  {author} {\bibinfo {author} {\bibfnamefont {J.~M.}\ \bibnamefont
  {Obrecht}}, \bibinfo {author} {\bibfnamefont {R.~J.}\ \bibnamefont {Wild}},\
  and\ \bibinfo {author} {\bibfnamefont {E.~A.}\ \bibnamefont {Cornell}},\
  }\bibfield  {title} {\bibinfo {title} {Measuring electric fields from surface
  contaminants with neutral atoms},\ }\href
  {https://doi.org/10.1103/PhysRevA.75.062903} {\bibfield  {journal} {\bibinfo
  {journal} {Phys. Rev. A}\ }\textbf {\bibinfo {volume} {75}},\ \bibinfo
  {pages} {062903} (\bibinfo {year} {2007})}\BibitemShut {NoStop}%
\bibitem [{\citenamefont {Savard}\ \emph {et~al.}(1997)\citenamefont {Savard},
  \citenamefont {O'Hara},\ and\ \citenamefont
  {Thomas}}]{savard_laser-noise-induced_1997}%
  \BibitemOpen
  \bibfield  {author} {\bibinfo {author} {\bibfnamefont {T.~A.}\ \bibnamefont
  {Savard}}, \bibinfo {author} {\bibfnamefont {K.~M.}\ \bibnamefont {O'Hara}},\
  and\ \bibinfo {author} {\bibfnamefont {J.~E.}\ \bibnamefont {Thomas}},\
  }\bibfield  {title} {\bibinfo {title} {Laser-noise-induced heating in far-off
  resonance optical traps},\ }\href {https://doi.org/10.1103/PhysRevA.56.R1095}
  {\bibfield  {journal} {\bibinfo  {journal} {Physical Review A}\ }\textbf
  {\bibinfo {volume} {56}},\ \bibinfo {pages} {R1095} (\bibinfo {year}
  {1997})}\BibitemShut {NoStop}%
\bibitem [{\citenamefont {Leibrandt}\ \emph {et~al.}(2007)\citenamefont
  {Leibrandt}, \citenamefont {Yurke},\ and\ \citenamefont
  {Slusher}}]{Leibrandt_2007}%
  \BibitemOpen
  \bibfield  {author} {\bibinfo {author} {\bibfnamefont {D.}~\bibnamefont
  {Leibrandt}}, \bibinfo {author} {\bibfnamefont {B.}~\bibnamefont {Yurke}},\
  and\ \bibinfo {author} {\bibfnamefont {R.}~\bibnamefont {Slusher}},\
  }\bibfield  {title} {\bibinfo {title} {{Modeling ion trap thermal noise
  decoherence}},\ }\href {https://doi.org/10.26421/QIC7.1-2-2} {\bibfield
  {journal} {\bibinfo  {journal} {Quant. Inf. Comput.}\ }\textbf {\bibinfo
  {volume} {7}},\ \bibinfo {pages} {052} (\bibinfo {year} {2007})}\BibitemShut
  {NoStop}%
\end{thebibliography}%

\newpage
\section*{Supplementary Information}
\beginsupplement
\subsection*{Transport and experimental sequences}
The electrical potentials needed in conjunction with the magnetic field for trapping an ion at a particular location $\mbf{r}_0 = (x_0,y_0,z_0)$ are generated by applying suitable voltages to the 25 independent electrodes of the trap chip. Using the software package pytrans \cite{Mordini_pytrans}, which applies convex optimization techniques, we calculate voltages that approximate a cylindrically symmetric electrical potential 
\begin{equation}
    \phi = \frac{m\omega_z^2}{2e}\left((z-z_0)^2 - \frac{(x-x_0)^2 + (y-y_0)^2}{2}\right)\ ,
\end{equation}
where $m$ is the ion mass and $\omega_z$ is the axial motional frequency along the direction $z$ along the magnetic field. When performing ion transport, we calculate sequences of potentials such that the equilibrium position is varied in steps of \SI{1}{\micro\meter}. Interpolation of the resulting voltage sequence leads to a smooth translation of the trap and to low-excitation transport of the ion. We furthermore calculate sets of voltages which produce purely electric fields at any desired location, with no potential curvature. Such fields are used for counteracting stray electric fields. 

The speed of transport in our apparatus is severely restricted due to strong low-pass filtering applied to the trap electrodes, with the strongest filtering applied to the five strip electrodes closest to the ion with a cutoff frequency of approximately \SI{7}{\kilo\hertz}. The execution time of transport lies between \SI{1}{\ms} and \SI{8}{\ms} depending on the distance of displacement, corresponding to speeds between \SIrange{2}{6}{\centi\meter\per\second}. At these speeds, we do not observe excitation of the motional modes during transport beyond what is expected from the heating rates at the respective ion--electrode distances \cite{jain_penning_2024}.

\begin{figure*}[ht!]
\resizebox{510pt}{!}
{\includegraphics{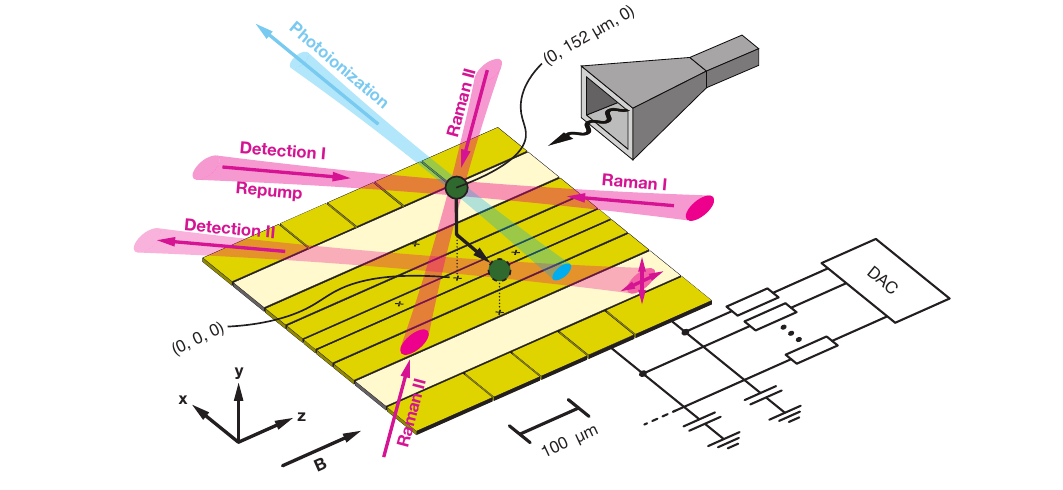}}
\caption{\label{fig:apparatus}\textbf{Trap chip and control fields. }Schematic view of the central region of the trap, including an indication of the coordinate system and its origin. A typical transport path of an ion from the ``cooling'' location at $\mbf{r}=(0,\SI{152}{\um}, 0)$ to a target location is shown. The propagation path of the laser beams used to cool, repump and detect the ion is drawn, as well as the Raman beams used for ground-state cooling and coherent manipulation. The path of the photo-ionization laser is shown as well as the microwave horn emitting radiation close to \SI{83.2}{\GHz} (not to scale). A schematic view of the digital-to-analog converter (DAC) producing the slowly varying trapping voltages is added, together with the RC low-pass filters placed directly below the trap. The outermost strip electrodes to which the axialization drive is applied are indicated in beige.}
\end{figure*}

Fig.~\ref{fig:apparatus} depicts the central region of the trap, together with illustrations of the laser beams and the microwave source used to control an ion. In this work, an experimental sequence always starts by Doppler cooling of the ion at the ``cooling'' location $(0, \SI{152}{\micro\meter}, 0)$ using the ``Detection I'' laser beam. Simultaneous to the cooling pulse, we apply an oscillating voltage at the bare cyclotron frequency $\omega_c=2\pi\times\SI{5.118}{\mega\hertz}$ to the axialization electrodes (colored in beige). The resulting coupling between the two radial modes of motion assists in cooling of the magnetron mode \cite{jain_penning_2024}. Subsequently, the ion is always prepared in the bright state using the ``Repump'' beam. A transport sequence is then applied which brings the ion to a target location. Any path can be taken in principle and we choose to always first move along the out-of-plane direction, bringing the ion to its target ion--electrode distance. Subsequently, we displace it in the $x$ or $z$ direction.

\subsubsection*{Sequence for measuring static electric fields}

When measuring static electric fields, the ion is first Doppler cooled and prepared in the bright state. Subsequently, a transport sequence brings it to a target location, while simultaneously applying correction fields. A secondary detection laser beam ``Detection II'' is aligned to the target position and fluorescence detection is performed. No mode-coupling axialization drive is applied during detection due to the ion potentially being far removed from the electric-field null of the axialization electrodes. The imaging system is adjusted such that the resulting fluorescence is focused onto the camera. The ion is subsequently transported back to the cooling location.

\subsubsection*{Sequence for measuring fluctuating electric fields}

A different experimental sequence is executed when performing heating rate measurements. Again starting at the ``cooling'' location to which all the necessary laser beams are aligned, the ion is first cooled to the motional ground state by successive Doppler cooling and continuous sideband cooling using the ``Raman I'' and ``Raman II'' lasers \cite{jain_penning_2024}. It is also prepared in the bright state. The ion is then transported to a target location where the displacement effect of the previously measured stray fields is canceled by applying correction fields. After a variable wait time, the ion is transported back to the starting location where a sideband probe pulse maps the motional excitation onto the spin, which is read out by fluorescence detection. In this measurement protocol, we additionally ensure that the motional frequency of the mode under investigation is equal at the starting and the target locations.

\subsubsection*{Sequence for measuring magnetic fields}

For the measurement of magnetic fields, the experimental sequence again starts by Doppler cooling, bright state preparation and transport to a target location while canceling stray fields. A microwave pulse is then performed using the horn antenna. When measuring the shifts of the static magnetic field, the frequency of the microwave radiation is scanned across the spin-flip resonance. In contrast, we fix the frequency to be on resonance when measuring the Rabi rate of the transition, while scanning the pulse duration. After subsequent transport back to the start, fluorescence detection is performed, revealing if the microwave drive has driven the ion into the dark state.

\subsection*{Imaging apparatus and calibration of the ion--electrode distance}

\begin{figure*}[ht!]
\resizebox{510pt}{!}{\includegraphics{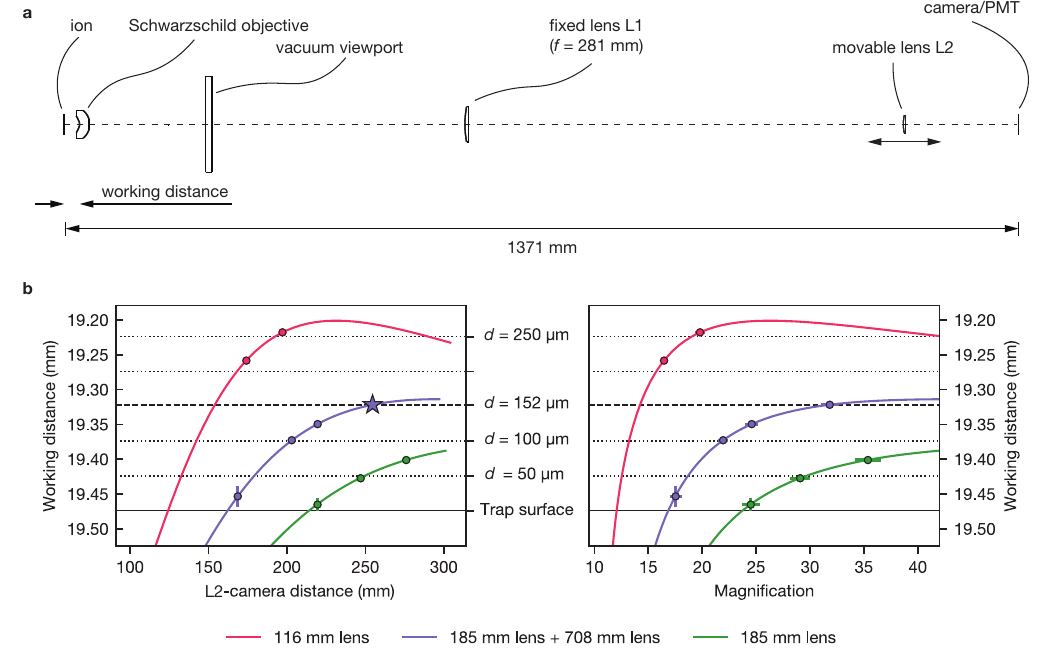}}
\caption{\label{fig:imaging_height_cal}\textbf{Imaging system and ion--electrode distance calibration. }\textbf{a. }Schematic view of the imaging system. From left to right, fluorescence light from the ion is collected by a Schwarzschild objective, passes a viewport and is then focused onto an EMCCD camera or a PMT by a fixed and a movable lens. \textbf{b. }Relationship between the position of the movable lens L2, the working distance at which the imaging focus occurs and the resulting magnification. The dashed line as well as the star marker indicate the reference working distance found by moving the ion onto the rf null at $d=\SI{152}{\micro\meter}$. The dotted lines show various ion-electrode distances relative to this reference, with the solid black line marking the expected location of the trap surface. Solid lines are ray-tracing simulations, where the three different focal lengths of L2 used in the experiment are chosen.}
\end{figure*}

We collect the ion fluorescence with a $0.55$ NA Schwarzschild objective. Two lenses focus the imaging light onto an electron-multiplying charge-coupled device (EMCCD) camera or photo-multiplier tube (PMT), where the second lens can be moved along the optical axis in order to bring ions into focus provided that the electrode--ion distance is smaller than approximately \SI{300}{\um}. A to-scale view of the imaging components is provided in fig.~\ref{fig:imaging_height_cal}a. Choosing from a variety of focal lengths of the movable lens L2 allows selecting the range of working distances across which an ion can be re-focused, as well as choosing the magnification. Maximizing the magnification is beneficial for measurements of static electric fields based on the detection of positional shifts. Ray-tracing simulations showing the relationship between the placement of the second lens and the resulting working distance and magnification are found in fig.~\ref{fig:imaging_height_cal}b for the three sets of lenses used in our apparatus.

We use the relationship between lens placement and the resulting position of the focus to verify the ion-electrode distance. 
First, an absolute position reference is obtained by moving the ion to a known ion--electrode distance, in this case to the field null of the axialization electrodes, which we simulate to be at a height of \SI{152}{\micro\meter} above the surface. If the ion is not at the field null, applying an oscillatory voltage at the cyclotron frequency leads to resonant excitation of the ion motion. A subsequent detection pulse registers a loss in fluorescence. We displace the ion until this excitation becomes minimal. Bringing the ion into focus and then measuring the position of L2 yields the working distance corresponding to an ion-electrode distance of $d=\SI{152}{\micro\meter}$, marked by a star symbol in fig.~\ref{fig:imaging_height_cal}b.

To calibrate other ion-electrode distances, we transport an ion to a desired nominal trap height as calculated using the electrostatic model of the trap, while correcting for stray fields. A pulse of the detection laser illuminates the ion and the resulting fluorescence is imaged. The movable lens L2 is placed such that the ion appears focused, with the resulting lens position translating to an estimate of the working distance as per the graphs in fig.~\ref{fig:imaging_height_cal}b. The shift compared to the reference at $d=\SI{152}{\micro\meter}$ yields an estimate for the position of the ion relative to the surface. As plotted in fig.~\ref{fig:imaging_height_cal}b, we carry out this procedure for nominal trap heights of \SIlist{50;75;100;125;152;210;256}{\um}. At each position, we choose a suitable focal length of the movable lens L2 such that the magnification is maximized. Measured ion-electrode distances are \SIlist{48(4);73(4);101(4);124(3);152(0);215(4);255(4)}{\micro\meter}, where error bars take into account uncertainties in the placement of all the imaging components, as well as an estimate of the error in finding the true focus. Two further data points are taken by illuminating the trap surface and bringing it into focus. Using this implementation of the imaging system, the ion cannot be imaged if placed at ion--electrode distances greater than $\approx \SI{300}{\um}$. Data presented in this work with the ion at greater heights is thus given without visual confirmation of the location, as well as without stray-field correction.

\subsection*{Measurement of static electric fields}
\label{supp:meas_el_field}

Due to the charged nature of trapped ions, external electric fields cause shifts in the position of an ion, which can be used to measure the field strength. We assume an ion to be trapped in the electrical potential $\phi_\mathrm{app}(\mbf{r})$ deliberately applied through the electrodes, superposed with an additional stray potential $\phi_\mathrm{stray}(\mbf{r})$, for example due to charging of the trap surface. Through electrostatic simulations, $\phi_\mathrm{app}(\mbf{r})$ is calculated to confine an ion at a desired position $\mbf{r}^*$ in the absence of stray electric fields, fulfilling $\nabla \phi_\mathrm{app}(\mbf{r}^*) = 0$. Adding the field due to $\phi_\mathrm{stray}$ will displace the ion from $\mbf{r}^*$, with the shift determined by the magnitude of the curvatures of $\phi_\mathrm{app}$.

By minimizing the position shifts when scaling the curvature of the applied potential, the stray field can be determined. To this end, we calculate sets of voltages that produce purely electric fields at a given position above the trap chip and use these to spatially translate the ion. We set an initial correction field $\mbf{E_1}=0$ and translate the ion to a target location where the applied trapping potential is $f_{\mathrm{ax,1}}^2\cdot\phi_\mathrm{app}$, with $f_{\mathrm{ax,1}}^2$ a scaling factor. Then, fluorescence detection is performed and the resulting position $\mbf{r_0}$ of the ion on the camera sensor recorded. Subsequently, we scale the applied potential by a different factor $f_{\mathrm{ax,2}}^2$ and additionally apply a correction field $\mbf{E_2}$, chosen such that the ion is again found at $\mbf{r_0}$. For both situations, the condition on the equilibrium position $\mbf{r_0}$ is given by

\begin{equation}\label{eq:stray_field_gradient}
    \begin{aligned}
        \nabla\phi_\mathrm{stray}(\mbf{r_0}) + (f_{\mathrm{ax,1}})^2\cdot\nabla\phi_\mathrm{app}(\mbf{r_0}) - \mbf{E_1} &= 0\\
        \nabla\phi_\mathrm{stray}(\mbf{r_0}) + (f_{\mathrm{ax,2}})^2\cdot\nabla\phi_\mathrm{app}(\mbf{r_0}) - \mbf{E_2}&= 0\ .
    \end{aligned}
\end{equation}

We take $\phi_\mathrm{app}$ to be a trapping potential resulting in an axial motional frequency of $\omega_z/2\pi = \SI{1}{\mega\hertz}$. Thus, $f_{\mathrm{ax,i}}$ can be interpreted as the nominal axial frequency in multiples of a megahertz. 

In this way, we can find both the stray field at position $\mbf{r_0}$ and the gradient of the applied potential by rewriting eq.~\ref{eq:stray_field_gradient} to find

\begin{equation}\label{eq:stray_field}
    \begin{aligned}
    \nabla \phi_\mathrm{app}(\mbf{r_0}) &= \frac{\mbf{E_2}- \mbf{E_1}}{f_{\mathrm{ax,2}}^2 - f_{\mathrm{ax,1}}^2}\\
    \nabla \phi_\mathrm{stray}(\mbf{r_0}) &= -\frac{f_{\mathrm{ax,1}}^2\mbf{E_2} - f_{\mathrm{ax,2}}^2\mbf{E_1}}{f_{\mathrm{ax,2}}^2-f_{\mathrm{ax,1}}^2}\ .
    \end{aligned}
\end{equation}

In a typical situation, one would find that the gradient of the applied potential $\nabla \phi_\mathrm{app}(\mbf{r_0})$ is not zero, meaning that the position $\mbf{r_0}$ does not coincide with the true trap center $\mbf{r^*}$. In this case, the measurement is restarted while incorporating the correction field by setting $\mbf{E_1}= -\nabla \phi_\mathrm{stray}(\mbf{r_0})$. By iterating this process, we eventually reach the position where $\mbf{E_2} = \mbf{E_1}$ and thus $\nabla \phi_\mathrm{app}(\mbf{r_0})=0$.

In this way, we both calibrate the positioning of the ion by removing any shifts due to stray fields as well as measuring said fields. Translating the ion over a grid of positions allows to create a map of the stray field from which the stray potential can be reconstructed.

\begin{figure*}[ht]
\resizebox{510pt}{!}{\includegraphics{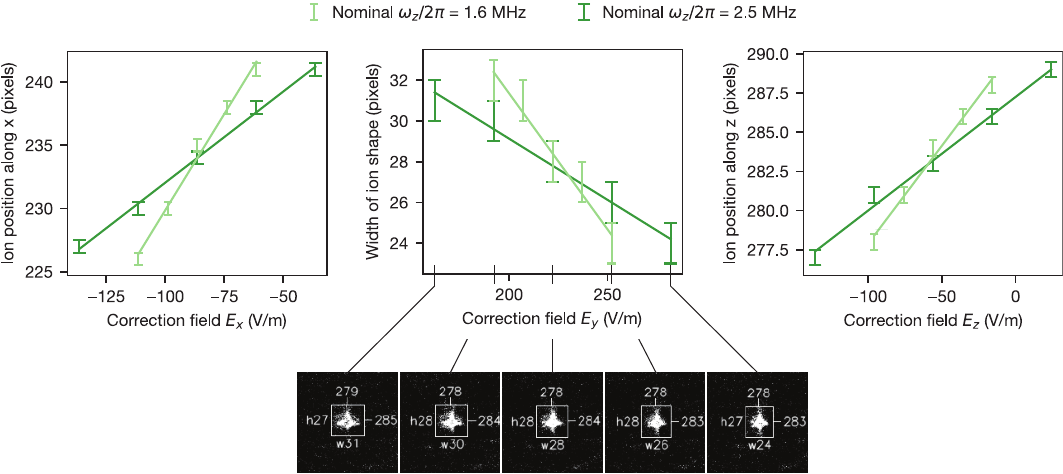}}
\caption{\label{fig:StrayFieldsIntercepts}\textbf{Measurement of the stray field at (0, 75 $\upmu$m, 0). }Position and width of the ion image on the camera sensor depending on applied correction fields. Light green indicates a nominal axial frequency of the applied potential of \SI{1.6}{\mega\hertz}, while \SI{2.5}{\mega\hertz} was used for data presented in dark green. Error bars are $\pm 0.5$ pixels for ion positions along $x$ and $z$. The middle panel shows the width of the ion image as the ion is moved through the focal plane by an out-of-plane correction field, where error bars of the width are $\pm 1$ pixel. The inset below shows the images obtained on the camera, demonstrating the change in aspect ratio. Solid lines are linear fits to the data.}
\end{figure*}

While the position of the ion in the directions perpendicular to the optical axis of the imaging apparatus, $x$ and $z$, can be easily read by determining the center coordinates of the point-spread function observed on the camera sensor, obtaining information about the position along the imaging axis is more challenging. Images of ions in our apparatus appear diamond-shaped due to aberrations of unknown origin. We find that as an ion is moved through the focal plane, the width of the point-spread function changes. Consequently, we use this as a substitute for the ion position in the out-of-plane direction.

As an example, fig.~\ref{fig:StrayFieldsIntercepts} shows a stray-field measurement at position $(0, \SI{75}{\micro\meter}, 0)$. The position and width of the point-spread function on the camera sensor are measured by fitting a contour to it using the OpenCV library \cite{opencv_library}, finding its center and extent. We apply trapping potentials with axial frequencies $\omega_z/2\pi$ of \SI{1.6}{\mega\hertz} or \SI{2.5}{\mega\hertz}, while varying the correction fields in the $x$, $y$ and $z$ directions. At some correction field $\mbf{E}_\mathrm{cross}$, a crossover point is observed where the ion position or shape is independent of the trapping potential strength. This signifies that this ion position corresponds to the true field null of the applied potential. Linear fits to the data yield $\mbf{E}_\mathrm{cross}$ and an error estimate. Since camera readings can only be specified in terms of integer pixels, error estimates for the $x$ and $z$-position are $\pm0.5$ pixels. For the out-of-plane direction, the error on the width of the point-spread function is set to $\pm 1$ pixel due to empirically observed fluctuations of the shape readout.

The sensitivity of this method is better understood by estimating it from the experimental parameters. Given the pixel size $p$ of the camera and the imaging magnification $M$, the minimal detectable position shift is $\pm p/(2M)$, which is approximately \SIrange{0.25}{0.5}{\micro\meter} in our apparatus, due to a pixel size of \SI{16}{\micro\meter} and magnifications in the range of \SIrange{15}{35}{}. The electric field that would cause such a shift in the axial direction is given by $\Delta E_z = mp\omega_z^2/(2eM)$, yielding an estimate for the smallest resolvable electric field in the axial direction. In the radial $x$-direction, the measurement is twice as sensitive due to the deconfining radial curvature being half the axially confining curvature, assuming a cylindrically symmetric trapping potential. Propagating the error through eq.~\ref{eq:stray_field} yields the uncertainty of the measured stray field. The sensitivity benefits from a low ion mass, low trapping potential strength, small camera pixels and a large magnification.

The measurement of out-of-plane fields relies on tracking de-focusing effects. The achievable precision depends on the depth-of-field of the imaging system, given by $\lambda/\mathrm{NA}^2$ \cite{Fundamentals_Light_Microscopy} in the limit of large NA. For $\lambda \simeq \SI{313}{\nano\meter}$ and an NA of 0.55, we obtain a depth of field of approximately \SI{1}{\micro\meter}.

To understand the impact of stray fields with magnitudes up to $\approx \SI{500}{\volt\per\meter}$ as found in this work, it is worth noting that the position of a \Beplus ion in a cylindrically symmetric trapping potential with an axial frequency of \SI{2.6}{\mega\hertz} shifts by \SI{20}{\micro\meter} in the axial direction and \SI{-40}{\micro\meter} in a radial direction if subjected to such a field strength.

\subsection*{Complete dataset of stray field measurements}

\begin{figure*}[ht]
\resizebox{510pt}{!}{\includegraphics{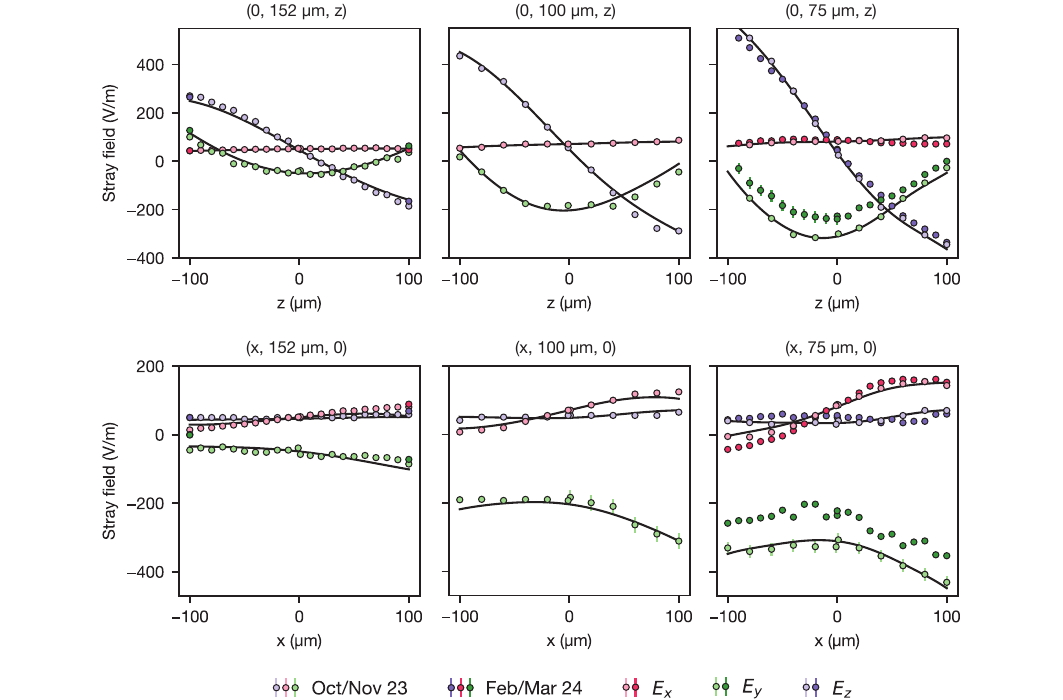}}
\caption{\label{fig:stray_fields_detail}\textbf{Detailed stray field data. }Full dataset of measured stray electric fields. The top row depicts fields measured at positions along the axial direction and at three ion-electrode distances, \SI{152}{\micro\meter}, \SI{100}{\micro\meter} and \SI{75}{\micro\meter}. The bottom row shows fields measured along the radial $x$-direction. The bulk of the data was taken in October and November 2023 (faded colors) and partially retaken in February and March 2024 (saturated colors) as a check of temporal stability. The data presented in a 3-d arrangement in fig.~\ref{fig:StrayFields}a. corresponds to the data taken in fall 2023. Solid black lines correspond to the electric fields produced by the fitted dipole density distribution shown in fig.~\ref{fig:StrayFields}b.}
\end{figure*}

The full dataset of measured stray fields as presented in fig.~\ref{fig:StrayFields}a can be found in fig.~\ref{fig:stray_fields_detail}. A part of the data was retaken about 4 months later as a check of temporal stability. We find that the out-of-plane component of the stray field increased by less than \SI{40}{\volt\per\meter} at a trap height of \SI{152}{\micro\meter} and by about \SI{100}{\volt\per\meter} at $d=\SI{75}{\micro\meter}$. The fields in the $x$ and $z$-directions appear to have remained largely constant despite continuous operation of the apparatus.

Error estimates on the measured static electric fields are obtained from linear fits to the ion position/width as a function of the correction field similar to the procedure shown in fig.~\ref{fig:StrayFieldsIntercepts}, yielding the error of the intercept. We perform such sensitivity scans at a few locations at each trap height used to measure static electric fields. This yields the error estimates for the data shown in fig.~\ref{fig:stray_fields_detail} and fig.~\ref{fig:StrayFields}a, resulting in estimated measurement inaccuracies in the range of \SIrange{2}{6}{\volt\per\meter} for fields along the $z$-direction, \SIrange{1}{3}{\volt\per\meter} for fields in the $x$-direction and \SIrange{10}{25}{\volt\per\meter} for out-of-plane fields.

\subsection*{Simulation of dipole densities on the trap surface}

We have measured static electric fields above the trap surface which are not caused by voltages applied externally to the trap electrodes. Such stray electric fields are common in ion traps \cite{Harlander_2010, Wang_2011, Lee_2024, auchterIndustriallyMicrofabricatedIon2022, Harter_2014, Doret_2012, Narayanan_2011} and are thought to be caused by several processes, such as laser-induced charging and deposition of neutral atoms on the surface. Assuming that the measured fields indeed originate from charge distributions on the trap surface, we attempt to find their spatial distribution.

Charging through laser light has been investigated in detail \cite{Harlander_2010, Wang_2011}, finding that the resulting electric fields take a dipolar form. It is assumed that the light excites electrons onto insulating patches on the trap surface, forming dipolar fields due to their interaction with the surface. A further mechanism leading to dipole fields is given by adsorbates on the trap surface distorting the electron distributions at the surface \cite{Leung_2003, McGuirk_2004, Tauschinsky_2010, Davtyan_2018, Sedlacek_2016, Thiele_2014}.

\begin{figure*}[ht]
\resizebox{420pt}{!}{\includegraphics{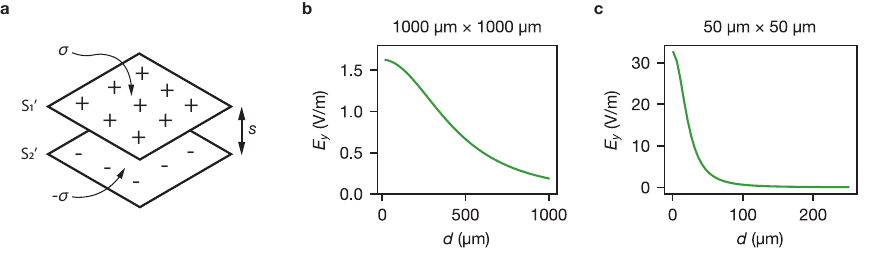}}
\caption{\label{fig:dipole_patch}\textbf{Dipole patch model. }\textbf{a. } Illustration of a rectangular patch with uniform dipole moment density. Two surfaces with opposite charge density $\pm\sigma$, with a distance $s$ separating them. \textbf{b. }The electric fields at a range of distances to the surface $d$ produced by a dipole patch with a dipole moment density $D=\SI{1e3}{\elementarycharge\angstrom\per\um\squared}$ and a size of $\SI{1}{\mm}\times\SI{1}{\mm}$. \textbf{c. }Out-of-plane fields produced by a $\SI{50}{\um}\times\SI{50}{\um}$ patch with the same $D$.}
\end{figure*}

We thus model the charge distribution on the trap surface by subdividing it into a number of small rectangular patches and assuming that each patch is covered in electric dipoles such that the dipole moment density across it is uniform. An illustration of a single patch is given in fig.~\ref{fig:dipole_patch}a, showing two surfaces $S_1^\prime$ and $S_2^\prime$ separated by $s$ and holding uniform but opposing charge densities $\pm\sigma$. The electrical potential generated by such a charge distribution is

\begin{align}
\Phi(\mbf{r}) = \frac{1}{4\uppi\epsilon_0}&\int_{S_1'} \frac{\sigma}{|\mbf{r}-\mbf{r'}|}dS_1'\\
- \frac{1}{4\uppi\epsilon_0}&\int_{S_2'} \frac{\sigma}{|\mbf{r}-\mbf{r'}|}dS_2'\ .
\end{align}

We consider the patch to hold a dipole moment density $D$, where we assume the two surfaces to be infinitesimally close while the charge density grows in such a way that

\begin{equation}
    \lim_{s\rightarrow0}(s\cdot\sigma) = D\ .
\end{equation}

The potential and electric fields at any position due to a dipole patch can be calculated numerically. As an example, fig.~\ref{fig:dipole_patch}b shows the out-of-plane electric field produced by an area of $\SI{1}{\mm}\times\SI{1}{\mm}$ covered with a dipole moment density of $D=\SI{1000}{\elementarycharge\angstrom\per\um\squared}$, while fig.~\ref{fig:dipole_patch}c shows the fields produced by a patch of $\SI{50}{\um}\times\SI{50}{\um}$ with the same value of $D$. Note that the electric field of a single dipole decays as $1/d^3$ with distance $d$, whereas fig.~\ref{fig:dipole_patch}b and c indicate a slower decay of the field produced by patches of finite size. To find a dipole distribution producing the measured stray fields, we segment a surface area of $\SI{1}{\mm}\times\SI{1}{\mm}$ situated at the trap center into $20\times20$ patches, each with an area of $\SI{50}{\um}\times\SI{50}{\um}$. By assigning a dipole moment density $D_i$ to the $i$-th patch, the overall electric field can be calculated. We perform a fit of this model to the data, finding the set of dipole densities $\{D_i\}$ which is plotted in fig.~\ref{fig:StrayFields}b. As this fit is under-constrained (240 measured field values vs. 400 free parameters), we add a regularization term $\sum_i D_i^2$ to the cost function to prevent over-fitting. Due to the resulting fit being biased towards a default state with $D_i=0$, we additionally allow the entire $\SI{1}{\mm}\times\SI{1}{\mm}$ area to hold a background dipole moment density. The fit yields minimal residuals when using $D_\mathrm{background}=\SI{180e3}{\elementarycharge\angstrom\per\um\squared}$. We hypothesize that this background stems from surface dipoles due to adsorbates on the surface. Without knowing the involved species, a realistic estimate for the dipole moment per adsorbed particle is in the range of \SIrange{1}{10}{\elementarycharge\angstrom} as found in \cite{Obrecht_2007}. The value of $D_\mathrm{background}$ then corresponds to a contaminant density of \SIrange{18e3}{180e3}{\per\um\squared}, well in line with previous investigations \cite{Davtyan_2018}.

Using a fit with patches of different sizes does not significantly alter the resulting dipole distribution, provided that the patches are chosen sufficiently small. Varying the total size of the considered surface area affects the resulting background dipole moment density but leaves the remaining features of the fit intact. Further investigation into the type and spatial distribution of the contaminants would be required to further refine the model. Alternatively, more data at across larger spatial ranges would be necessary.

\subsection*{Motional heating due to electric-field noise}

If $r_{\lambda 0}$ is the zero-point rms spread of the wave function, $\boldsymbol{\gamma}_{\lambda}$ is the normalized mode vector and $\hat{a}^{\dagger}_{\lambda} (\hat{a}_{\lambda})$ is the creation (annihilation) operator associated with the mode $\lambda$, we can write the position vector of the ion as 
\begin{equation}\label{eq:mode_decomp}
    \begin{aligned}
        \rvecop = -\i \sum_{\lambda}\sum_{\nu} r_{\lambda 0} \left[ \gamma^{\ast}_{\lambda \nu} \hat{a}^{\dagger}_{\lambda} - \gamma_{\lambda \nu} \hat{a}_{\lambda}\right] \mbf{e}_{\nu}.
    \end{aligned}
\end{equation}
Here the index $\nu$ refers to the axes ($x$, $y$ and $z$) defined by the coordinate system of the trap such that $\mbf{e}_{\nu}$ is the unit vector along that axis. We consider the case when the noisy electric field is along any general direction, so that $\mbf{E}(t) = E_x (t) \mbf{e}_x + E_y (t) \mbf{e}_y + E_z (t) \mbf{e}_z$. The resulting interaction Hamiltonian is 
\begin{equation}\label{eq:motheatinggen}
    \begin{aligned}
        \H^{\prime} 
        &= -e \rvecop \cdot \mbf{E} \left(\rvecop, \omega, t \right)\\
        &= \i e \sum_{\lambda}\sum_{\nu} r_{\lambda 0} E_{\nu} \left[ \gamma^{\ast}_{\lambda \nu} \hat{a}^{\dagger}_{\lambda} - \gamma_{\lambda \nu} \hat{a}_{\lambda}\right].
    \end{aligned}
\end{equation}
Ignoring the spatial variation of the electric field over the motion of the ion and assuming that the noisy field is perturbative compared to the trapping electric field, the heating rate $\dot{\ols{n}}_{\lambda}$ of each mode can be derived as \cite{savard_laser-noise-induced_1997}
\begin{equation}\label{eq:HeatingRateGen}
    \begin{aligned}
        \dot{\ols{n}}_{\lambda} 
        &= \frac{e^2 r^2_{\lambda 0}}{2 \hbar^2} \sum_{\nu} \left| \gamma_{\lambda \nu} \right|^2 S_{E_{\nu}} (\omega_{\lambda}),
    \end{aligned}
\end{equation}
where
\begin{equation}
    S_E(\omega) = 4 \int^{\infty}_0 \d \tau \langle E_{\nu}(t^{\prime}) E_{\nu}(t^{\prime} + \tau) \rangle \e^{\i \omega \tau}
\end{equation}
is the spectral noise density of the electric field along $\nu$.

For a Penning trap with cylindrically symmetric potential, equation \ref{eq:HeatingRateGen} simplifies to
\begin{equation}
    \begin{aligned}
        \dot{\ols{n}}_z &= \frac{e^2}{4 \hbar m \omega_z} S_E(\omega_z) 
    \end{aligned}
\end{equation}
for the axial mode, and 
\begin{equation}
    \begin{aligned}
        \dot{\ols{n}}_{\pm} 
        &= \frac{e^2}{4 \hbar m \left( \omega_+ - \omega_- \right)} S_E(\omega_{\pm})
    \end{aligned}
\end{equation}
for the radial modes.

\subsection*{Avoiding external noise sources}
\label{sec:noise_spikes}

\begin{figure*}[ht]
\centering
\resizebox{510pt}{!}{\includegraphics{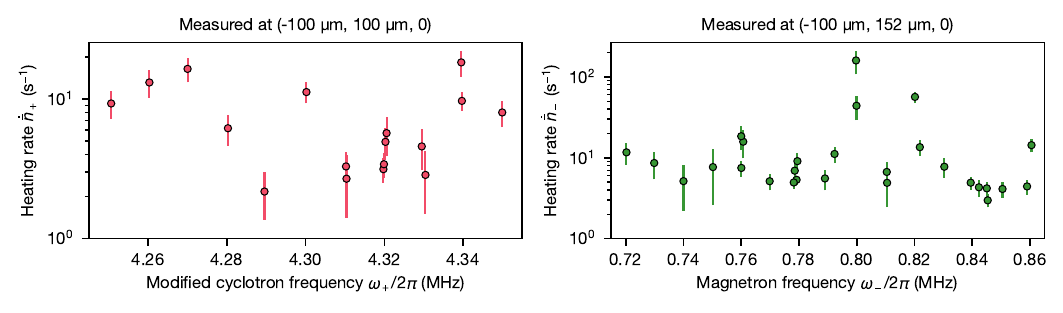}}
\caption{\label{fig:hr_discrete_noise}\textbf{External noise sources. }Heating rates of the cyclotron and magnetron modes as a function of the mode frequencies, revealing the presence of noise at discrete frequencies. Cyclotron heating rates are measured at a position of $(0, \SI{-100}{\micro\meter}, \SI{100}{\micro\meter})$, while the ion was placed at $(0, \SI{-100}{\micro\meter}, \SI{152}{\micro\meter})$ when measuring magnetron heating rates.}
\end{figure*}

To ensure comparability, all heating rates presented in this work are taken at the same motional frequencies, except when determining frequency scaling exponents. Axial heating rates are measured at $\omega_z = 2\pi\times\SI{2.6}{\mega\hertz}$ and we ensure the absence of external noise by taking comparison measurements with the electrodes detached from the voltage sources. For the radial modes, we instead perform heating rate measurements across a fine-grained grid of motional frequencies, shown in fig.~\ref{fig:hr_discrete_noise}. Noise at discrete frequencies likely caused by external technical equipment is detected in this way and can be avoided. For this purpose, we choose to measure all magnetron heating rates at $\omega_- = 2\pi\times\SI{0.845}{\mega\hertz}$ and cyclotron heating rates at $\omega_+ = 2\pi\times\SI{4.32}{\mega\hertz}$. Some of the responsible noise sources have been identified, such as a number of optical fiber amplifiers contributing broadband noise above $\simeq \SI{3}{\mega\hertz}$ or the EMCCD camera causing discrete noise at a number of frequencies below \SI{1}{\mega\hertz}, while the source of other noise spikes is unknown.

\subsection*{Modeling of the measured electric-field noise}

The distance-dependent heating rates shown in fig.~\ref{fig:hr}a exhibit a more complex behavior than the power-law scaling $d^{-\beta}$ expected from microscopic processes on the trap surface. Furthermore, we have shown in appendix~\ref{sec:noise_spikes} that the radial modes are affected by external noise sources at discrete frequencies, posing the question if also broadband noise is present. We thus consider that the measured heating rates are caused by multiple noise sources, causing a deviation from the expected scaling laws of surface noise. This hypothesis is further corroborated by the change in frequency-scaling depending on position presented in fig.~\ref{fig:hr}b.

\subsubsection*{Thermal noise}

One source of noise which is guaranteed to be present is Johnson noise \cite{Monroe1995}. This type of noise arises due to thermal fluctuations of charge carriers in resistive constituents of the circuitry attached to a trap electrode, such as resistors in filter networks or the bulk electrode material and wiring. We calculate its expected magnitude depending on position and frequency and infer the resulting heating rates of all motional modes. The spectral density of the voltage noise at electrode $i$ produced by this process is given by

\begin{equation}
\label{eq:johnson_noise}
    S_{\mathrm{V},i}(\omega, T) = 4k_\mathrm{B}T\cdot\operatorname{Re}\left(Z_i(\omega,T)\right)\ ,
\end{equation}

where $k_\mathrm{B}$ is the Boltzmann constant and $Z_i$ is the complex impedance seen by the electrode at a frequency $\omega$ and temperature $T$. Following \citeauthor{Leibrandt_2007}, $S_{\mathrm{V},i}$ is converted to an electric-field noise spectral density $S_{\mathrm{E_\nu},i}$ along the spatial direction $\nu$ and at the location $\mbf{r_0}$. To find this conversion, we calculate the electric field $E_{\nu,i}(\mbf{r_0})$ generated when applying a voltage $V_i$ to electrode $i$ and define a characteristic distance $d_{\nu,i}$ through

\begin{equation}
    E_{\nu,i}(\mbf{r_0}) = \frac{V_i}{d_{\nu,i}(\mbf{r_0})}\ .
\end{equation}

The spectral density at $\mbf{r_0}$ is then found by writing

\begin{equation}
    S_{E_\nu,i}(\omega,T, \mbf{r_0}) = \frac{S_{V,i}(\omega,T)}{(d_{\nu,i}(\mbf{r_0}))^2}\ .
\end{equation}

In our apparatus, the major contribution to the impedance causing Johnson noise is due to the last stage of filters placed on a PCB directly below the trap chip, held at a temperature of \SI{6.5}{\kelvin}. Each electrode is equipped with a resistor-capacitor circuit, with the strongest filtering applied to the five strip electrodes closest to the ion, using $R=\SI{1}{\kilo\ohm}$ and $C=\SI{22}{\nano\farad}$. The axially segmented control electrodes are equipped with filters consisting of $R=\SI{10}{\kilo\ohm}$ and $C=\SI{560}{\pico\farad}$. The axialization electrodes are equipped with the weakest filtering using a \SI{1}{\kilo\ohm}/\SI{560}{\pico\farad} stage, with the electrodes additionally co-wired after the filters. We include resistive elements within the connection between the filters and the trap electrodes, estimated to be no more than \SI{0.25}{\ohm}.

The noise spectral density in both the axial and radial direction is converted to heating rates $\dot{\bar{n}}_{\lambda,\ \mathrm{Js}}$ where $\lambda\in\{z,+,-\}$ denotes the motional mode. Fig.~\ref{fig:noise_models} shows the contributions to the total measured noise. Some striking features caused by the filter network and the electrode geometry are visible. Since only the axially segmented electrodes produce a non-negligible axial electric field, the axial Johnson noise decreases close to the surface. Furthermore, a local minimum in the radial Johnson noise is visible around $d=\SI{152}{\um}$ due to the two axialization electrodes being co-wired and exhibiting an electric-field null.

\subsubsection*{Correlated technical noise}

Motivated by the evidence of the radial modes being affected by external noise (see fig.~\ref{fig:hr_discrete_noise}), we model technical noise reaching the trap through electrical connections. Noise originating from the individual DAC channels is often considered as a source \cite{winelandExperimentalIssuesCoherent1998}. However, DAC-noise is filtered by the cryogenic filters described above as well as fourth-order Butterworth filters (\SI{10}{\kilo\hertz} cutoff) inserted between the DAC and the vacuum apparatus. The resulting noise suppression is so large that such uncorrelated DAC noise is an unlikely cause of the measured heating rates. 

Instead, we consider voltage fluctuations which are correlated across all electrodes. Such noise may arise due to electromagnetic pickup in the trap wiring or due to fluctuations between ground levels in the apparatus. We assume that voltage noise with a spectral density $S_{V,\mathrm{corr}}$ impinges equally on each wire leading to the trap, such that the noise passes through the final stage of cryogenic filters. We further assume for simplicity that the noise is in-phase across all wires and remains so after the filters.

The noise spectral density experienced by the ion in this situation is found by propagating the voltage noise through the trap filters. We find the electric field $E_\mathrm{corr}(\omega, \mbf{r_0})$ experienced by an ion at position $\mbf{r_0}$ when the set of voltages $V_i = V\cdot|T_i(\omega)|$ is applied to the electrodes. $V$ is the amplitude of a noise voltage common to all wires and $T_i$ denotes the transfer function of the trap filter of electrode $i$. The resulting characteristic distance

\begin{equation}
    d_{\nu,i}(\omega,\mbf{r_0}) = \frac{V}{E_{\mathrm{corr},\nu}(\omega,\mbf{r_0})}
\end{equation}

leads to the electric-field spectral density along $\nu\in\{x,y,z\}$

\begin{equation}
\label{eq:heating_rate_model_axial}
    S_{E_\nu, \mathrm{corr}}(\omega, \mbf{r_0}) = \frac{S_{V,\mathrm{corr}}}{(d_{\nu,i}(\omega,\mbf{r_0}))^2}\ .
\end{equation}

Note that this simple model of correlated noise can only explain radial noise. Due to the topology of the filter network, the symmetry of the trap and the assumption of in-phase noise, such fluctuations cannot produce axially polarized noise.

\subsubsection*{Overall models}

\begin{figure*}[ht]
\centering
\resizebox{510pt}{!}{\includegraphics{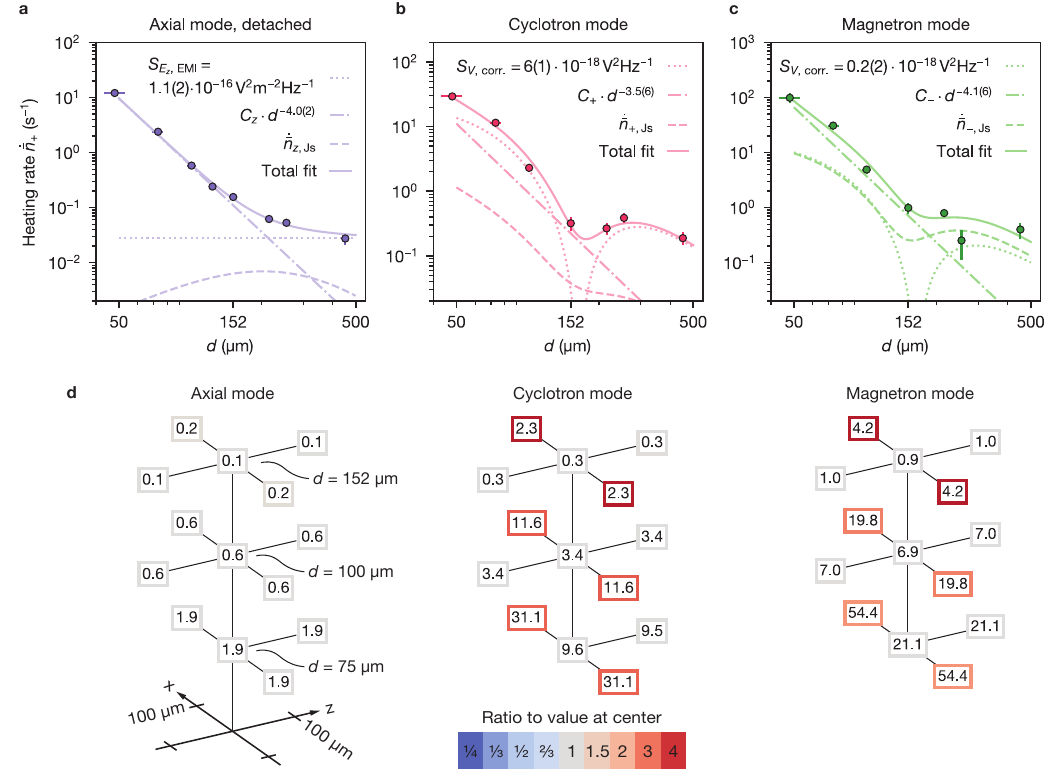}}
\caption{\label{fig:noise_models}\textbf{Models of the measured heating rates. }\textbf{a. }Heating rates of the axial mode taken with the trap electrodes detached, depending on the ion--electrode distance $d$ as presented in fig.~\ref{fig:hr}a. Overlaid is the fit of the model given in eq.~\ref{eq:heating_rate_model_axial} including the contributions by the individual sources of noise as described in the text. From the fit component with power-law scaling, assumed to be surface noise, a scaling exponent $\beta=4.0(2)$ is extracted. \textbf{b. } Distance-dependent heating rates of the cyclotron mode of motion as shown in fig.~\ref{fig:hr}a and fit of the model given in eq.~\ref{eq:heating_rate_model_radial}. Correlated voltage fluctuations of \SI{6(1)e-18}{\volt\squared\per\hertz} at the cyclotron frequency of \SI{4.32}{\MHz} are inferred to be impinging on the connections to the trap electrodes. The scaling exponent of the surface noise component is extracted to be $\beta=3.5(6)$. \textbf{c. } Magnetron heating rates as given in fig.~\ref{fig:hr}a and fit of the model given in eq.~\ref{eq:heating_rate_model_radial}. Correlated voltage noise of \SI{0.2(2)e-18}{\volt\squared\per\hertz} at the magnetron frequency of \SI{0.845}{\MHz} is inferred from the fit, as well as surface noise with a scaling exponent $\beta=4.1(6)$. \textbf{d. }Evaluation of the three fitted heating rate models on the grid of positions used for the measurement of heating rates in 3-d presented in fig.~\ref{fig:hr}d, plotted in the same style.}
\end{figure*}

We attempt to fit the radial heating rates of mode $\lambda\in\{+,-\}$ with the function

\begin{equation}
\label{eq:heating_rate_model_radial}
    \dot{\bar{n}}_\lambda(d) = \dot{\bar{n}}_{\lambda,\,\mathrm{Js}}(d) + C_\lambda\cdot d^{-\beta} + \dot{\bar{n}}_{\lambda,\,\mathrm{corr.}}(d)\ ,
\end{equation}

which includes Johnson noise and a term with power-law scaling of order $\beta$ meant to capture surface noise. Finally, we add a contribution $\dot{\bar{n}}_{\lambda,\,\mathrm{corr.}}(d)$ caused by correlated noise with voltage spectral density $S_{V,\mathrm{corr}}$. The free parameters of this model are $C_\lambda$, $\beta$ and $S_{V,\mathrm{corr}}$. The resulting fits to the cyclotron and magnetron data are shown in fig.~\ref{fig:noise_models}b and c. 

Focusing on the fit of the cyclotron heating rates, we observe that correlated technical noise is dominant for most of the considered range of distances, while Johnson noise contributes negligibly. Our model of correlated noise predicts no effect on the ion at a distance slightly above $d=\SI{152}{\um}$. Similar to the locally minimal level of Johnson noise at $d=\SI{152}{\um}$, this feature is caused by the fact that the two axialization electrodes are equipped with the weakest filters and thus carry the majority of the noise, leading to an effect similar to an electric-field null of two strip electrodes. The fit favors a power-law noise component with exponent $\beta=3.5(6)$, consistent with the theoretical expectation of $\beta=4$ and experimental evidence of noise above metallic surfaces. The magnetron mode is less afflicted by correlated noise, similar in magnitude to the expected Johnson noise. Surface noise appears to be the largest contribution to the fit for $d<\SI{152}{\um}$. Note that the inferred correlated voltage fluctuations are at a level around \SI{1e-18}{\volt\squared\per\hertz}, comparable to the noise pickup of a loop of wire with \SI{10}{\cm} diameter in an unshielded indoors environment \cite{brownnuttIontrapMeasurementsElectricfield2015}.

Correlated technical noise cannot generate axially polarized noise due to the assumption of all noise being in-phase across electrodes, as well as due to the symmetry of the trap and filter network. However, the measured axial heating rates deviate from power-law behavior and approach a constant value at increasing distances to the surface. When modeling the axial heating rates, we thus use the fit function 

\begin{equation}
\label{eq:heating_rate_model_ER}
    \dot{\bar{n}}_z(d) = \dot{\bar{n}}_{z,\,\mathrm{Js}}(d) + C_z\cdot d^{-\beta} + \dot{\bar{n}}_{z,\,\mathrm{EMI}}\ ,
\end{equation}

consisting of a term due to Johnson noise, surface noise with power-law distance scaling and a location-independent contribution $\dot{\bar{n}}_{z,\,\mathrm{EMI}}$. A noise source causing this last term could be direct electromagnetic interference (EMI) from the environment of the apparatus \cite{brownnuttIontrapMeasurementsElectricfield2015}. Fitting this model to the data yields a good fit when assuming EMI noise at a level of \SI{1.1(2)e-16}{\volt\squared\per\meter\squared\per\hertz}, with the result shown in fig.~\ref{fig:noise_models}a. Noise levels in an indoors environment can be six orders of magnitude higher \cite{brownnuttIontrapMeasurementsElectricfield2015}, however detailed simulations would be necessary to find the shielding factor of the experimental apparatus. At distances $d<\SI{152}{\um}$, noise with power-law scaling dominates the data and we extract $\beta=4.0(2)$, again consistent with expectations.

Finally, we evaluate the obtained noise models on the 3-d grid of positions used in fig.~\ref{fig:hr}d. The resulting heating rates are shown in fig.~\ref{fig:noise_models}d, where we have used the models with the same fit parameters found for the fits to the distance-dependent data shown in fig.~\ref{fig:noise_models}a-c. A comparison to the measured values as shown in fig.~\ref{fig:hr}d reveals that the models broadly recover the observed structure for the cyclotron and magnetron data, namely an increase in noise when displaced in either direction along the $x$-axis. This can be attributed to the reduced distance to the weakly filtered axialization electrodes. We take this agreement as further indication that technical noise of correlated nature is indeed present. 

\subsection*{Complete dataset of magnetic field measurements}

\begin{figure*}[ht]
\resizebox{510pt}{!}{\includegraphics{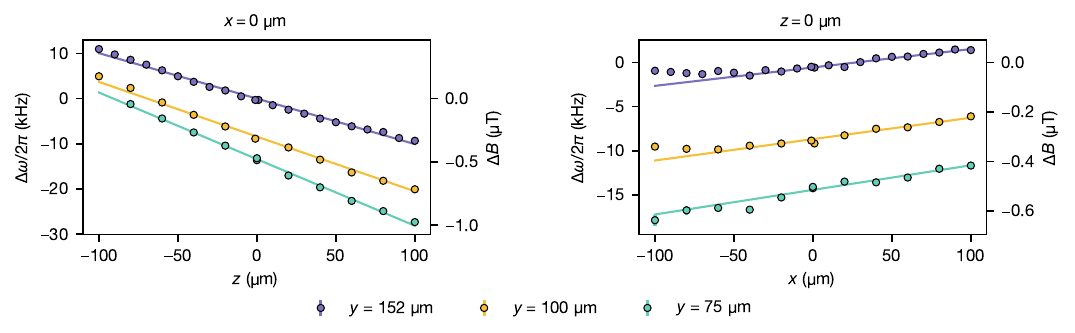}}
\caption{\label{fig:mag_gradient_detail}\textbf{Detailed measurements of the static magnetic field. }Full dataset of measured magnetic field shifts as a function of location. The left-hand panel shows data taken at locations $(0, y, z)$ for three ion-electrode distances $y$ and with the axial position $z$ ranging from \SIrange{-100}{100}{\micro\meter}. Shifts of the resonance frequency are relative to the measured carrier frequency at position $(0, \SI{152}{\micro\meter}, 0)$. The right-hand panel shows data taken at locations $(x, y, 0)$. Almost all error bars are smaller than the markers. Solid lines indicate linear fits to the data, yielding the magnetic field gradients as detailed in fig.~\ref{fig:mmw}.}
\end{figure*}

\begin{figure*}[ht]
\resizebox{510pt}{!}{\includegraphics{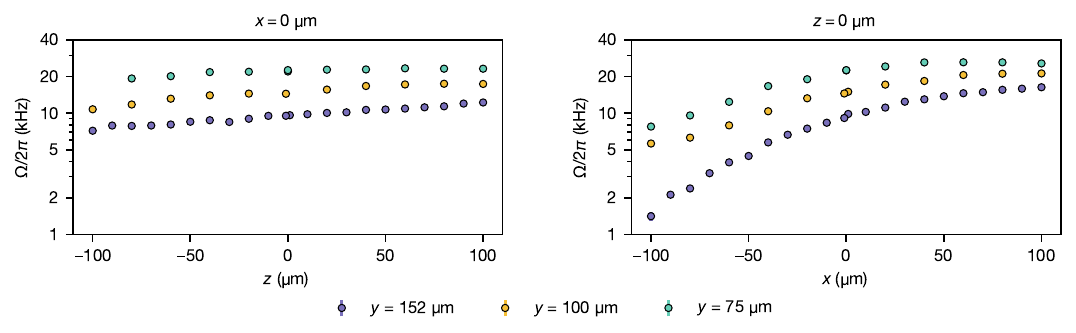}}
\caption{\label{fig:mmw_power_detail}\textbf{Detailed microwave Rabi rate measurements. }Full dataset of the measured Rabi rates as a function of location. The left-hand panel shows data taken at locations $(0, y, z)$ for three ion-electrode distances $y$ and with the axial position $z$ ranging from \SIrange{-100}{100}{\micro\meter}. The right-hand panel shows data taken at locations $(x, y, 0)$ for the same choices of $d$. Almost all error bars are smaller than the markers.}
\end{figure*}

The full set of magnetic-field gradient data as presented in fig.~\ref{fig:mmw}c is shown in fig.~\ref{fig:mag_gradient_detail}. Both the shift in the carrier resonance frequency $\Delta\omega$ and the corresponding magnetic-field shift $\Delta B$ are shown depending on position, relative to the values measured at position $(0, \SI{152}{\micro\meter}, 0)$. The displacement effects of stray electric fields has been corrected throughout these measurements. Fig.~\ref{fig:mmw_power_detail} shows the Rabi rates when driving the qubit transition with the microwave field near \SI{83.2}{\giga\hertz}, as presented in fig.~\ref{fig:mmw}b and d.

\end{document}